\documentclass[11pt]{article}
\usepackage[margin=0.75in]{geometry}
\usepackage[utf8]{inputenc}
\usepackage[english]{babel}
\usepackage[T1]{fontenc}
\usepackage{amsthm}
\usepackage{amsmath,amssymb}
\usepackage{graphicx}           
\usepackage{hyperref}
\usepackage{subcaption}
\usepackage{bigints}
\usepackage{relsize}
\usepackage{authblk}
\usepackage{tcolorbox}
\usepackage{thm-restate}
\hypersetup{
    colorlinks=true,
    linkcolor=blue,
    citecolor=magenta,
    filecolor=magenta,      
    urlcolor=blue,
    pdftitle={},
    pdfpagemode=FullScreen,
    }
\usepackage{soul}
\usepackage{mathtools}

\usepackage{tikz}
\usepackage{quantikz}
\usetikzlibrary{shapes}
\usetikzlibrary{decorations.pathreplacing,calligraphy}
\usepackage{xcolor}
\theoremstyle{definition}
\usepackage{float}
\makeatletter
\newlength\min@xx

\makeatother

\usepackage{lipsum}

\newtheorem{definition}{Definition}

\newcommand{\IK}[1]{{\color{blue}(IK) #1}}
\newcommand{\JP}[1]{{\color{red}(JP) #1}}

\begin{document}

\title{Complementarity and the unitarity of the black hole $S$-matrix}
\author[1]{Isaac H. Kim\thanks{\small \texttt{ikekim@ucdavis.edu}}}
\affil[1]{\textit{Department of Computer Science, UC Davis,  Davis, CA 95616, USA}}
\author[2,3]{\normalsize John Preskill}
\affil[2]{\textit{Institute for Quantum Information and Matter and Walter Burke Institute for Theoretical Physics,
California Institute of Technology,
Pasadena, CA 91125, USA}}
\affil[3]{
\textit{AWS Center for Quantum Computing,
Pasadena, CA 91125, USA}
}
\date{}
\maketitle
\begin{abstract}
Recently, Akers et al.~proposed a non-isometric holographic map from the interior of a black hole to its exterior. Within this model, we study properties of the black hole $S$-matrix, which are in principle accessible to observers who stay outside the black hole.  Specifically, we investigate a scenario in which an infalling agent interacts with radiation both outside and inside the black hole. Because the holographic map involves postselection, the unitarity of the $S$-matrix is not guaranteed in this scenario, but we find that unitarity is satisfied to very high precision if suitable conditions are met. If the internal black hole dynamics is described by a pseudorandom unitary transformation, and if the operations performed by the infaller have computational complexity scaling polynomially with the black hole entropy, then the $S$-matrix is unitary up to corrections that are superpolynomially small in the black hole entropy. Furthermore, while in principle quantum computation  assisted by postselection can be very powerful, we find under similar assumptions that the $S$-matrix of an evaporating black hole has polynomial computational complexity.
\end{abstract}

\section{Introduction}
\label{sec:introduction}

Recently, Akers et al.~proposed a solvable model of black hole complementarity~\cite{Akers2022}. This model describes the physics of an evaporating black hole from two perspectives: the effective picture and the fundamental picture. The effective picture describes physics seen by someone who falls into the black hole and experiences effective field theory on a weakly curved background. The fundamental picture describes physics seen by someone far outside the black hole, to whom the black hole is a fast scrambling quantum mechanical system interacting with its surroundings. To illuminate the properties of the black hole interior and to explain how information escapes from the interior, Akers et al.~proposed a non-isometric map from the effective picture to the fundamental picture, which can be realized as a random unitary transformation followed by postselection. They argued that this map has desirable features; for example, it follows that deviations from unitarity can be negligibly small in the effective description, and that entropy in the fundamental description can be computed by applying the quantum extremal surface formula ~\cite{Ryu2006,Faulkner2013,Engelhardt2015} in the effective description.

A central thesis of Akers et al.~is that the fundamental description can be precisely unitary, while the effective description is indistinguishable from unitary to observers who perform operations of subexponential complexity. This idea is appealing, but due to the postselection inherent in the non-isometric map, it is not obvious whether it is consistent. In particular, the authors of Ref.~\cite{Akers2022} considered observers who interact with either the exterior or the interior of the black hole, but not with both. Suppose instead that, after a black hole forms, an infalling agent interacts with radiation both outside and inside the black hole. When this black hole evaporates completely, how are the incoming and outgoing quantum states related? If the resulting process were flagrantly non-unitary due to the agent's actions combined with the non-isometric map relating the interior and exterior, this would cast doubt on the proposal in Ref.~\cite{Akers2022}. In this paper we investigate just such a scenario. 

In the context of the black hole final state proposal~\cite{Horowitz2004}, postselection can result in non-unitary and retro-causal effects ~\cite{Lloyd2014}. In the proposal of Ref.~\cite{Akers2022}, postselection is a feature of the holographic map rather than a physical process occurring in the bulk. Nevertheless, although the physical setting is different, similar technical issues arise as we will discuss. 

Aside from the issue of unitarity, the computational complexity of the black hole $S$-matrix is also of great interest. Aaronson showed that polynomial-time quantum computation, accompanied by postselection, can solve any problem in the complexity class PP~\cite{Aaronson2005}, which includes the well-known class NP. One wonders, then, whether the postselection postulated in Ref.~\cite{Akers2022} endows an evaporating black hole with unreasonable computational power, in violation of the quantum extended Chuch-Turing thesis~\cite{Deutsch1985,Susskind2020}. We investigate this issue as well. 

We formulate thought experiments in which the unitarity and the complexity of the black hole $S$-matrix can be assessed by a single observer living outside the black hole. First, following Ref.~\cite{Akers2022}, we model the black hole dynamics by a Haar-random unitary transformation, and conclude that the deviations from unitarity induced by the infaller are negligible. Then, more realistically, we assume that the black hole dynamics is pseudorandom~\cite{Ji2018,Kim2020}, and infer that deviations from unitarity are small if operations performed by the infaller have complexity polynomial in the black hole entropy. Moreover, under related assumptions, we show that the computational complexity of the black hole $S$-matrix remains polynomial in its entropy. Thus, from the perspective of an observer staying outside, the known laws of physics remain intact to high precision. That the proposal in Ref.~\cite{Akers2022} passes these nontrivial tests adds credence to its validity. 

Briefly summarized, our work goes beyond the analysis of Akers et al.~\cite{Akers2022} in three significant ways. First, we consider the consequences of infalling agents that interact with both the exterior and interior of the black hole. This modification could also be viewed as taking into account the interactions between left-moving and right-moving modes in the interior.\footnote{We thank Daniel Harlow for pointing this out.} Second, while the internal dynamics of the black hole was modeled as a Haar-random unitary in Ref.~\cite{Akers2022}, we reach similar conclusions by modeling the dynamics more realistically as a computationally efficient pseudorandom unitary transformation~\cite{Ji2018,Kim2020}. Third, we show that, even in the presence of infalling agents, the non-isometric map of Ref.~\cite{Akers2022} does not enable the black hole to perform operations of superpolynomial complexity. 

We should clarify what we mean by a computation of the ``black hole $S$-matrix'' in our work. The $S$-matrix is normally construed as the unitary transformation that maps asymptotic incoming states to asymptotic outgoing states. We are particularly interested in a nonasymptotic setting, in which a incoming state is mapped to the joint quantum state of a partially evaporated large black hole and the degrees of freedom outside the black hole horizon. According to the ``central dogma'' of black hole physics, this map, too, should be precisely unitary in the fundamental picture, and in an abuse of language we call this map the black hole $S$-matrix even though the outgoing state is not asymptotic. We study this regime because, since the black hole in the outgoing state has a large entropy, the effective picture of its interior should be trustworthy. 

In the case where we model the black hole's internal dynamics as a Haar-random unitary transformation, the computations we perform are similar to computations in Ref.~\cite{Akers2022}, but our motivation and interpretation are different. We consider how actions of an infalling agent in the effective description induce deviations from exact unitarity in the $S$-matrix. We find that these deviations are very small, and so infer that correspondingly small adjustments in the effective description can restore exact unitarity in the fundamental picture. If the deviations had been large instead, we would have expected correspondingly large deviations from unitarity in the asymptotic $S$-matrix as well, which would be detectable in principle by observers outside the black hole. In that case, we would have concluded that a more extensive modification of the holographic map is needed to ensure compatibility with the central dogma. In contrast, Akers et al.~investigated whether deviations from the predictions of quantum mechanics are detectable in the effective description, and concluded that these deviations are negligible if quantum states and observables in the effective description have reasonable computational complexity.

The rest of the paper is structured as follows. In Section~\ref{sec:setup}, we briefly review Ref.~\cite{Akers2022} and provide a simplified model that we use in our thought experiments, which are formulated in Section~\ref{sec:issues}. In Section~\ref{sec:adversarial} we describe an example to demonstrate that the unitarity of the black hole $S$-matrix cannot be guaranteed unconditionally in the presence of an infaller. This example, however, is not a realistic model of the black hole because the internal black hole dynamics is not described by a scrambling unitary. In Section~\ref{sec:average}, we go to the opposite extreme and show that, provided that the internal black hole dynamics is described by a Haar-random unitary, unitarity of the black hole $S$-matrix can be preserved with high precision and high probability, even in the presence of an infaller. The drawback of this model is that Haar-random unitaries generally have exponentially large complexity, while in contrast a realistic black hole is expected to be described by a unitary whose complexity is polynomial in its entropy. In Section~\ref{sec:pseudorandomness}, we show that unitarity of the $S$-matrix can also be ensured when the black hole dynamics has polynomial complexity. This is achieved under the assumption that the internal black hole dynamics is pseudorandom~\cite{Ji2018} and also that the infaller performs an operation with complexity scaling polynomially in the entropy of the 
black hole. Moreover, under similar assumptions, we show that the complexity of the black hole $S$-matrix is polynomial in the black hole entropy. In Section~\ref{sec:constraint_infaller}, we discuss some pathologies that may arise if we allow the infaller to perform operations of exponential complexity in the interior of the black hole. Specifically, such an infaller can induce a sizable violation of unitarity that can be observed from outside the black hole. We summarize our conclusions in Section~\ref{sec:conclusion}.

\section{Modeling the black hole interior}
\label{sec:setup}

\subsection{Effective and fundamental pictures}
We now present a simplified version of the model in Ref.~\cite{Akers2022}. In their model, there are two complementary pictures, referred to as the ``effective picture'' and the ``fundamental picture.'' The effective picture describes the viewpoint of observers who may encounter the interior of the black hole. The fundamental picture describes the viewpoint of observers who stay outside the black hole. 

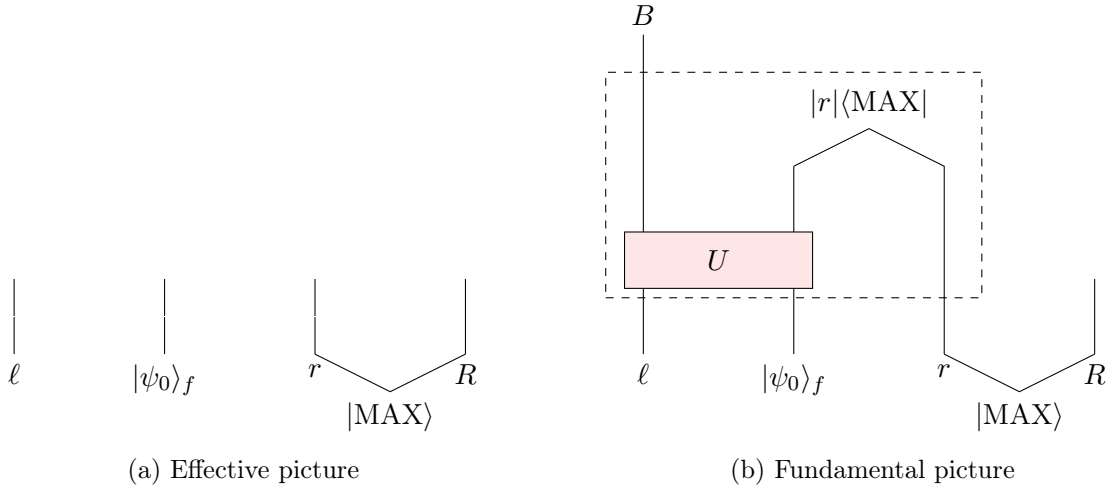
\begin{figure}[h]
\centering
\begin{subfigure}{0.4\textwidth}
    \begin{tikzpicture}
    \draw[] (0,0) -- (0,1);
    \draw[] (2,0) -- (2,1);
    \draw[] (4,0) -- (4,1);
    \draw[] (6,0) -- (6,1);
    \draw[] (4,0) -- (5, -0.5) -- (6,0);
    \draw[white] (-0.5, 0.5) -- (4.5, 0.5) -- (4.5, 4) -- (-0.5, 4) -- cycle;

    \node[above] at (3, 3) {\color{white} $\sqrt{|r|} \langle \text{MAX}|$};
    \node[below] at (0,0) {$\ell$};
    \node[below] at (2,0) {$|\psi_0\rangle_f$};
    \node[below] at (4,0) {$r$};
    \node[below] at (6,0) {$R$};
    \node[below] at (5,-0.5) {$|\text{MAX}\rangle$};
    \node[above] at (0, 4.25) {\color{white}$B$};
    \end{tikzpicture}
    \caption{Effective picture}
\end{subfigure}
\hspace{1cm}
\begin{subfigure}{0.4\textwidth}
    \begin{tikzpicture}
    \draw[] (0,0) -- (0,1);
    \draw[] (2,0) -- (2,1);
    \draw[] (4,0) -- (4,1);
    \draw[] (6,0) -- (6,1);
    \draw[] (4,0) -- (5, -0.5) -- (6,0);
    \draw[] (0, 1.5) -- (0, 4.25);
    \draw[] (2, 1.5) -- (2, 2.5);
    \draw[] (2, 2.5) -- (3, 3) -- (4, 2.5) -- (4,1);
    
    \draw[fill=red!10!white] (-0.25, 0.875) -- (2.25, 0.875) -- (2.25, 1.625) -- (-0.25, 1.625) -- cycle;
    
    \draw[dashed] (-0.5, 0.75) -- (4.5, 0.75) -- (4.5, 3.75) -- (-0.5, 3.75) -- cycle;
    
    \node[above] at (0, 4.25) {$B$};
    \node[above] at (3, 3) {$|r| \langle \text{MAX}|$};
    \node[] at (1,1.25) {$U$};
    \node[below] at (0,0) {$\ell$};
    \node[below] at (2,0) {$|\psi_0\rangle_f$};
    \node[below] at (4,0) {$r$};
    \node[below] at (6,0) {$R$};
    \node[below] at (5,-0.5) {$|\text{MAX}\rangle$};
    \end{tikzpicture}
    \caption{Fundamental picture}
\end{subfigure}
    \caption{A simplified version of the model in Ref.~\cite{Akers2022}. The part enclosed by dashed lines is the holograhic map $V_H$.}
    \label{fig:model_no_observer}
\end{figure}

Let us discuss the structure of the Hilbert space in these two different pictures. In the effective picture (see Fig.~\ref{fig:model_no_observer}a), the Hilbert space has the structure
\begin{equation}
\mathcal{H}_{\ell} \otimes \mathcal{H}_r \otimes \mathcal{H}_f \otimes \mathcal{H}_R
\end{equation}
where $\ell$ and $r$ represent the left-moving and the right-moving modes in the black hole interior,\footnote{These are the modes that move in the radial direction either inwardly (left) or outwardly (right).} 
and $R$ is a radiation reservoir outside the black hole. The system $f$ accounts for additional fixed degrees of freedom which do not play an essential role in our analysis. We may think of $\ell$ as keeping track of the infalling matter that gravitationally collapses to form the black hole, as well as additional matter that might have fallen in while the black hole was evaporating. In this effective picture, the Hawking radiation arises from entanglement between the right-moving interior modes in $r$ and the exterior radiation modes in $R$, as captured in Fig.~\ref{fig:model_no_observer}a where $|\textrm{MAX}\rangle$ denotes a maximally entangled state of $rR$.

In the fundamental picture, we have 
\begin{equation}
\mathcal{H}_B \otimes \mathcal{H}_R,
\end{equation}
where $R$ is again the exterior radiation reservoir and $B$ describes the microscopic state of the black hole as viewed from outside. Hence, its dimension $|B|$ is given by $e^{S_B}$, where $S_B$ is the black hole's Bekenstein-Hawking entropy. 

A puzzling feature of black hole physics is that in the effective picture, as the black hole evaporates, the Hilbert space dimension $|\ell r f|$ of the black hole's interior grows monotonically, as does the entanglement entropy of $rR$. Meanwhile, the dimension $|B|$ of the black hole in the fundamental description decreases, unless additional matter falls in during the evaporation process. Eventually, the two pictures conflict when the entanglement entropy of $rR$ in the effective picture exceeds the black hole entropy $S_B$ in the fundamental description. Akers et al. proposed to resolve this conflict by defining an appropriate dictionary (a ``holographic map'') relating the internal black hole degrees of freedom in the effective picture to the black hole microstates in the fundamental description. 

This map $V_H:  \mathcal{H}_\ell \otimes \mathcal{H}_{f}\otimes \mathcal{H}_r \to \mathcal{H}_B$, depicted inside the dotted line shown in Fig.~\ref{fig:model_no_observer}b, is defined as
\begin{equation}
V_H = (I_B \otimes |r|\langle \text{MAX}|_{r'r}) U,
\end{equation}
where $U: \mathcal{H}_{\ell} \otimes \mathcal{H}_f \to \mathcal{H}_B \otimes \mathcal{H}_{r'}$ describes the scrambling unitary dynamics of the black hole. Here $|r| = \dim (\mathcal{H}_r)$, $r'$ is an auxiliary Hilbert space whose dimension matches that of $r$, and $|\text{MAX}\rangle_{r'r}$ denotes the maximally entangled state of $r'$ and $r$. Note that, upon applying the holographic map to a state in the effective picture, we may interpret $U$ as a unitary process in the fundamental picture mapping an infalling state of matter to the black hole $B$ and its emitted Hawking radiation $R$:
\begin{equation}
\centering
    \begin{tikzpicture}[baseline={([yshift=-.5ex]current bounding box.center)}]
    \draw[] (0,0) -- (0,1);
    \draw[] (2,0) -- (2,1);
    \draw[] (4,0) -- (4,1);
    \draw[] (6,0) -- (6,1);
    \draw[] (4,0) -- (5, -0.5) -- (6,0);
    \draw[] (0, 1.5) -- (0, 2.5);
    \draw[] (2, 1.5) -- (2, 2.5);
    \draw[] (2, 2.5) -- (3, 3) -- (4, 2.5) -- (4,1);
    
    \draw[fill=red!10!white] (-0.25, 0.875) -- (2.25, 0.875) -- (2.25, 1.625) -- (-0.25, 1.625) -- cycle;
    
    \node[above] at (0, 2.5) {$B$};
    \node[above] at (3, 3) {$|r| \langle \text{MAX}|$};
    \node[] at (1,1.25) {$U$};
    \node[below] at (0,0) {$\ell$};
    \node[below] at (2,0) {$|\psi_0\rangle_f$};
    \node[below] at (4,0) {$r$};
    \node[below] at (6,0) {$R$};
    \node[below] at (5,-0.5) {$|\text{MAX}\rangle$};
    
    \node[] at (7.5, 1.75) {$=$};
    \begin{scope}[xshift= 10cm]
    \draw[] (0,0) -- (0,1);
    \draw[] (2,0) -- (2,1);
    \draw[] (0, 1.5) -- (0, 2.5);
    \draw[] (2, 1.5) -- (2, 2.5);
    
    \draw[fill=red!10!white] (-0.25, 0.875) -- (2.25, 0.875) -- (2.25, 1.625) -- (-0.25, 1.625) -- cycle;
    
    \draw[white, dashed] (-0.5, 0.5) -- (4.5, 0.5) -- (4.5, 4) -- (-0.5, 4) -- cycle;
    
    \node[above] at (0, 2.5) {$B$};   
    \node[above] at (2, 2.5) {$R$};
    \node[] at (1,1.25) {$U$};
    \node[below] at (0,0) {$\ell$};
    \node[below] at (2,0) {$|\psi_0\rangle_f$};
    \end{scope}
    \end{tikzpicture}.
    \label{eq:effective_to_fundamental_noinfaller}
\end{equation}

A crucial feature of the holographic map is that it is non-isometric, capable of mapping a very-high-dimensional black hole interior in the effective description to a much-lower-dimensional black hole system $B$ in the fundamental description. This reduction in dimension is achieved by projecting the state of $r'r$ to a particular maximally entangled state. We may interpret this projection as an orthogonal measurement of $r'r$ in a maximally entangled basis, followed by postselection onto one particular outcome of the measurement. In effect, this postselected measurement outcome teleports quantum information from the black hole interior into the radiation system $R$, providing an intuitively appealing explanation for how information escapes from inside the black hole. If such a measurement were really performed, the desired outcome would occur with probability $1/|r|^2$, and the factor of $|r|$ in the map $V_H$ ensures that the output of the map is a properly normalized state. Importantly, this postsection is not to be regarded as an actual physical process occurring inside the black hole; instead it is a property of the holographic map that reconciles the effective and fundamantal descriptions. 

Akers et al. considered a more refined model of black hole dynamics, in which $U$ is decomposed as a sequence of many unitary transformations, each representing the evolution of the black hole during a short interval of time~\cite{Akers2022}. For our purposes this refinement is not needed, so we have simplified their model by replacing the product of many unitary transformations by the single unitary $U$. Though the model is far from a fully realistic description of black hole dynamics, Akers et al.~argue persuasively that it captures relevant features of the realistic case; see Ref.~\cite{Akers2022} for a detailed discussion.

\subsection{Testing the model}

The premise underlying the model of Ref.~\cite{Akers2022} is that evolution in the fundamental picture is precisely unitary, what has been called the ``central dogma'' of black hole physics~\cite{almheiri2021entropy}. The model aims to reconcile this central dogma with the assertion that effective quantum field theory accurately describes the experience of an observer who falls into a black hole and interacts with its interior (at least until the observer reaches regions of high spacetime curvature close to the singularity). 

Our goal in this paper is to test the robustness of the conclusions of Ref.~\cite{Akers2022} under well-motivated modifications to the effective picture beyond those explicitly considered in \cite{Akers2022}. Specifically, we will investigate deviations from exact unitarity that arise when the holographic map is applied to the modified effective picture. To be clear, if large deviations from unitarity were found, we would regard this not as evidence against the central dogma, but rather as a reason to question the modified effective picture, the holographic map, or both. 

We will see that reasonable modifications to the effective picture map to deviations from unitarity in the fundamental picture which are superpolynomially small in the black hole entropy. Our attitude is that such small deviations are acceptable, under the presumption that a minuscule tweak in the holographic map could restore exact unitarity in the fundamental picture. On the other hand, if significantly larger deviations from unitarity had been found under the same conditions, we would have concluded that a more extensive overhaul of the holographic map may be needed. 

\subsection{Interactions of an infaller with the radiation}
\label{sec:infaller}

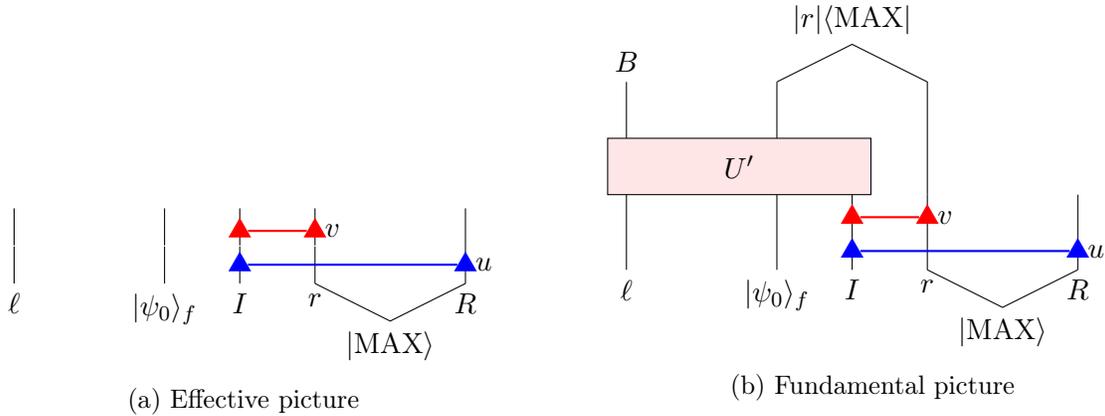
\begin{figure}[h]
    \centering
    \begin{subfigure}{0.4\textwidth}
    \begin{tikzpicture}[triangle/.style = {fill=blue!20, regular polygon, regular polygon sides=3 },
    node rotated/.style = {rotate=180},
    border rotated/.style = {shape border rotate=180}]
    \draw[] (0,0) -- (0,1);
    \draw[] (2,0) -- (2,1);
    \draw[] (4,0) -- (4,1);
    \draw[] (6,0) -- (6,1);
    \draw[] (4,0) -- (5, -0.5) -- (6,0);
    \draw[] (3,0) -- (3,1);
    \draw[white] (-0.5, 0.5) -- (4.5, 0.5) -- (4.5, 4) -- (-0.5, 4) -- cycle;

    \node[fill=blue, regular polygon, regular polygon sides=3, minimum size=0.35cm, inner sep=0pt] (u1) at (3, 0.25) {};
    \node[fill=blue, regular polygon, regular polygon sides=3, minimum size=0.35cm, inner sep=0pt] (u2) at (6, 0.25) {};
    \node[fill=red, regular polygon, regular polygon sides=3, minimum size=0.35cm, inner sep=0pt] (v1) at (3, 0.7) {};
    \node[fill=red, regular polygon, regular polygon sides=3, minimum size=0.35cm, inner sep=0pt] (v2) at (4, 0.7) {};
    
    \node[right] at (u2) {$u$};
    \node[right] at (v2) {$v$};
    
    \draw[blue, thick] (u1) -- (u2);
    \draw[red, thick] (v1) -- (v2);

    \node[above] at (3, 3) {\color{white} $\sqrt{|r|} \langle \text{MAX}|$};
    \node[below] at (0,0) {$\ell$};
    \node[below] at (2,0) {$|\psi_0\rangle_f$};
    \node[below] at (3,0) {$I$};
    \node[below] at (4,0) {$r$};
    \node[below] at (6,0) {$R$};
    \node[below] at (5,-0.5) {$|\text{MAX}\rangle$};
    \end{tikzpicture}
    \caption{Effective picture}
    \label{fig:effective_with_observer}
\end{subfigure}
\hspace{1cm}
\begin{subfigure}{0.4\textwidth}
\begin{tikzpicture}
    \draw[] (0,0) -- (0,1);
    \draw[] (2,0) -- (2,1);
    \draw[] (4,0) -- (4,1);
    \draw[] (6,0) -- (6,1);
    \draw[] (4,0) -- (5, -0.5) -- (6,0);
    \draw[] (0, 1.5) -- (0, 2.5);
    \draw[] (2, 1.5) -- (2, 2.5);
    \draw[] (2, 2.5) -- (3, 3) -- (4, 2.5) -- (4,1);
    \draw[] (3,0) -- (3,1);
    
    \draw[fill=red!10!white] (-0.25, 1) -- (3.25, 1) -- (3.25, 1.75) -- (-0.25, 1.75) -- cycle;
    
     \node[fill=blue, regular polygon, regular polygon sides=3, minimum size=0.35cm, inner sep=0pt] (u1) at (3, 0.25) {};
    \node[fill=blue, regular polygon, regular polygon sides=3, minimum size=0.35cm, inner sep=0pt] (u2) at (6, 0.25) {};
    \node[fill=red, regular polygon, regular polygon sides=3, minimum size=0.35cm, inner sep=0pt] (v1) at (3, 0.7) {};
    \node[fill=red, regular polygon, regular polygon sides=3, minimum size=0.35cm, inner sep=0pt] (v2) at (4, 0.7) {};
    
    \node[right] at (u2) {$u$};
    \node[right] at (v2) {$v$};
    
    \draw[blue, thick] (u1) -- (u2);
    \draw[red, thick] (v1) -- (v2);
    
    \node[above] at (0, 2.5) {$B$};
    \node[above] at (3, 3) {$|r| \langle \text{MAX}|$};
    \node[] at (1.5,1.375) {$U'$};
    \node[below] at (0,0) {$\ell$};
    \node[below] at (2,0) {$|\psi_0\rangle_f$};
    \node[below] at (3,0) {$I$};
    \node[below] at (4,0) {$r$};
    \node[below] at (6,0) {$R$};
    \node[below] at (5,-0.5) {$|\text{MAX}\rangle$};
    \end{tikzpicture}
    \caption{Fundamental picture}
    \label{fig:fundamental_with_observer}
\end{subfigure}
    \caption{A model with an infaller. The infaller can interact with the exterior (blue gate) or interior (red gate) radiation modes.}
    \label{fig:model_with_observer}
\end{figure}

Concretely, we will modify the model described in Fig.~\ref{fig:model_no_observer} by adding a party who falls into the black hole (an ``infaller''). Until the infaller closely approaches the black hole singularity (where the semiclassical effective picture breaks down) we expect the interactions of the infaller with other degrees of freedom to be well-approximated by effective quantum field theory on a semiclassical spacetime background. We accommodate the infaller by adding another tensor subfactor to the Hilbert space, denoted $I$. Thus, the Hilbert space in the effective picture enlarges to become
\begin{equation}
\mathcal{H} = \mathcal{H}_\ell \otimes \mathcal{H}_r \otimes \mathcal{H}_f \otimes \mathcal{H}_I \otimes \mathcal{H}_R.
\end{equation}
The infaller interacts unitarily first with $R$ while still ouside the black hole, and then with $r$ after crossing the event horizon. We call the corresponding unitary transformations $u$ acting on $IR$ and $v$ acting on $Ir$; see Fig.~\ref{fig:effective_with_observer}.

How should we map this effective picture to the fundamental picture? The answer depends on whether the infaller has already entered the black hole or not. If not, we can assume that the scrambling unitary $U$ --- a simple model of black hole's internal dynamics --- acts trivially on the infaller. On the other hand, if the infaller has entered the black hole, the scrambling unitary should act on the infaller as well. We thus modify the model by replacing $U$ by a new unitary $U'$ that acts on $I$ as well as $\ell f$, as shown in Fig.~\ref{fig:fundamental_with_observer}. 

Throughout this paper, we will be agnostic about precisely how the infaller's interaction with $R$ and $r$ modifies the infaller's own memory; that is why we speak of an ``infaller'' rather than an ``infalling observer.'' Under the holographic map, the infaller's memory gets scrambled so that from the fundamental viewpoint the information encoded in this memory winds up shared between $B$ and $R$, just like the rest of the quantum information that fell into the black hole. The important thing is that the infaller is a system that interacts with $R$ and then $r$, and we are interested in how those interactions impact the black hole's evolution as seen by an agent who stays outside the black hole, as we describe in the following section. 

\subsection{A challenge to unitarity}

A potential issue with the model in Fig.~\ref{fig:model_with_observer} is that this model may not be unitary in the fundamental picture. To see why, let us ``straighten out'' some of the edges:
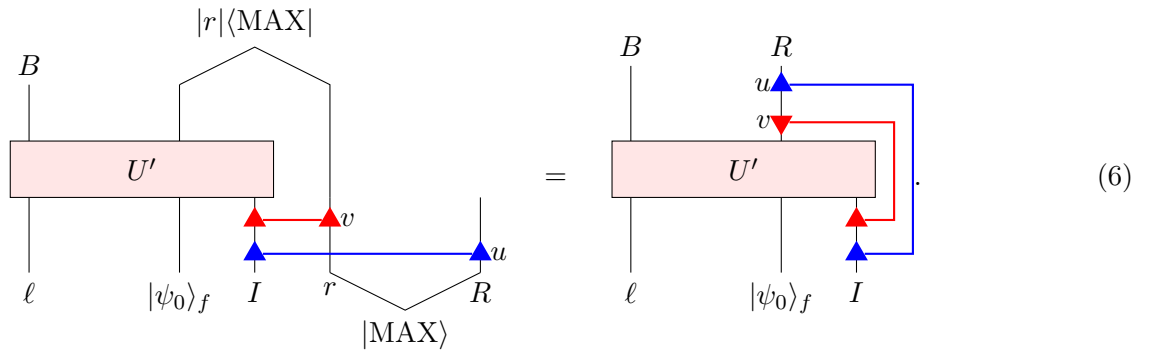
\begin{equation}
\begin{tikzpicture}[baseline={([yshift=-.5ex]current bounding box.center)}]
    \draw[] (0,0) -- (0,1);
    \draw[] (2,0) -- (2,1);
    \draw[] (4,0) -- (4,1);
    \draw[] (6,0) -- (6,1);
    \draw[] (4,0) -- (5, -0.5) -- (6,0);
    \draw[] (0, 1.5) -- (0, 2.5);
    \draw[] (2, 1.5) -- (2, 2.5);
    \draw[] (2, 2.5) -- (3, 3) -- (4, 2.5) -- (4,1);
    \draw[] (3,0) -- (3,1);
    
    \draw[fill=red!10!white] (-0.25, 1) -- (3.25, 1) -- (3.25, 1.75) -- (-0.25, 1.75) -- cycle;
    
     \node[fill=blue, regular polygon, regular polygon sides=3, minimum size=0.35cm, inner sep=0pt] (u1) at (3, 0.25) {};
    \node[fill=blue, regular polygon, regular polygon sides=3, minimum size=0.35cm, inner sep=0pt] (u2) at (6, 0.25) {};
    \node[fill=red, regular polygon, regular polygon sides=3, minimum size=0.35cm, inner sep=0pt] (v1) at (3, 0.7) {};
    \node[fill=red, regular polygon, regular polygon sides=3, minimum size=0.35cm, inner sep=0pt] (v2) at (4, 0.7) {};
    
    \node[right] at (u2) {$u$};
    \node[right] at (v2) {$v$};
    
    \draw[blue, thick] (u1) -- (u2);
    \draw[red, thick] (v1) -- (v2);
    
    \node[above] at (0, 2.5) {$B$};
    \node[above] at (3, 3) {$|r| \langle \text{MAX}|$};
    \node[] at (1.5,1.375) {$U'$};
    \node[below] at (0,0) {$\ell$};
    \node[below] at (2,0) {$|\psi_0\rangle_f$};
    \node[below] at (3,0) {$I$};
    \node[below] at (4,0) {$r$};
    \node[below] at (6,0) {$R$};
    \node[below] at (5,-0.5) {$|\text{MAX}\rangle$};
    
    \node[] at (7, 1.25) {$=$};
    
    \begin{scope}[xshift=8cm]
    \draw[] (0,0) -- (0,1);
    \draw[] (2,0) -- (2,1);
    \draw[] (0, 1.5) -- (0, 2.75);
    \draw[] (2, 1.5) -- (2, 2.75);
    \draw[] (3,0) -- (3,1);
    
    \draw[fill=red!10!white] (-0.25, 1) -- (3.25, 1) -- (3.25, 1.75) -- (-0.25, 1.75) -- cycle;
    
     \node[fill=blue, regular polygon, regular polygon sides=3, minimum size=0.35cm, inner sep=0pt] (u1) at (3, 0.25) {};
    \node[fill=blue, regular polygon, regular polygon sides=3, minimum size=0.35cm, inner sep=0pt] (u2) at (2, 2.5) {};
    \node[fill=red, regular polygon, regular polygon sides=3, minimum size=0.35cm, inner sep=0pt] (v1) at (3, 0.7) {};
    \node[fill=red, regular polygon, regular polygon sides=3, minimum size=0.35cm, inner sep=0pt, rotate=180] (v2) at (2, 2) {};
    
    \node[left] at (u2) {$u$};
    \node[left] at (v2) {$v$};
    
    \draw[blue, thick] (u1) -- ++ (0.75, 0) -- ++ (0,2.25) -- (u2);
    \draw[red, thick] (v1) -- ++ (0.5, 0) -- ++ (0, 1.3) --  (v2);
    
    \node[above] at (2, 2.75) {$R$};
    \node[above] at (0, 2.75) {$B$};
    \node[] at (1.5,1.375) {$U'$};
    \node[below] at (0,0) {$\ell$};
    \node[below] at (2,0) {$|\psi_0\rangle_f$};
    \node[below] at (3,0) {$I$};
    \end{scope}.
    \end{tikzpicture}.
    \label{eq:straightening}
\end{equation}
We immediately see two issues. First, the red unitary is partially transposed. Generally speaking, a partial transpose of a unitary is not a unitary.\footnote{This fact can be verified for randomly generated two-qubit unitaries.} Secondly, the order in which $u$, $v$, and $U'$ act on $I$ is the opposite of the order in which they act on $R$. Generally speaking, even if the partially transposed $v$ is a unitary, it is not clear if the combined effect of the $u$ and $v$, in conjunction with $U'$, remains unitary. 

Similar issues were raised by Lloyd and Preskill~\cite{Lloyd2014} and by Gottesman and Preskill~\cite{gottesman2004comment} for the black hole final state proposal~\cite{Horowitz2004}. In that proposal, the projection of $r'r$ onto a maximally entangled state is regarded as an actual physical process occurring inside the black hole. Here, instead, the projection is a feature of the holographic map rather than a physical process. Nevertheless, we should consider how the actions of the infaller in the effective description, together with the holographic map, affect phenomena that are accessible to exterior observers. 

\section{Detecting nonunitarity}
\label{sec:issues}
Since we will be studying potential deviations from unitarity in the black hole $S$-matrix, it is important to establish that such deviations are really operationally meaningful. For that purpose, we will formulate a thought experiment illustrating how an observer outside a black hole could verify the failure of unitarity. The observer who conducts this experiment can manipulate the Hawking radiation emitted by the black hole using a quantum computer, but we will insist that operations performed by the observer have computational complexity scaling polynomially with the entropy of the black hole. Here we adopt the computer scientist's credo that operations of polynomial complexity are physically plausible, while operations of superpolynomial complexity are physically unreasonable. Indeed, we expect that operations of superpolynomial complexity could undermine foundations of effective field theory on a semiclassical background~\cite{Harlow2013,Kim2020}, and are therefore beyond the scope of our discussion. 

Having set the ground rules, let us now discuss the setup of the thought experiment. We consider an observer who can create two identical black holes and can send identical robots into the black holes. Let us label these black holes and robots as $1$ and $2$; robot $1$ falls into the black hole $1$ and robot $2$ falls into the black hole $2$. The robots shall be initialized to the states $|\phi_1\rangle$ and $|\phi_2\rangle$, respectively, prior to being sent into the black holes. Each robot is programmed to interact with the radiation in the exterior ($R$) and in the interior ($r$) via  unitary transformations $u$ and $v$. Each robot can be regarded as an infaller as in Fig.~\ref{fig:model_with_observer}, programmed to perform these specific tasks. 

Now two identical black holes are created each in the same pure state, and start to emit Hawking radiation. When the black holes are old enough to have radiated away more than half of their initial entropy, the exterior observer sends in the robots to the respective black holes; then both black holes evaporate completely. The observer collects all the Hawking radiation emitted by black hole 1, both before and after robot 1 fell in, and deposits it in a quantum memory, acquiring a quantum state $\rho_1$. 
The observer does the same for black hole 2, acquiring the state $\rho_2$ in a second quantum memory. If this entire process is unitary, then the states $\rho_1= |\psi_1\rangle \langle \psi_1|$ and $\rho_2= |\psi_2\rangle \langle \psi_2|$ are pure, and furthermore their overlap $|\langle\psi_1|\psi_2\rangle|^2$ matches $|\langle \phi_1  | \phi_2\rangle|^2$. A detected deviation of the overlap from $|\langle \phi_1  | \phi_2\rangle|^2$ would indicate that unitarity is violated. 

\begin{figure}[h]
\centering
    \begin{quantikz}
    \lstick{\ket{0}} & \gate{H} & \ctrl{2} & \gate{H} & \meter{} \\
    \lstick{\ket{\psi_1}}& \qw & \swap{1} & \qw & \qw \\
    \lstick{\ket{\psi_2}}& \qw & \targX{} & \qw & \qw
    \end{quantikz}
    \caption{Circuit for the swap test estimating the overlap of  $|\psi_1\rangle$ and $|\psi_2\rangle$.}
    \label{fig:swap_test}
\end{figure}
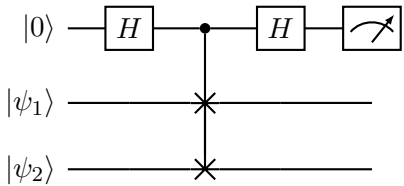

The overlap can be estimated using the \emph{swap test}. Suppose we want to compute the overlap of two pure states $|\psi_1\rangle$ and $|\psi_2\rangle$. In the swap test, one executes a swap operation on the two states controlled by a single qubit as in Fig.~\ref{fig:swap_test}, where $H$ denotes the Hadamard gate. When the control qubit is measured in the standard basis, the outcome $|0\rangle$ occurs with probability $\frac{1}{2}\left(1 + |\langle \psi_1  | \psi_2\rangle|^2\right)$. By repeating this experiment sufficiently many times, then, we can estimate the overlap with a small statistical error. If $|\psi_1\rangle$ and $|\psi_2\rangle$ are $n$-qubit states, the controlled-swap gate can be decomposed into $O(n)$ one-qubit and two-qubit gates. Thus, the overlap can be computed efficiently.

Coming back to our physical setup, the entire unitarity test can be conducted efficiently. Since the robots are small compared to the black holes, their initialization is efficient. A black hole evaporates in a time that scales polynomially with the entropy of the black hole that initially forms, and the number $n$ of radiation qubits after the evaporation process is complete also scales polynomially with this entropy. Since the swap test has complexity scaling linearly with $n$, as we have just seen, the complete experiment has complexity polynomial in the initial black hole entropy. 

Now suppose that $V$ is a linear map which takes pure states to pure states, but is not isometric. Its non-isometric nature can be verified by using the swap test to estimate state overlaps. Let $|\phi_i\rangle$ be an eigenstate of $V^{\dagger}V$ with eigenvalue $\lambda_i$, where $\lambda_1\neq \lambda_2$. Upon applying $V$ to these states, we obtain $|\psi_i\rangle = V|\phi_i\rangle / \|V|\phi_i\rangle \|$, assuming $V$ does not annihilate $|\phi_i\rangle$.\footnote{Here we used a prescription in which the state obtained after applying $V$ is normalized; otherwise the result of the swap test may not lie in the interval $[0, 1]$, which is nonsensical.} Then we find 
\begin{equation}
    |\langle \psi_2| \psi_1\rangle|^2 = \frac{|\langle\phi_2|V^\dagger V|\phi_1\rangle|^2}{|\langle\phi_2|V^\dagger V|\phi_2\rangle|\cdot|\langle\phi_1|V^\dagger V|\phi_1\rangle|}= \frac{\lambda_1^2}{\lambda_1 \lambda_2}|\langle \phi_2|\phi_1\rangle|^2\ne |\langle \phi_2|\phi_1\rangle|^2.
\end{equation}
Thus the swap test detects that the overlap of the output states differs from the overlap of the input states.

\section{Violation of unitarity: a simple example}
\label{sec:adversarial}

We emphasize again that the circuit shown in Eq.~\eqref{eq:straightening} is not necessarily unitary. To be concrete, suppose that $f$ is one-dimensional, that $\ell$, $I$, $B$, and $R$ all have dimension $d$, that $U'$ and $u$ are identity maps, and that $v$ is the swap operation. Then the right-hand-side of Eq.~\eqref{eq:straightening} is the partial transpose of the swap; represented diagrammatically, this partial transpose is
\begin{equation}
\label{eq:swap_partial_transpose}
    \begin{tikzpicture}[baseline={([yshift=-.5ex]current bounding box.center)}]
    \draw[] (0, 0) -- ++ (0, 1) -- ++ (1, 0.5) -- ++ (0,1);
    \node[fill=white, draw=white] () at (0.5, 1.25) {};
    \draw[] (1, 0) -- ++ (0, 1) -- ++ (-1, 0.5) -- ++ (0,1);
    \node[below] at (0,0) {$1$};
    \node[below] at (1,0) {$2$};
    \end{tikzpicture}
    \quad
    \stackrel{T_1}{\longrightarrow}
    \quad 
    \begin{tikzpicture}[baseline={([yshift=-.5ex]current bounding box.center)}]
    \draw[] (0, 0) -- ++ (0, 0.5) -- ++ (0.5, 0.5) -- ++ (0.5,-0.5) -- ++ (0, -0.5);
    \draw[] (0, 2.5) -- ++ (0, -0.5) -- ++ (0.5, -0.5) -- ++ (0.5, 0.5) -- ++ (0, 0.5);
    \node[below] at (0,0) {$1$};
    \node[below] at (1,0) {$2$};
    \end{tikzpicture},
\end{equation}
where the bottom two legs are the input and the top two legs are the output; under transposition applied to the first leg, the input and and output are interchanged.  Thus, the partial transpose of swap is $d|\text{MAX}\rangle \langle \text{MAX}|$, where $|\text{MAX}\rangle$ is a normalized maximally entangled state. Evidently, this map is not isometric --- in particular, it annihilates any state orthogonal to $|\text{MAX}\rangle$. 

This simple example clearly illustrates that the circuit in Eq.~\eqref{eq:straightening} need not be unitary, but on the other hand, choosing $U'$ to be the identity does not capture the chaotic nature of black hole dynamics. In Section~\ref{sec:average}, following Ref.~\cite{Akers2022}, we analyze the departure from unitarity when $U'$ is a Haar-random unitary transformation, finding that in that case the violation of unitarity is very strongly suppressed with very high probability. Then in Section~\ref{sec:pseudorandomness}, we find similar supression under the more realistic assumption that $U'$ is pseudorandom, if $u$, $v$ have computational complexity scaling polynomially with the black hole entropy. 

\section{Average-case analysis}
\label{sec:average}
For a realistic black hole, we expect the unitary $U'$ to be a scrambling unitary~\cite{Hayden2007}. A simple model of scrambling unitary is a Haar-random unitary. In this section, we investigate whether the process in Eq.~\eqref{eq:straightening} is unitary \emph{on average}, and we also characterize the fluctuations from the average behavior. 

To be concrete, we view the operator in Eq.~\eqref{eq:straightening} as a linear map $V(U')$ that depends on $U'$, and seek to compute the average of $V(U')^\dagger V(U')$. To that end, we can use the formula~\cite{Collins2003}
\begin{equation}
    \int  U_{ij} U^{\dagger}_{j' i' } dU = \frac{1}{d} \delta_{i i'}\delta_{j j'},
\end{equation}
where $d$ is the dimension of the Hilbert space that $U$ acts on and the integration is over the Haar measure on the unitary group. This expression admits a convenient graphical representation:
\begin{equation}
   \int dU
   \left(
   \begin{tikzpicture}[baseline={([yshift=-.5ex]current bounding box.center)}]
   \node[rectangle, draw=black, minimum size= 0.75cm] (U) at (0,0) {$U$};
   \node[rectangle, draw=black, minimum size= 0.75cm] (Udag) at (0, 2) {$U^{\dagger}$};
   
    \draw[] (U) -- ++ (0, 0.75);
    \draw[] (U) -- ++ (0, -0.75);
    \draw[] (Udag) -- ++ (0, 0.75);
    \draw[] (Udag) -- ++ (0, -0.75);
    
    \node[right] at (0, -0.75) {$j$};
    \node[right] at (0, 0.75) {$i$};
    \node[right] at (0, 1.25) {$i'$};
    \node[right] at (0, 2.75) {$j'$};
   \end{tikzpicture}
   \right) = 
   \frac{1}{d}\,\,
   \begin{tikzpicture}[baseline={([yshift=-.5ex]current bounding box.center)}]
   \node[rectangle, draw=white, minimum size= 0.75cm] (U) at (0,0) {\color{white} $U$};
   \node[rectangle, draw=white, minimum size= 0.75cm] (Udag) at (0, 2) {\color{white} $U^{\dagger}$};
   
    \draw[] (U) -- ++ (0, 0.75);
    \draw[] (U) -- ++ (0, -0.75);
    \draw[] (Udag) -- ++ (0, 0.75);
    \draw[] (Udag) -- ++ (0, -0.75);
    
    \node[right] at (0, -0.75) {$j$};
    \node[right] at (0, 0.75) {$i$};
    \node[right] at (0, 1.25) {$i'$};
    \node[right] at (0, 2.75) {$j'$};
    
    \draw[] (0, 2.375) -- ++ (-0.75, 0) -- ++ (0, -2.75) -- ++ (0.75, 0);   
    \draw[] (0, 1.6125) -- ++ (-0.375, 0) -- ++ (0, -1.25) -- ++ (0.375, 0);
   \end{tikzpicture}
   . \label{eq:haar_average}
\end{equation}
This leads to an expression for the Haar average of $V(U')^\dagger V(U')$:
\begin{equation}
\label{eq:unitarity_average}
\begin{aligned}
\int dU'
    \left(
    \begin{tikzpicture}[scale=0.8, baseline={([yshift=-.5ex]current bounding box.center)}]
    \draw[] (0,0) -- (0,1);
    \draw[] (2,0) -- (2,1);
    \draw[] (0, 1.5) -- (0, 2.75);
    \draw[] (2, 1.5) -- (2, 2.75);
    \draw[] (3,0) -- (3,1);
    
    \draw[fill=red!10!white] (-0.25, 1) -- (3.25, 1) -- (3.25, 1.75) -- (-0.25, 1.75) -- cycle;
    
     \node[fill=blue, regular polygon, regular polygon sides=3, minimum size=0.35cm, inner sep=0pt] (u1) at (3, 0.25) {};
    \node[fill=blue, regular polygon, regular polygon sides=3, minimum size=0.35cm, inner sep=0pt] (u2) at (2, 2.5) {};
    \node[fill=red, regular polygon, regular polygon sides=3, minimum size=0.35cm, inner sep=0pt] (v1) at (3, 0.7) {};
    \node[fill=red, regular polygon, regular polygon sides=3, minimum size=0.35cm, inner sep=0pt, rotate=180] (v2) at (2, 2) {};
    
    \node[left] at (u2) {$u\phantom{l}$};
    \node[left] at (v2) {$v\phantom{l}$};
    
    \draw[blue, thick] (u1) -- ++ (0.75, 0) -- ++ (0,2.25) -- (u2);
    \draw[red, thick] (v1) -- ++ (0.5, 0) -- ++ (0, 1.3) --  (v2);
    
    \node[] at (1.5,1.375) {$U'$};
    \node[below] at (0,0) {$\ell$};
    \node[below] at (2,0) {$|\psi_0\rangle_f$};
    \node[below] at (3,0) {$I$};
    \begin{scope}[yscale=-1, yshift=-5.5cm]
    
    \draw[] (0,0) -- (0,1);
    \draw[] (2,0) -- (2,1);
    \draw[] (0, 1.5) -- (0, 2.75);
    \draw[] (2, 1.5) -- (2, 2.75);
    \draw[] (3,0) -- (3,1);
    
    \draw[fill=red!10!white] (-0.25, 1) -- (3.25, 1) -- (3.25, 1.75) -- (-0.25, 1.75) -- cycle;
    
    \node[fill=blue, regular polygon, regular polygon sides=3, minimum size=0.35cm, inner sep=0pt, rotate=180] (u1) at (3, 0.25) {};
    \node[fill=blue, regular polygon, regular polygon sides=3, minimum size=0.35cm, inner sep=0pt, rotate=180] (u2) at (2, 2.5) {};
    \node[fill=red, regular polygon, regular polygon sides=3, minimum size=0.35cm, inner sep=0pt, rotate=180] (v1) at (3, 0.7) {};
    \node[fill=red, regular polygon, regular polygon sides=3, minimum size=0.35cm, inner sep=0pt] (v2) at (2, 2) {};
    
    \node[left] at (u2) {$u^{\dagger}$};
    \node[left] at (v2) {$v^{\dagger}$};
    
    \draw[blue, thick] (u1) -- ++ (0.75, 0) -- ++ (0,2.25) -- (u2);
    \draw[red, thick] (v1) -- ++ (0.5, 0) -- ++ (0, 1.3) --  (v2);
    
    \node[] at (1.5,1.375) {${U'}^{\dagger}$};
    \node[above] at (0,0) {$\ell$};
    \node[above] at (2,0) {$\langle\psi_0|_f$};
    \node[above] at (3,0) {$I$};
    \end{scope}
    \end{tikzpicture}
    \right)
    &= \frac{1}{|\ell| |I||f|}
    \begin{tikzpicture}[scale=0.8, baseline={([yshift=-.5ex]current bounding box.center)}]
    \draw[] (0,0) -- (0,1);
    \draw[] (2,0) -- (2,1);
    \draw[] (0, 1.5) -- (0, 2.75);
    \draw[] (2, 1.5) -- (2, 2.5);
    \draw[] (3,0) -- (3,1);

     \node[fill=blue, regular polygon, regular polygon sides=3, minimum size=0.35cm, inner sep=0pt] (u1) at (3, 0.25) {};
    \node[fill=blue, regular polygon, regular polygon sides=3, minimum size=0.35cm, inner sep=0pt] (u2) at (2, 2.5) {};
    \node[fill=red, regular polygon, regular polygon sides=3, minimum size=0.35cm, inner sep=0pt] (v1) at (3, 0.7) {};
    \node[fill=red, regular polygon, regular polygon sides=3, minimum size=0.35cm, inner sep=0pt, rotate=180] (v2) at (2, 2) {};
    
    \node[left] at (u2) {$u\phantom{l}$};
    \node[left] at (v2) {$v\phantom{l}$};
    
    \draw[blue, thick] (u1) -- ++ (0.75, 0) -- ++ (0,2.25) -- (u2);
    \draw[red, thick] (v1) -- ++ (0.5, 0) -- ++ (0, 1.3) --  (v2);
    
    \node[below] at (0,0) {$\ell$};
    \node[below] at (2,0) {$|\psi_0\rangle_f$};
    \node[below] at (3,0) {$I$};
    \begin{scope}[yscale=-1, yshift=-5.5cm]
    
    \draw[] (0,0) -- (0,1);
    \draw[] (2,0) -- (2,1);
    \draw[] (0, 1.5) -- (0, 2.75);
    \draw[] (2, 1.5) -- (2, 3);
    \draw[] (3,0) -- (3,1);
    
    \node[fill=blue, regular polygon, regular polygon sides=3, minimum size=0.35cm, inner sep=0pt, rotate=180] (u1) at (3, 0.25) {};
    \node[fill=blue, regular polygon, regular polygon sides=3, minimum size=0.35cm, inner sep=0pt, rotate=180] (u2) at (2, 2.5) {};
    \node[fill=red, regular polygon, regular polygon sides=3, minimum size=0.35cm, inner sep=0pt, rotate=180] (v1) at (3, 0.7) {};
    \node[fill=red, regular polygon, regular polygon sides=3, minimum size=0.35cm, inner sep=0pt] (v2) at (2, 2) {};
    
    \node[left] at (u2) {$u^{\dagger}$};
    \node[left] at (v2) {$v^{\dagger}$};
    
    \draw[blue, thick] (u1) -- ++ (0.75, 0) -- ++ (0,2.25) -- (u2);
    \draw[red, thick] (v1) -- ++ (0.5, 0) -- ++ (0, 1.3) --  (v2);
    
    \node[above] at (0,0) {$\ell$};
    \node[above] at (2,0) {$\langle\psi_0|_f$};
    \node[above] at (3,0) {$I$};
    \end{scope}
    
    \draw[] (0, 1) --++ (-0.25, 0.25) --++ (0, 3) --++ (0.25, 0.25);    
    \draw[] (0, 1.5) --++ (0.5, 0) --++ (0, 2.5) --++ (-0.5, 0);
    
    \draw[] (2,1) --++ (-1.25, 0.25) --++ (0, 3) --++ (1.25, 0.25);
    \draw[] (2, 1.5) --++ (-0.75, 0) --++ (0, 2.5) --++ (0.75, 0);
    
    \draw[] (3,1) --++ (0, 3.5);
    \end{tikzpicture} \\
    &= 
    \frac{|B|}{|\ell| |I| |f|}
        \begin{tikzpicture}[scale=0.8, baseline={([yshift=-.5ex]current bounding box.center)}]
    \draw[] (2, 1.5) --++ (-0.75, 0) --++ (0, 2.5) --++ (0.75, 0) -- cycle;
    \draw[] (0,0) -- (0,1);
    \draw[] (0, 1.5) -- (0, 2.5);
    \draw[] (3,0) -- (3,1);

     \node[fill=blue, regular polygon, regular polygon sides=3, minimum size=0.35cm, inner sep=0pt] (u1) at (3, 0.25) {};
    \node[fill=blue, regular polygon, regular polygon sides=3, minimum size=0.35cm, inner sep=0pt] (u2) at (2, 2.5) {};
    \node[fill=red, regular polygon, regular polygon sides=3, minimum size=0.35cm, inner sep=0pt] (v1) at (3, 0.7) {};
    \node[fill=red, regular polygon, regular polygon sides=3, minimum size=0.35cm, inner sep=0pt, rotate=180] (v2) at (2, 2) {};
    
    \node[left] at (u2) {$u\phantom{l}$};
    \node[left] at (v2) {$v\phantom{l}$};
    
    \draw[blue, thick] (u1) -- ++ (0.75, 0) -- ++ (0,2.25) -- (u2);
    \draw[red, thick] (v1) -- ++ (0.5, 0) -- ++ (0, 1.3) --  (v2);
    
    \node[below] at (0,0) {$\ell$};
    \node[below] at (3,0) {$I$};
    \begin{scope}[yscale=-1, yshift=-5.5cm]
    
    \draw[] (0,0) -- (0,1);
    \draw[] (0, 1.5) -- (0, 2.5);
    \draw[] (3,0) -- (3,1);
    
    \node[fill=blue, regular polygon, regular polygon sides=3, minimum size=0.35cm, inner sep=0pt, rotate=180] (u1) at (3, 0.25) {};
    \node[fill=blue, regular polygon, regular polygon sides=3, minimum size=0.35cm, inner sep=0pt, rotate=180] (u2) at (2, 2.5) {};
    \node[fill=red, regular polygon, regular polygon sides=3, minimum size=0.35cm, inner sep=0pt, rotate=180] (v1) at (3, 0.7) {};
    \node[fill=red, regular polygon, regular polygon sides=3, minimum size=0.35cm, inner sep=0pt] (v2) at (2, 2) {};
    
    \node[left] at (u2) {$u^{\dagger}$};
    \node[left] at (v2) {$v^{\dagger}$};
    
    \draw[blue, thick] (u1) -- ++ (0.75, 0) -- ++ (0,2.25) -- (u2);
    \draw[red, thick] (v1) -- ++ (0.5, 0) -- ++ (0, 1.3) --  (v2);
    
    \node[above] at (0,0) {$\ell$};
    \node[above] at (3,0) {$I$};
    \end{scope}
    
    \draw[] (0, 1) --++ (0,3.5);   
    
    \draw[] (3,1) --++ (0, 3.5);
    \end{tikzpicture}\\
    &= I_{\ell} \otimes I_I,
\end{aligned}
\end{equation}
where in the last line we used the following identity:
\begin{equation}
\begin{tikzpicture}[scale=0.8, baseline={([yshift=-.5ex]current bounding box.center)}]
    \draw[] (2, 1.5) --++ (-0.75, 0) --++ (0, 2.5) --++ (0.75, 0) -- cycle;
    \draw[] (3,0) -- (3,1);
    
     \node[fill=blue, regular polygon, regular polygon sides=3, minimum size=0.35cm, inner sep=0pt] (u1) at (3, 0.25) {};
    \node[fill=blue, regular polygon, regular polygon sides=3, minimum size=0.35cm, inner sep=0pt] (u2) at (2, 2.5) {};
    \node[fill=red, regular polygon, regular polygon sides=3, minimum size=0.35cm, inner sep=0pt] (v1) at (3, 0.7) {};
    \node[fill=red, regular polygon, regular polygon sides=3, minimum size=0.35cm, inner sep=0pt, rotate=180] (v2) at (2, 2) {};
    
    \node[left] at (u2) {$u\phantom{l}$};
    \node[left] at (v2) {$v\phantom{l}$};
    
    \draw[blue, thick] (u1) -- ++ (0.75, 0) -- ++ (0,2.25) -- (u2);
    \draw[red, thick] (v1) -- ++ (0.5, 0) -- ++ (0, 1.3) --  (v2);
    
    \node[below] at (3,0) {$I$};
    \begin{scope}[yscale=-1, yshift=-5.5cm]
    \draw[] (3,0) -- (3,1);
    
    \node[fill=blue, regular polygon, regular polygon sides=3, minimum size=0.35cm, inner sep=0pt, rotate=180] (u1) at (3, 0.25) {};
    \node[fill=blue, regular polygon, regular polygon sides=3, minimum size=0.35cm, inner sep=0pt, rotate=180] (u2) at (2, 2.5) {};
    \node[fill=red, regular polygon, regular polygon sides=3, minimum size=0.35cm, inner sep=0pt, rotate=180] (v1) at (3, 0.7) {};
    \node[fill=red, regular polygon, regular polygon sides=3, minimum size=0.35cm, inner sep=0pt] (v2) at (2, 2) {};
    
    \node[left] at (u2) {$u^{\dagger}$};
    \node[left] at (v2) {$v^{\dagger}$};
    
    \draw[blue, thick] (u1) -- ++ (0.75, 0) -- ++ (0,2.25) -- (u2);
    \draw[red, thick] (v1) -- ++ (0.5, 0) -- ++ (0, 1.3) --  (v2);
    
    \node[above] at (3,0) {$I$};
    \end{scope}
    \draw[] (3,1) --++ (0, 3.5);
    \end{tikzpicture}
=
\begin{tikzpicture}[scale=0.8, baseline={([yshift=-.5ex]current bounding box.center)}]
    \draw[] (3,1) --++ (0, 3.5);
    \draw[] (2, 1.25) --++ (-0.75, 0) --++ (0, 2.5) --++ (0.75, 0) -- cycle;
    \draw[] (3,0) -- (3,1);
    
    \node[fill=blue, regular polygon, regular polygon sides=3, minimum size=0.35cm, inner sep=0pt] (u1) at (3, 1.75) {};
    \node[fill=blue, regular polygon, regular polygon sides=3, minimum size=0.35cm, inner sep=0pt] (u2) at (2, 1.75) {};
    \node[fill=red, regular polygon, regular polygon sides=3, minimum size=0.35cm, inner sep=0pt] (v1) at (3, 2.25) {};
    \node[fill=red, regular polygon, regular polygon sides=3, minimum size=0.35cm, inner sep=0pt] (v2) at (1.25, 2.25) {};
    
    \node[left] at (u2) {$u\phantom{l}$};
    \node[left] at (v2) {$v\phantom{l}$};
    
    \draw[blue, thick] (u1) -- (u2);
    \draw[red, thick] (v1) --  (v2);
    
    \node[below] at (3,0) {$I$};
    \begin{scope}[yscale=-1, yshift=-5.cm]
    \draw[] (3,0) -- (3,1);
    
    \node[fill=blue, regular polygon, regular polygon sides=3, minimum size=0.35cm, inner sep=0pt, rotate=180] (u1) at (3, 1.75) {};
    \node[fill=blue, regular polygon, regular polygon sides=3, minimum size=0.35cm, inner sep=0pt, rotate=180] (u2) at (2, 1.75) {};
    \node[fill=red, regular polygon, regular polygon sides=3, minimum size=0.35cm, inner sep=0pt, rotate=180] (v1) at (3, 2.25) {};
    \node[fill=red, regular polygon, regular polygon sides=3, minimum size=0.35cm, inner sep=0pt, rotate=180] (v2) at (1.25, 2.25) {};
    
    \node[left] at (u2) {$u^{\dagger}$};
    \node[left] at (v2) {$v^{\dagger}$};
    
    \draw[blue, thick] (u1) -- (u2);
    \draw[red, thick] (v1) -- (v2);
    
    \node[above] at (3,0) {$I$};
    \end{scope}
    \end{tikzpicture}= |R| I_I.
\end{equation}
Thus, on average, the unitarity of the black hole $S$-matrix is preserved, no matter how $u$ and $v$ are chosen.

We can also compute the fluctuation about this mean value. Without loss of generality, consider pure states $|\varphi\rangle_{\ell I}, |\varphi' \rangle_{\ell I} \in \mathcal{H}_{\ell}\otimes \mathcal{H}_I$. As we have just seen, the random variable $\langle\varphi| V(U')^{\dagger}V(U')|\varphi'\rangle$
has mean value $ \langle \varphi|\varphi'\rangle$, while its variance is
\begin{equation}
\begin{aligned}
    \int dU' \left| \langle \varphi| V(U')^{\dagger }V(U')|\varphi'\rangle - \langle \varphi|\varphi'\rangle \right|^2 &= \int dU' |\langle\varphi| V(U')^{\dagger}V(U')|\varphi'\rangle|^2 - |\langle\varphi |\varphi'\rangle |^2.
\end{aligned}
    \label{eq:fluctuation_inner_product}
\end{equation}
This integral can be computed using the expression for the $(2,2)$ moment of Haar measure~\cite{Collins2003}:
\begin{equation}
    \int U_{i_1j_1} U_{i_2 j_2} U^\dagger_{j_1' i_1'} U^{\dagger}_{j_2'i_2'} dU= \frac{1}{d^2-1} \delta_{i_1 i_1'} \delta_{i_2 i_2'} \delta_{j_1 j_1'} \delta_{j_2 j_2'} - \frac{1}{d(d^2-1)} \delta_{i_1 i_2'}\delta_{i_2 i_1'} \delta_{j_1 j_2'} \delta_{j_2 j_1'}.
\end{equation}
Diagrammatically, this can be expressed as
\begin{equation}
   \int dU
   \left(
   \begin{tikzpicture}[baseline={([yshift=-.5ex]current bounding box.center)}]
   \node[rectangle, draw=black, minimum size= 0.75cm] (U) at (0,0) {$U$};
   \node[rectangle, draw=black, minimum size= 0.75cm] (Udag) at (0, 2) {$U^{\dagger}$};
   
    \draw[] (U) -- ++ (0, 0.75);
    \draw[] (U) -- ++ (0, -0.75);
    \draw[] (Udag) -- ++ (0, 0.75);
    \draw[] (Udag) -- ++ (0, -0.75);
    
    \node[right] at (0, -0.75) {$j_1$};
    \node[right] at (0, 0.75) {$i_1$};
    \node[right] at (0, 1.25) {$i_1'$};
    \node[right] at (0, 2.75) {$j_1'$};
    \begin{scope}[xshift=1.5cm]
   \node[rectangle, draw=black, minimum size= 0.75cm] (U) at (0,0) {$U$};
   \node[rectangle, draw=black, minimum size= 0.75cm] (Udag) at (0, 2) {$U^{\dagger}$};
   
    \draw[] (U) -- ++ (0, 0.75);
    \draw[] (U) -- ++ (0, -0.75);
    \draw[] (Udag) -- ++ (0, 0.75);
    \draw[] (Udag) -- ++ (0, -0.75);
    
    \node[right] at (0, -0.75) {$j_2$};
    \node[right] at (0, 0.75) {$i_2$};
    \node[right] at (0, 1.25) {$i_2'$};
    \node[right] at (0, 2.75) {$j_2'$};
    
    \end{scope}
   \end{tikzpicture}
   \right) = 
   \frac{1}{d^2-1} \left(
   \begin{tikzpicture}[baseline={([yshift=-.5ex]current bounding box.center)}]
   \node[rectangle, draw=white, minimum size= 0.75cm] (U) at (0,0) {\color{white} $U$};
   \node[rectangle, draw=white, minimum size= 0.75cm] (Udag) at (0, 2) {\color{white} $U^{\dagger}$};
   
    \draw[] (U) -- ++ (0, 0.75);
    \draw[] (U) -- ++ (0, -0.75);
    \draw[] (Udag) -- ++ (0, 0.75);
    \draw[] (Udag) -- ++ (0, -0.75);
    
    \node[right] at (0, -0.75) {$j_1$};
    \node[right] at (0, 0.75) {$i_1$};
    \node[right] at (0, 1.25) {$i_1'$};
    \node[right] at (0, 2.75) {$j_1'$};
    
    \draw[] (0, 2.375) -- ++ (-0.75, 0) -- ++ (0, -2.75) -- ++ (0.75, 0);   
    \draw[] (0, 1.6125) -- ++ (-0.375, 0) -- ++ (0, -1.25) -- ++ (0.375, 0);
    \begin{scope}[xshift=1.5cm]
   \node[rectangle, draw=white, minimum size= 0.75cm] (U) at (0,0) {\color{white} $U$};
   \node[rectangle, draw=white, minimum size= 0.75cm] (Udag) at (0, 2) {\color{white} $U^{\dagger}$};
   
    \draw[] (U) -- ++ (0, 0.75);
    \draw[] (U) -- ++ (0, -0.75);
    \draw[] (Udag) -- ++ (0, 0.75);
    \draw[] (Udag) -- ++ (0, -0.75);
    
    \node[right] at (0, -0.75) {$j_2$};
    \node[right] at (0, 0.75) {$i_2$};
    \node[right] at (0, 1.25) {$i_2'$};
    \node[right] at (0, 2.75) {$j_2'$};
    
    \draw[] (0, 2.375) -- ++ (-0.75, 0) -- ++ (0, -2.75) -- ++ (0.75, 0);    
    \draw[] (0, 1.6125) -- ++ (-0.375, 0) -- ++ (0, -1.25) -- ++ (0.375, 0);
    
    \end{scope}
   \end{tikzpicture}
   \right)
   -\frac{1}{d(d^2-1)}
   \left(
   \begin{tikzpicture}[baseline={([yshift=-.5ex]current bounding box.center)}]
   \node[rectangle, draw=white, minimum size= 0.75cm] (U) at (0,0) {\color{white} $U$};
   \node[rectangle, draw=white, minimum size= 0.75cm] (Udag) at (0, 2) {\color{white} $U^{\dagger}$};
   
    \draw[] (U) -- ++ (0, 0.75);
    \draw[] (Udag) -- ++ (0, -0.75);

    \node[left] at (0, 0.75) {$i_1$};
    \node[left] at (0, 1.25) {$i_1'$};

    \draw[] (0, 1.6125) -- (1.5, 0.3625);
    \node[circle, fill=white, minimum size = 0.2cm, inner sep = 0pt] () at (0.75, 0.9875) {};
    \draw[] (0, 0.3625) -- (1.5, 1.6125);
    \begin{scope}[xshift=1.5cm]
   \node[rectangle, draw=white, minimum size= 0.75cm] (U) at (0,0) {\color{white} $U$};
   \node[rectangle, draw=white, minimum size= 0.75cm] (Udag) at (0, 2) {\color{white} $U^{\dagger}$};
   
    \draw[] (U) -- ++ (0, 0.75);
    \draw[] (Udag) -- ++ (0, -0.75);
    
    \node[right] at (0, 0.75) {$i_2$};
    \node[right] at (0, 1.25) {$i_2'$};
       
    \end{scope}
    \begin{scope}[xshift=1.75cm]
       \node[rectangle, draw=white, minimum size= 0.75cm] (U) at (0,0) {\color{white} $U$};
   \node[rectangle, draw=white, minimum size= 0.75cm] (Udag) at (0, 2) {\color{white} $U^{\dagger}$};
   
    \draw[] (U) -- ++ (0, -0.75);
    \draw[] (Udag) -- ++ (0, 0.75);
    
    \node[left] at (0, -0.75) {$j_1$};
    \node[left] at (0, 2.75) {$j_1'$};

    \begin{scope}[xshift=1.5cm]
   \node[rectangle, draw=white, minimum size= 0.75cm] (U) at (0,0) {\color{white} $U$};
   \node[rectangle, draw=white, minimum size= 0.75cm] (Udag) at (0, 2) {\color{white} $U^{\dagger}$};
   
    \draw[] (U) -- ++ (0, -0.75);
    \draw[] (Udag) -- ++ (0, 0.75);
    
    \node[right] at (0, -0.75) {$j_2$};
    \node[right] at (0, 2.75) {$j_2'$};
       
    \draw plot [smooth] coordinates { (-1.5, 2.375) (0, -0.375)};
        \node[circle, fill=white, minimum size = 0.2cm, inner sep = 0pt] () at (-0.75, 0.9875) {};
    \draw plot [smooth] coordinates {(-1.5, -0.375)   (0, 2.375)};
    \end{scope}
    
    \end{scope}
   \end{tikzpicture}
   \right)
   . \label{eq:haar_average_second}
\end{equation}

Applied to our setup, we get:
\begin{equation}
\begin{aligned}
\int dU'
    \left(
    \begin{tikzpicture}[scale=0.8, baseline={([yshift=-.5ex]current bounding box.center)}]
    \draw[] (0,0) -- (0,1);
    \draw[] (2,0) -- (2,1);
    \draw[] (0, 1.5) -- (0, 2.75);
    \draw[] (2, 1.5) -- (2, 2.5);
    \draw[] (3,0) -- (3,1);
    
    \draw[fill=red!10!white] (-0.25, 1) -- (3.25, 1) -- (3.25, 1.75) -- (-0.25, 1.75) -- cycle;
    
     \node[fill=blue, regular polygon, regular polygon sides=3, minimum size=0.35cm, inner sep=0pt] (u1) at (3, 0.25) {};
    \node[fill=blue, regular polygon, regular polygon sides=3, minimum size=0.35cm, inner sep=0pt] (u2) at (2, 2.5) {};
    \node[fill=red, regular polygon, regular polygon sides=3, minimum size=0.35cm, inner sep=0pt] (v1) at (3, 0.7) {};
    \node[fill=red, regular polygon, regular polygon sides=3, minimum size=0.35cm, inner sep=0pt, rotate=180] (v2) at (2, 2) {};
    
    \node[left] at (u2) {$u\phantom{l}$};
    \node[left] at (v2) {$v\phantom{l}$};
    
    \draw[blue, thick] (u1) -- ++ (0.75, 0) -- ++ (0,2.25) -- (u2);
    \draw[red, thick] (v1) -- ++ (0.5, 0) -- ++ (0, 1.3) --  (v2);
    
    \node[] at (1.5,1.375) {$U'$};
    \node[below] at (0,0) {$\ell$};
    \node[below] at (2,0) {$|\psi_0\rangle_f$};
    \node[below] at (3,0) {$I$};
    \begin{scope}[yscale=-1, yshift=-5.5cm]
    
    \draw[] (0,0) -- (0,1);
    \draw[] (2,0) -- (2,1);
    \draw[] (0, 1.5) -- (0, 2.75);
    \draw[] (2, 1.5) -- (2, 2.5);
    \draw[] (3,0) -- (3,1);
    
    \draw[fill=red!10!white] (-0.25, 1) -- (3.25, 1) -- (3.25, 1.75) -- (-0.25, 1.75) -- cycle;
    
    \node[fill=blue, regular polygon, regular polygon sides=3, minimum size=0.35cm, inner sep=0pt, rotate=180] (u1) at (3, 0.25) {};
    \node[fill=blue, regular polygon, regular polygon sides=3, minimum size=0.35cm, inner sep=0pt, rotate=180] (u2) at (2, 2.5) {};
    \node[fill=red, regular polygon, regular polygon sides=3, minimum size=0.35cm, inner sep=0pt, rotate=180] (v1) at (3, 0.7) {};
    \node[fill=red, regular polygon, regular polygon sides=3, minimum size=0.35cm, inner sep=0pt] (v2) at (2, 2) {};
    
    \node[left] at (u2) {$u^{\dagger}$};
    \node[left] at (v2) {$v^{\dagger}$};
    
    \draw[blue, thick] (u1) -- ++ (0.75, 0) -- ++ (0,2.25) -- (u2);
    \draw[red, thick] (v1) -- ++ (0.5, 0) -- ++ (0, 1.3) --  (v2);
    
    \node[] at (1.5,1.375) {${U'}^{\dagger}$};
    \node[above] at (0,0) {$\ell$};
    \node[above] at (2,0) {$\langle \psi_0|_f$};
    \node[above] at (3,0) {$I$};
    \end{scope}
    \end{tikzpicture}
    \right)^{\otimes 2}
    = \left(1- \frac{1}{(|\ell | |I| |f|)^2} \right)^{-1}(I_{\ell}\otimes I_R)^{\otimes 2} + \Delta,
\end{aligned}
\end{equation}
where 
\begin{equation}
    \Delta = -\frac{|B|}{d(d^2-1)}\text{SWAP}_{\ell \leftrightarrow \ell'}\left(
    \begin{tikzpicture}[scale=0.8, baseline={([yshift=-.5ex]current bounding box.center)}]
    \draw[] (2, 1.5) -- (2, 2.75);
    \draw[] (3,0) -- (3,1);

     \node[fill=blue, regular polygon, regular polygon sides=3, minimum size=0.35cm, inner sep=0pt] (u1) at (3, 0.25) {};
    \node[fill=blue, regular polygon, regular polygon sides=3, minimum size=0.35cm, inner sep=0pt] (u2) at (2, 2.5) {};
    \node[fill=red, regular polygon, regular polygon sides=3, minimum size=0.35cm, inner sep=0pt] (v1) at (3, 0.7) {};
    \node[fill=red, regular polygon, regular polygon sides=3, minimum size=0.35cm, inner sep=0pt, rotate=180] (v2) at (2, 2) {};
    
    \node[left] at (u2) {$u\phantom{l}$};
    \node[left] at (v2) {$v\phantom{l}$};
    
    \node[circle, fill=black, inner sep=0pt, minimum size =0.2cm] (vertex1) at (3,1) {};
    \node[circle, fill=black, inner sep=0pt, minimum size =0.2cm] (vertex2) at (2,1.5) {};
    \node[above] at (vertex1) {$4$};
    \node[below] at (vertex2) {$3$};
    
    \draw[blue, thick] (u1) -- ++ (0.75, 0) -- ++ (0,2.25) -- (u2);
    \draw[red, thick] (v1) -- ++ (0.5, 0) -- ++ (0, 1.3) --  (v2);
    
    \node[below] at (3,0) {$I$};
    \begin{scope}[yscale=-1, yshift=-5.5cm]
    \draw[] (2, 1.5) -- (2, 2.75);
    \draw[] (3,0) -- (3,1);

    \node[fill=blue, regular polygon, regular polygon sides=3, minimum size=0.35cm, inner sep=0pt, rotate=180] (u1) at (3, 0.25) {};
    \node[fill=blue, regular polygon, regular polygon sides=3, minimum size=0.35cm, inner sep=0pt, rotate=180] (u2) at (2, 2.55) {};
    \node[fill=red, regular polygon, regular polygon sides=3, minimum size=0.35cm, inner sep=0pt, rotate=180] (v1) at (3, 0.7) {};
    \node[fill=red, regular polygon, regular polygon sides=3, minimum size=0.35cm, inner sep=0pt] (v2) at (2, 2) {};
    
    \node[left] at (u2) {$u^{\dagger}$};
    \node[left] at (v2) {$v^{\dagger}$};
    
    \node[circle, fill=black, inner sep=0pt, minimum size =0.2cm] (vertex1) at (3,1) {};
    \node[circle, fill=black, inner sep=0pt, minimum size =0.2cm] (vertex2) at (2,1.5) {};
    \node[below] at (vertex1) {$1$};
    \node[above] at (vertex2) {$2$};
    
    \draw[blue, thick] (u1) -- ++ (0.75, 0) -- ++ (0,2.25) -- (u2);
    \draw[red, thick] (v1) -- ++ (0.5, 0) -- ++ (0, 1.3) --  (v2);
    
    \node[above] at (3,0) {$I$};
    \end{scope}
    \begin{scope}[xshift=4cm]
    \draw[] (2, 1.5) -- (2, 2.75);
    \draw[] (3,0) -- (3,1);
  
     \node[fill=blue, regular polygon, regular polygon sides=3, minimum size=0.35cm, inner sep=0pt] (u1) at (3, 0.25) {};
    \node[fill=blue, regular polygon, regular polygon sides=3, minimum size=0.35cm, inner sep=0pt] (u2) at (2, 2.5) {};
    \node[fill=red, regular polygon, regular polygon sides=3, minimum size=0.35cm, inner sep=0pt] (v1) at (3, 0.7) {};
    \node[fill=red, regular polygon, regular polygon sides=3, minimum size=0.35cm, inner sep=0pt, rotate=180] (v2) at (2, 2) {};
    
    \node[left] at (u2) {$u\phantom{l}$};
    \node[left] at (v2) {$v\phantom{l}$};
    
    \node[circle, fill=black, inner sep=0pt, minimum size =0.2cm] (vertex1) at (3,1) {};
    \node[circle, fill=black, inner sep=0pt, minimum size =0.2cm] (vertex2) at (2,1.5) {};
    \node[above] at (vertex1) {$1$};
    \node[below] at (vertex2) {$2$};

    \draw[blue, thick] (u1) -- ++ (0.75, 0) -- ++ (0,2.25) -- (u2);
    \draw[red, thick] (v1) -- ++ (0.5, 0) -- ++ (0, 1.3) --  (v2);
    
    \node[below] at (3,0) {$I'$};
    \begin{scope}[yscale=-1, yshift=-5.5cm]
    
    \draw[] (2, 1.5) -- (2, 2.75);
    \draw[] (3,0) -- (3,1);

    \node[fill=blue, regular polygon, regular polygon sides=3, minimum size=0.35cm, inner sep=0pt, rotate=180] (u1) at (3, 0.25) {};
    \node[fill=blue, regular polygon, regular polygon sides=3, minimum size=0.35cm, inner sep=0pt, rotate=180] (u2) at (2, 2.5) {};
    \node[fill=red, regular polygon, regular polygon sides=3, minimum size=0.35cm, inner sep=0pt, rotate=180] (v1) at (3, 0.7) {};
    \node[fill=red, regular polygon, regular polygon sides=3, minimum size=0.35cm, inner sep=0pt] (v2) at (2, 2) {};
    
    \node[left] at (u2) {$u^{\dagger}$};
    \node[left] at (v2) {$v^{\dagger}$};
    
    \node[circle, fill=black, inner sep=0pt, minimum size =0.2cm] (vertex1) at (3,1) {};
    \node[circle, fill=black, inner sep=0pt, minimum size =0.2cm] (vertex2) at (2,1.5) {};
    \node[below] at (vertex1) {$4$};
    \node[above] at (vertex2) {$3$};

    \draw[blue, thick] (u1) -- ++ (0.75, 0) -- ++ (0,2.25) -- (u2);
    \draw[red, thick] (v1) -- ++ (0.5, 0) -- ++ (0, 1.3) --  (v2);
    
    \node[above] at (3,0) {$I'$};
    \end{scope}
    \end{scope}
    \end{tikzpicture}  
    \right);
    \label{eq:def_delta}
\end{equation}
here $d= |\ell| |I| |f| = |B| |R|$, $\ell'$ is a copy of $\ell$ and similarly $I'$ is a copy of $I$. SWAP${}_{\ell \leftrightarrow \ell'}$ is the swap operation between $\ell$ and $\ell'$ and the black dots with the same integer are contracted with each other. Thus the expression for the variance reduces to
\begin{equation}
\begin{aligned}
    \int dU' \left|  \langle \varphi| V(U')^{\dagger }V(U')|\varphi'\rangle-\langle \varphi|\varphi'\rangle \right|^2 =\left(\langle \varphi |  \otimes |\langle\varphi'| \right) \Delta \left( |\varphi'\rangle \otimes |\varphi\rangle \right) - \frac{1}{d^2-1} |\langle \varphi| \varphi'\rangle|^2. 
\end{aligned}
\label{eq:fluctuation_2}
\end{equation}
Note that the second term in Eq.~\eqref{eq:fluctuation_2} is bounded above by $1/(d^2-1)$, which is exponentially small in the black hole entropy $\log |B|$. Thus, to argue that the variance is small it suffices to show that the first term is small.

To bound the first term in Eq.~\eqref{eq:fluctuation_2}, let us suppose at first that $|\varphi\rangle = |\varphi_I\rangle\otimes |\varphi_{\ell}\rangle$ and $|\varphi'\rangle = |\varphi_I'\rangle \otimes |\varphi_{\ell}'\rangle$ are product states; after bounding the variance for such product states, we will be able to extend the computation to entangled states of $I\ell$. Using the definition of $\Delta$ in Eq.~\eqref{eq:def_delta}, we get
\begin{equation}
    \langle \varphi| \otimes \langle \varphi'|)\Delta (|\varphi'\rangle\otimes |\varphi \rangle = 
    -\frac{|B|}{d(d^2-1)}\left(
    \begin{tikzpicture}[scale=0.8, baseline={([yshift=-.5ex]current bounding box.center)}]
    \draw[] (2, 1.5) -- (2, 2.75);
    \draw[] (3,0) -- (3,1);

     \node[fill=blue, regular polygon, regular polygon sides=3, minimum size=0.35cm, inner sep=0pt] (u1) at (3, 0.25) {};
    \node[fill=blue, regular polygon, regular polygon sides=3, minimum size=0.35cm, inner sep=0pt] (u2) at (2, 2.5) {};
    \node[fill=red, regular polygon, regular polygon sides=3, minimum size=0.35cm, inner sep=0pt] (v1) at (3, 0.7) {};
    \node[fill=red, regular polygon, regular polygon sides=3, minimum size=0.35cm, inner sep=0pt, rotate=180] (v2) at (2, 2) {};
    
    \node[left] at (u2) {$u\phantom{l}$};
    \node[left] at (v2) {$v\phantom{l}$};
    
    \node[circle, fill=black, inner sep=0pt, minimum size =0.2cm] (vertex1) at (3,1) {};
    \node[circle, fill=black, inner sep=0pt, minimum size =0.2cm] (vertex2) at (2,1.5) {};
    \node[above] at (vertex1) {$4$};
    \node[below] at (vertex2) {$3$};

    \draw[blue, thick] (u1) -- ++ (0.75, 0) -- ++ (0,2.25) -- (u2);
    \draw[red, thick] (v1) -- ++ (0.5, 0) -- ++ (0, 1.3) --  (v2);
    
    \node[below] at (3,0) {$|\varphi_I' \rangle$};
    \begin{scope}[yscale=-1, yshift=-5.5cm]
    \draw[] (2, 1.5) -- (2, 2.75);
    \draw[] (3,0) -- (3,1);

    \node[fill=blue, regular polygon, regular polygon sides=3, minimum size=0.35cm, inner sep=0pt, rotate=180] (u1) at (3, 0.25) {};
    \node[fill=blue, regular polygon, regular polygon sides=3, minimum size=0.35cm, inner sep=0pt, rotate=180] (u2) at (2, 2.5) {};
    \node[fill=red, regular polygon, regular polygon sides=3, minimum size=0.35cm, inner sep=0pt, rotate=180] (v1) at (3, 0.7) {};
    \node[fill=red, regular polygon, regular polygon sides=3, minimum size=0.35cm, inner sep=0pt] (v2) at (2, 2) {};
    
    \node[left] at (u2) {$u^{\dagger}$};
    \node[left] at (v2) {$v^{\dagger}$};
    
    \node[circle, fill=black, inner sep=0pt, minimum size =0.2cm] (vertex1) at (3,1) {};
    \node[circle, fill=black, inner sep=0pt, minimum size =0.2cm] (vertex2) at (2,1.5) {};
    \node[below] at (vertex1) {$1$};
    \node[above] at (vertex2) {$2$};
    
    \draw[blue, thick] (u1) -- ++ (0.75, 0) -- ++ (0,2.25) -- (u2);
    \draw[red, thick] (v1) -- ++ (0.5, 0) -- ++ (0, 1.3) --  (v2);
    \draw[dashed] (1.375, 2.25) -- (3.625, 2.25) -- (3.625, 0.5) -- (1.375, 0.5) -- cycle;
    
    \node[above] at (3,0) {$\langle \varphi_I|$};
    \end{scope}
    \begin{scope}[xshift=4cm]
    \draw[] (2, 1.5) -- (2, 2.75);
    \draw[] (3,0) -- (3,1);
  
     \node[fill=blue, regular polygon, regular polygon sides=3, minimum size=0.35cm, inner sep=0pt] (u1) at (3, 0.25) {};
    \node[fill=blue, regular polygon, regular polygon sides=3, minimum size=0.35cm, inner sep=0pt] (u2) at (2, 2.5) {};
    \node[fill=red, regular polygon, regular polygon sides=3, minimum size=0.35cm, inner sep=0pt] (v1) at (3, 0.7) {};
    \node[fill=red, regular polygon, regular polygon sides=3, minimum size=0.35cm, inner sep=0pt, rotate=180] (v2) at (2, 2) {};
    
    \node[left] at (u2) {$u\phantom{l}$};
    \node[left] at (v2) {$v\phantom{l}$};
    
    \node[circle, fill=black, inner sep=0pt, minimum size =0.2cm] (vertex1) at (3,1) {};
    \node[circle, fill=black, inner sep=0pt, minimum size =0.2cm] (vertex2) at (2,1.5) {};
    \node[above] at (vertex1) {$1$};
    \node[below] at (vertex2) {$2$};
    \draw[dashed] (1.375, 2.25) -- (3.625, 2.25) -- (3.625, 0.5) -- (1.375, 0.5) -- cycle;

    \draw[blue, thick] (u1) -- ++ (0.75, 0) -- ++ (0,2.25) -- (u2);
    \draw[red, thick] (v1) -- ++ (0.5, 0) -- ++ (0, 1.3) --  (v2);
    
    \node[below] at (3,0) {$|\varphi_I\rangle$};
    \begin{scope}[yscale=-1, yshift=-5.5cm]
    
    \draw[] (2, 1.5) -- (2, 2.75);
    \draw[] (3,0) -- (3,1);

    \node[fill=blue, regular polygon, regular polygon sides=3, minimum size=0.35cm, inner sep=0pt, rotate=180] (u1) at (3, 0.25) {};
    \node[fill=blue, regular polygon, regular polygon sides=3, minimum size=0.35cm, inner sep=0pt, rotate=180] (u2) at (2, 2.5) {};
    \node[fill=red, regular polygon, regular polygon sides=3, minimum size=0.35cm, inner sep=0pt, rotate=180] (v1) at (3, 0.7) {};
    \node[fill=red, regular polygon, regular polygon sides=3, minimum size=0.35cm, inner sep=0pt] (v2) at (2, 2) {};
    
    \node[left] at (u2) {$u^{\dagger}$};
    \node[left] at (v2) {$v^{\dagger}$};

    \node[circle, fill=black, inner sep=0pt, minimum size =0.2cm] (vertex1) at (3,1) {};
    \node[circle, fill=black, inner sep=0pt, minimum size =0.2cm] (vertex2) at (2,1.5) {};
    \node[below] at (vertex1) {$4$};
    \node[above] at (vertex2) {$3$};

    \draw[blue, thick] (u1) -- ++ (0.75, 0) -- ++ (0,2.25) -- (u2);
    \draw[red, thick] (v1) -- ++ (0.5, 0) -- ++ (0, 1.3) --  (v2);
    
    \node[above] at (3,0) {$\langle \varphi_I'|$};
    \end{scope}
    \end{scope}
    \end{tikzpicture}\right).
    \label{eq:pushing1}
\end{equation}
This diagram can be simplified by ``pushing'' the unitaries enclosed in the dashed lines. For instance, consider the unitaries $v^{\dagger}$ and $v$ in Eq.~\eqref{eq:pushing1}. Recalling that the dots labeled by the same integers are connected with each other, these two unitaries can be cancelled in the following way.
\begin{equation}
\begin{aligned}
\begin{tikzpicture}[scale=0.8, baseline={([yshift=-.5ex]current bounding box.center)}]
\draw[] (0,0) -- (0,2);
\draw[] (2,0) -- (2,2);
\node[circle, fill=black, inner sep=0pt, minimum size =0.2cm] (vertex2) at (0,1) {};
\node[circle, fill=black, inner sep=0pt, minimum size =0.2cm] (vertex1) at (2,1) {};
\node[left] at (vertex2) {$2$};
\node[right] at (vertex1) {$1$};
\node[fill=red, regular polygon, regular polygon sides=3, minimum size=0.35cm, inner sep=0pt] (v1) at (2, 0.5) {};
\node[fill=red, regular polygon, regular polygon sides=3, minimum size=0.35cm, inner sep=0pt, rotate=180] (v2) at (0, 1.5) {};
\draw[red, thick] (v1)-- (v2);
\node[left] at (v2) {$v\phantom{l}$};

\node[fill=white, minimum size=0.25cm, inner sep=0pt] () at (1,1) {};

\node[fill=red, regular polygon, regular polygon sides=3, minimum size=0.35cm, inner sep=0pt, rotate=180] (v1d) at (2, 1.5) {};
\node[fill=red, regular polygon, regular polygon sides=3, minimum size=0.35cm, inner sep=0pt] (v2d) at (0, 0.5) {};
\draw[red, thick] (v1d)-- (v2d);
\node[left] at (v2d) {$v^{\dagger}$};
\end{tikzpicture}
&= 
\begin{tikzpicture}[scale=0.8, baseline={([yshift=-.5ex]current bounding box.center)}]
\draw[] (0,0) -- (0,2);
\draw[] (2,0) -- (2,2);
\draw[] (2,0) -- (2.5,0) -- (2.5,2) -- (3,2) -- (3,0);
\node[circle, fill=black, inner sep=0pt, minimum size =0.2cm] (vertex2) at (0,1) {};
\node[circle, fill=black, inner sep=0pt, minimum size =0.2cm] (vertex1) at (2,1) {};
\node[left] at (vertex2) {$2$};
\node[right] at (vertex1) {$1$};
\node[fill=red, regular polygon, regular polygon sides=3, minimum size=0.35cm, inner sep=0pt] (v1) at (2, 0.5) {};
\node[fill=red, regular polygon, regular polygon sides=3, minimum size=0.35cm, inner sep=0pt, rotate=180] (v2) at (0, 1.5) {};
\draw[red, thick] (v1)-- (v2);
\node[left] at (v2) {$v\phantom{l}$};

\node[fill=white, minimum size=0.25cm, inner sep=0pt] () at (1,1) {};

\node[fill=red, regular polygon, regular polygon sides=3, minimum size=0.35cm, inner sep=0pt, rotate=180] (v1d) at (2, 1.5) {};
\node[fill=red, regular polygon, regular polygon sides=3, minimum size=0.35cm, inner sep=0pt] (v2d) at (0, 0.5) {};
\draw[red, thick] (v1d)-- (v2d);
\node[left] at (v2d) {$v^{\dagger}$};
\end{tikzpicture}\\
&= 
\begin{tikzpicture}[scale=0.8, baseline={([yshift=-.5ex]current bounding box.center)}]
\draw[] (0,0) -- (0,2);
\draw[] (2,0) -- (2,2);
\draw[] (2,0) -- (2.5,0) -- (2.5,2) -- (3,2) -- (3,0);
\node[circle, fill=black, inner sep=0pt, minimum size =0.2cm] (vertex2) at (0,1) {};
\node[circle, fill=black, inner sep=0pt, minimum size =0.2cm] (vertex1) at (2,1) {};
\node[left] at (vertex2) {$2$};
\node[right] at (vertex1) {$1$};
\node[fill=red, regular polygon, regular polygon sides=3, minimum size=0.35cm, inner sep=0pt, rotate=180] (v1) at (2.5, 1.5) {};
\node[fill=red, regular polygon, regular polygon sides=3, minimum size=0.35cm, inner sep=0pt, rotate=180] (v2) at (0, 1.5) {};
\draw[red, thick] (v1)-- (v2);
\node[left] at (v2) {$v\phantom{l}$};

\node[fill=white, minimum size=0.25cm, inner sep=0pt] () at (1,1) {};

\node[fill=red, regular polygon, regular polygon sides=3, minimum size=0.35cm, inner sep=0pt] (v1d) at (2.5, 0.5) {};
\node[fill=red, regular polygon, regular polygon sides=3, minimum size=0.35cm, inner sep=0pt] (v2d) at (0, 0.5) {};
\draw[red, thick] (v1d)-- (v2d);
\node[left] at (v2d) {$v^{\dagger}$};
\end{tikzpicture}\\
&= 
\begin{tikzpicture}[scale=0.8, baseline={([yshift=-.5ex]current bounding box.center)}]
\draw[] (0,0) -- (0,2);
\draw[] (2,0) -- (2,2);
\draw[] (2,0) -- (2.5,0) -- (2.5,2) -- (3,2) -- (3,0);
\node[circle, fill=black, inner sep=0pt, minimum size =0.2cm] (vertex2) at (0,1) {};
\node[circle, fill=black, inner sep=0pt, minimum size =0.2cm] (vertex1) at (2,1) {};
\node[left] at (vertex2) {$2$};
\node[right] at (vertex1) {$1$};
\end{tikzpicture}
\\
&= 
\begin{tikzpicture}[scale=0.8, baseline={([yshift=-.5ex]current bounding box.center)}]
\draw[] (0,0) -- (0,2);
\draw[] (2,0) -- (2,2);
\node[circle, fill=black, inner sep=0pt, minimum size =0.2cm] (vertex2) at (0,1) {};
\node[circle, fill=black, inner sep=0pt, minimum size =0.2cm] (vertex1) at (2,1) {};
\node[left] at (vertex2) {$2$};
\node[right] at (vertex1) {$1$};
\end{tikzpicture},
\end{aligned}
\end{equation}
Applying the same identity over the dots $3$ and $4$, we remove the other occurrence of $v$ and $v^\dagger$ from the diagram, obtaining
\begin{equation}
 (\langle \varphi| \otimes \langle \varphi'|) \Delta (|\varphi'\rangle \otimes |\varphi\rangle)= \,\,
 -\frac{|B|}{d(d^2-1)} \left(
    \begin{tikzpicture}[scale=0.8, baseline={([yshift=-.5ex]current bounding box.center)}]
    \draw[] (2, 1.5) -- (2, 2.75);
    \draw[] (3,0) -- (3,1);

     \node[fill=blue, regular polygon, regular polygon sides=3, minimum size=0.35cm, inner sep=0pt] (u1) at (3, 0.25) {};
    \node[fill=blue, regular polygon, regular polygon sides=3, minimum size=0.35cm, inner sep=0pt] (u2) at (2, 2.5) {};
    
    \node[left] at (u2) {$u\phantom{l}$};
    
    \node[circle, fill=black, inner sep=0pt, minimum size =0.2cm] (vertex1) at (3,1) {};
    \node[circle, fill=black, inner sep=0pt, minimum size =0.2cm] (vertex2) at (2,1.5) {};
    \node[above] at (vertex1) {$4$};
    \node[below] at (vertex2) {$3$};
    
    \draw[blue, thick] (u1) -- ++ (0.75, 0) -- ++ (0,2.25) -- (u2);
    
    \node[below] at (3,0) {$|\varphi_I'\rangle$};
    \begin{scope}[yscale=-1, yshift=-5.5cm]
    \draw[] (2, 1.5) -- (2, 2.75);
    \draw[] (3,0) -- (3,1);

    \node[fill=blue, regular polygon, regular polygon sides=3, minimum size=0.35cm, inner sep=0pt, rotate=180] (u1) at (3, 0.25) {};
    \node[fill=blue, regular polygon, regular polygon sides=3, minimum size=0.35cm, inner sep=0pt, rotate=180] (u2) at (2, 2.5) {};
    
    \node[left] at (u2) {$u^{\dagger}$};
    
    \node[circle, fill=black, inner sep=0pt, minimum size =0.2cm] (vertex1) at (3,1) {};
    \node[circle, fill=black, inner sep=0pt, minimum size =0.2cm] (vertex2) at (2,1.5) {};
    \node[below] at (vertex1) {$1$};
    \node[above] at (vertex2) {$2$};
    
    \draw[blue, thick] (u1) -- ++ (0.75, 0) -- ++ (0,2.25) -- (u2);
    
    \node[above] at (3,0) {$\langle \varphi_I|$};
    \end{scope}
    \begin{scope}[xshift=4cm]
    \draw[] (2, 1.5) -- (2, 2.75);
    \draw[] (3,0) -- (3,1);
  
     \node[fill=blue, regular polygon, regular polygon sides=3, minimum size=0.35cm, inner sep=0pt] (u1) at (3, 0.25) {};
    \node[fill=blue, regular polygon, regular polygon sides=3, minimum size=0.35cm, inner sep=0pt] (u2) at (2, 2.5) {};
    
    \node[left] at (u2) {$u\phantom{l}$};
    
    \node[circle, fill=black, inner sep=0pt, minimum size =0.2cm] (vertex1) at (3,1) {};
    \node[circle, fill=black, inner sep=0pt, minimum size =0.2cm] (vertex2) at (2,1.5) {};
    \node[above] at (vertex1) {$1$};
    \node[below] at (vertex2) {$2$};

    \draw[blue, thick] (u1) -- ++ (0.75, 0) -- ++ (0,2.25) -- (u2);
    
    \node[below] at (3,0) {$|\varphi_I\rangle$};
    \begin{scope}[yscale=-1, yshift=-5.5cm]
    
    \draw[] (2, 1.5) -- (2, 2.75);
    \draw[] (3,0) -- (3,1);

    \node[fill=blue, regular polygon, regular polygon sides=3, minimum size=0.35cm, inner sep=0pt, rotate=180] (u1) at (3, 0.25) {};
    \node[fill=blue, regular polygon, regular polygon sides=3, minimum size=0.35cm, inner sep=0pt, rotate=180] (u2) at (2, 2.5) {};
    
    \node[left] at (u2) {$u^{\dagger}$};
    
    \node[circle, fill=black, inner sep=0pt, minimum size =0.2cm] (vertex1) at (3,1) {};
    \node[circle, fill=black, inner sep=0pt, minimum size =0.2cm] (vertex2) at (2,1.5) {};
    \node[below] at (vertex1) {$4$};
    \node[above] at (vertex2) {$3$};

    \draw[blue, thick] (u1) -- ++ (0.75, 0) -- ++ (0,2.25) -- (u2);
    
    \node[above] at (3,0) {$\langle \varphi_I'|$};
    \end{scope}
    \end{scope}
    \end{tikzpicture}\right).
    \label{eq:pushing2}
\end{equation}

We can also push the unitaries $u^{\dagger}$ and $u$, obtaining
\begin{equation}
\begin{aligned}
 (\langle \varphi| \otimes \langle \varphi'|) \Delta (|\varphi'\rangle \otimes |\varphi\rangle) &=\, - \frac{|B|}{d(d^2-1)}\left(
    \begin{tikzpicture}[scale=0.8, baseline={([yshift=-.5ex]current bounding box.center)}]
    \draw[] (2, 1.5) -- (2, 2.75);
    \draw[] (3,0) -- (3,1);

    \node[circle, fill=black, inner sep=0pt, minimum size =0.2cm] (vertex1) at (3,1) {};
    \node[circle, fill=black, inner sep=0pt, minimum size =0.2cm] (vertex2) at (2,1.5) {};
    \node[above] at (vertex1) {$4$};
    \node[below] at (vertex2) {$3$};

    \node[below] at (3,0) {$|\varphi_I'\rangle$};
    \begin{scope}[yscale=-1, yshift=-5.5cm]
    \draw[] (2, 1.5) -- (2, 2.75);
    \draw[] (3,0) -- (3,1);

    \node[circle, fill=black, inner sep=0pt, minimum size =0.2cm] (vertex1) at (3,1) {};
    \node[circle, fill=black, inner sep=0pt, minimum size =0.2cm] (vertex2) at (2,1.5) {};
    \node[below] at (vertex1) {$1$};
    \node[above] at (vertex2) {$2$};

    \node[above] at (3,0) {$\langle \varphi_I|$};
    \end{scope}
    \begin{scope}[xshift=4cm]
    \draw[] (2, 1.5) -- (2, 2.75);
    \draw[] (3,0) -- (3,1);
  
     \node[fill=blue, regular polygon, regular polygon sides=3, minimum size=0.35cm, inner sep=0pt] (u1) at (3, 0.25) {};
    \node[fill=blue, regular polygon, regular polygon sides=3, minimum size=0.35cm, inner sep=0pt] (u2) at (2, 2.5) {};
    \node[fill=blue, regular polygon, regular polygon sides=3, minimum size=0.35cm, inner sep=0pt, rotate=180] (v1) at (3, 0.7) {};
    \node[fill=blue, regular polygon, regular polygon sides=3, minimum size=0.35cm, inner sep=0pt, rotate=180] (v2) at (2, 2) {};
    
    \node[left] at (u2) {$u\phantom{l}$};
    \node[left] at (v2) {$u^{\dagger}$};
    
    \node[circle, fill=black, inner sep=0pt, minimum size =0.2cm] (vertex1) at (3,1) {};
    \node[circle, fill=black, inner sep=0pt, minimum size =0.2cm] (vertex2) at (2,1.5) {};
    \node[above] at (vertex1) {$1$};
    \node[below] at (vertex2) {$2$};

    \draw[blue, thick] (u1) -- ++ (0.75, 0) -- ++ (0,2.25) -- (u2);
    \draw[blue, thick] (v1) -- ++ (0.5, 0) -- ++ (0, 1.3) --  (v2);
    
    \node[below] at (3,0) {$|\varphi_I\rangle$};
    \begin{scope}[yscale=-1, yshift=-5.5cm]
    
    \draw[] (2, 1.5) -- (2, 2.75);
    \draw[] (3,0) -- (3,1);

    \node[fill=blue, regular polygon, regular polygon sides=3, minimum size=0.35cm, inner sep=0pt, rotate=180] (u1) at (3, 0.25) {};
    \node[fill=blue, regular polygon, regular polygon sides=3, minimum size=0.35cm, inner sep=0pt, rotate=180] (u2) at (2, 2.5) {};
    \node[fill=blue, regular polygon, regular polygon sides=3, minimum size=0.35cm, inner sep=0pt] (v1) at (3, 0.7) {};
    \node[fill=blue, regular polygon, regular polygon sides=3, minimum size=0.35cm, inner sep=0pt] (v2) at (2, 2) {};
    
    \node[left] at (u2) {$u^{\dagger}$};
    \node[left] at (v2) {$u\phantom{l}$};
    
    \node[circle, fill=black, inner sep=0pt, minimum size =0.2cm] (vertex1) at (3,1) {};
    \node[circle, fill=black, inner sep=0pt, minimum size =0.2cm] (vertex2) at (2,1.5) {};
    \node[below] at (vertex1) {$4$};
    \node[above] at (vertex2) {$3$};

    \draw[blue, thick] (u1) -- ++ (0.75, 0) -- ++ (0,2.25) -- (u2);
    \draw[blue, thick] (v1) -- ++ (0.5, 0) -- ++ (0, 1.3) --  (v2);
    
    \node[above] at (3,0) {$\langle \varphi_I'|$};
    \end{scope}
    \end{scope}
    \end{tikzpicture} \right)\\
    &=\,
    -\frac{|B|}{d(d^2-1)} \left(
    \begin{tikzpicture}[scale=0.8, baseline={([yshift=-.5ex]current bounding box.center)}]
    \draw[] (3,0) -- (3,1);
    \node[circle, fill=black, inner sep=0pt, minimum size =0.2cm] (vertex1) at (3,1) {};
    \node[above] at (vertex1) {$4$};
    
    \node[below] at (3,0) {$|\varphi_I'\rangle$};
    \begin{scope}[yscale=-1, yshift=-5.5cm]
    \draw[] (3,0) -- (3,1);
    \node[circle, fill=black, inner sep=0pt, minimum size =0.2cm] (vertex1) at (3,1) {};
    \node[below] at (vertex1) {$1$};
    
    \node[above] at (3,0) {$\langle \varphi_I|$};
    \end{scope}
    \begin{scope}[xshift=4cm]
    \draw[] (2, 1.5) -- (2, 2.75);
    \draw[] (3,0) -- (3,1);
  
     \node[fill=blue, regular polygon, regular polygon sides=3, minimum size=0.35cm, inner sep=0pt] (u1) at (3, 0.25) {};
    \node[fill=blue, regular polygon, regular polygon sides=3, minimum size=0.35cm, inner sep=0pt] (u2) at (2, 2.5) {};
    \node[fill=blue, regular polygon, regular polygon sides=3, minimum size=0.35cm, inner sep=0pt, rotate=180] (v1) at (3, 0.7) {};
    \node[fill=blue, regular polygon, regular polygon sides=3, minimum size=0.35cm, inner sep=0pt, rotate=180] (v2) at (2, 2) {};
    
    \node[left] at (u2) {$u\phantom{l}$};
    \node[left] at (v2) {$u^{\dagger}$};
    
    \node[circle, fill=black, inner sep=0pt, minimum size =0.2cm] (vertex1) at (3,1) {};
    \node[] (vertex2) at (2,1.5) {};
    \node[above] at (vertex1) {$1$};

    \draw[blue, thick] (u1) -- ++ (0.75, 0) -- ++ (0,2.25) -- (u2);
    \draw[blue, thick] (v1) -- ++ (0.5, 0) -- ++ (0, 1.3) --  (v2);
    
    \node[below] at (3,0) {$|\varphi_I\rangle$};
    \begin{scope}[yscale=-1, yshift=-5.5cm]
    
    \draw[] (2, 1.5) -- (2, 2.75);
    \draw[] (3,0) -- (3,1);

    \node[fill=blue, regular polygon, regular polygon sides=3, minimum size=0.35cm, inner sep=0pt, rotate=180] (u1) at (3, 0.25) {};
    \node[fill=blue, regular polygon, regular polygon sides=3, minimum size=0.35cm, inner sep=0pt, rotate=180] (u2) at (2, 2.5) {};
    \node[fill=blue, regular polygon, regular polygon sides=3, minimum size=0.35cm, inner sep=0pt] (v1) at (3, 0.7) {};
    \node[fill=blue, regular polygon, regular polygon sides=3, minimum size=0.35cm, inner sep=0pt] (v2) at (2, 2) {};
    
    \node[left] at (u2) {$u^{\dagger}$};
    \node[left] at (v2) {$u\phantom{l}$};
    
    \node[circle, fill=black, inner sep=0pt, minimum size =0.2cm] (vertex1) at (3,1) {};
    \node[] (vertex2) at (2,1.5) {};
    \node[below] at (vertex1) {$4$};
    \draw[] (2, 1.5) -- ++ (-1, 0) -- ++ (0, 2.5) -- ++ (1,0);
    \node[left] () at (1, 2.5) {$R$};

    \draw[blue, thick] (u1) -- ++ (0.75, 0) -- ++ (0,2.25) -- (u2);
    \draw[blue, thick] (v1) -- ++ (0.5, 0) -- ++ (0, 1.3) --  (v2);
    
    \node[above] at (3,0) {$\langle \varphi_I'|$};
    \end{scope}
    \end{scope}
    \end{tikzpicture}\right),
    \end{aligned}
\end{equation}
where the closed loop represents a partial trace over $R$. Now we can straighten out the legs using the algebraic identity
\begin{equation}
    \langle \phi| A |\phi'\rangle = \langle (\phi')^*| A^T |(\phi)^*\rangle,
\end{equation}
where $A^T$ is a transpose of $A$ (in some basis) and $|(\phi)^*\rangle$ is an entry-wise complex conjugation of $|\phi\rangle$ (in the same basis). We thus see
\begin{equation}
    \begin{aligned}
        |(\langle \varphi|\otimes \langle \varphi'|)\Delta (|\varphi'\rangle \otimes |\varphi\rangle)| = 
        \left|\,\,\,\frac{|B|}{d(d^2-1)}\left(
    \begin{tikzpicture}[scale=0.8, baseline={([yshift=-.5ex]current bounding box.center)}]
    \draw[] (2, 0.5) -- (2, 2.5);
    \draw[] (3,0.25) -- (3,1.75);
  
    \node[fill=blue, regular polygon, regular polygon sides=3, minimum size=0.35cm, inner sep=0pt, rotate=180] (u1) at (3, 1.25) {};
    \node[fill=blue, regular polygon, regular polygon sides=3, minimum size=0.35cm, inner sep=0pt] (u2) at (2, 1.25) {};
    \node[fill=blue, regular polygon, regular polygon sides=3, minimum size=0.35cm, inner sep=0pt] (v1) at (3, 0.75) {};
    \node[fill=blue, regular polygon, regular polygon sides=3, minimum size=0.35cm, inner sep=0pt, rotate=180] (v2) at (2, 0.75) {};
    
    \node[left] at (u2) {$u\phantom{l}$};
    \node[left] at (v2) {$u^{\dagger}$};
    
    \node[] (vertex2) at (2,1.5) {};

    \draw[blue, thick] (u1) -- (u2);
    \draw[blue, thick] (v1) --  (v2);
    
    \node[below] at (3,0.25) {$|(\varphi_I)^*\rangle$};
    \node[above] at (3,1.75) {$\langle(\varphi_I)^*|$};
    \begin{scope}[yscale=-1, yshift=-5cm]
    
    \draw[] (2, 0.5) -- (2, 2.5);
    \draw[] (3,0.25) -- (3,1.75);

    \node[fill=blue, regular polygon, regular polygon sides=3, minimum size=0.35cm, inner sep=0pt] (u1) at (3, 1.25) {};
    \node[fill=blue, regular polygon, regular polygon sides=3, minimum size=0.35cm, inner sep=0pt, rotate=180] (u2) at (2, 1.25) {};
    \node[fill=blue, regular polygon, regular polygon sides=3, minimum size=0.35cm, inner sep=0pt, rotate=180] (v1) at (3, 0.75) {};
    \node[fill=blue, regular polygon, regular polygon sides=3, minimum size=0.35cm, inner sep=0pt] (v2) at (2, 0.75) {};
    
    \node[left] at (u2) {$u^{\dagger}$};
    \node[left] at (v2) {$u\phantom{l}$};

    \draw[] (2, 0.5) -- ++ (-1, 0) -- ++ (0, 4) -- ++ (1,0);
    
    \node[left] () at (1, 2.5) {$R$};

    \draw[blue, thick] (u1)  -- (u2);
    \draw[blue, thick] (v1) --  (v2);
    
    \node[above] at (3,0.25) {$\langle (\varphi_I')^*|$};
    \node[below] at (3, 1.75) {$|(\varphi_I')^*\rangle$};
    \end{scope}
    \end{tikzpicture}\,\,\,\right)
        \right|.
    \end{aligned}
    \label{eq:pushing3}
\end{equation}
While this is a considerable simplification, this diagram contains partially transposed unitaries. 

In fact, we can convert the diagrammatic expression in Eq.~\eqref{eq:pushing3} to the one which does not contain partially transposed unitaries. Recalling the definition of the partial transpose, we note the following identities:
\begin{equation}
    \begin{tikzpicture}[scale=0.8, baseline={([yshift=-.5ex]current bounding box.center)}]
    \node[fill=blue, regular polygon, regular polygon sides=3, minimum size=0.35cm, inner sep=0pt] (u1) at (0, 0) {};
    \node[fill=blue, regular polygon, regular polygon sides=3, minimum size=0.35cm, inner sep=0pt, rotate=180] (u2) at (1, 0) {};
    
    \draw[] (u1) -- (u2);
    \draw[] (u1) -- ++ (0, 0.5);
    \draw[] (u1) -- ++ (0, -0.5);
    \draw[] (u2) -- ++ (0, 0.5);
    \draw[] (u2) -- ++ (0, -0.5);
    \node[left] () at (u1) {$u\,$};
    \end{tikzpicture}
    = 
    \begin{tikzpicture}[scale=0.8, baseline={([yshift=-.5ex]current bounding box.center)}]    
    \draw[] (0, -1.25) -- (0, 1.25);
    \node[fill=blue, regular polygon, regular polygon sides=3, minimum size=0.35cm, inner sep=0pt] (u1) at (0, 0) {};
    \node[fill=blue, regular polygon, regular polygon sides=3, minimum size=0.35cm, inner sep=0pt] (u2) at (1, 0) {};
    
    \draw[] (u1) -- (u2);
    \draw[] plot [smooth, tension=0.8] coordinates {(1,0.125) (1.125, 0.375) (1.5, 0.25) (1.25, -0.25) (1.125, -0.625) (1.05, -1) (1,-1.25)};
    \node[fill=white, draw=white] () at (1.2, -0.6) {};
    \draw[] plot [smooth, tension=0.8] coordinates {(1, 0.125) (1.05, -0.5) (1.5, -0.875) (1.875, -0.375) (1.75, 0.25) (1.25, 0.625) (1.05, 1) (1, 1.25)};
    \node[left] () at (u1) {$u\,$};
    \end{tikzpicture}
    \qquad \textrm{ and } \qquad
    \begin{tikzpicture}[scale=0.8, baseline={([yshift=-.5ex]current bounding box.center)}]
    \node[fill=blue, regular polygon, regular polygon sides=3, minimum size=0.35cm, inner sep=0pt, rotate=180] (u1) at (0, 0) {};
    \node[fill=blue, regular polygon, regular polygon sides=3, minimum size=0.35cm, inner sep=0pt] (u2) at (1, 0) {};
    
    \draw[] (u1) -- (u2);
    \draw[] (u1) -- ++ (0, 0.5);
    \draw[] (u1) -- ++ (0, -0.5);
    \draw[] (u2) -- ++ (0, 0.5);
    \draw[] (u2) -- ++ (0, -0.5);
    \node[left] () at (u1) {$u^{\dagger}$};
    \end{tikzpicture}
    = 
    \begin{tikzpicture}[scale=0.8, baseline={([yshift=-.5ex]current bounding box.center)}]
    \node[fill=blue, regular polygon, regular polygon sides=3, minimum size=0.35cm, inner sep=0pt, rotate=180] (u1) at (0, 0) {};
    \node[fill=blue, regular polygon, regular polygon sides=3, minimum size=0.35cm, inner sep=0pt, rotate=180] (u2) at (1, 0) {};
    \draw[] (0, -1.25) -- (0, 1.25);
    \draw[] (u1) -- (u2);
    \draw[] plot [smooth, tension=0.8] coordinates {(1,0.125) (1.125, 0.375) (1.5, 0.25) (1.25, -0.25) (1.125, -0.625) (1.05, -1) (1,-1.25)};
    \node[fill=white, draw=white] () at (1.2, -0.6) {};
    \draw[] plot [smooth, tension=0.8] coordinates {(1, 0.125) (1.05, -0.5) (1.5, -0.875) (1.875, -0.375) (1.75, 0.25) (1.25, 0.625) (1.05, 1) (1, 1.25)};
    \node[left] () at (u1) {$u^{\dagger}$};
    \end{tikzpicture}
    .
\end{equation}
We thus arrive at the following identities:
\begin{equation}
\begin{aligned}
    \begin{tikzpicture}[scale=0.8, baseline={([yshift=-.5ex]current bounding box.center)}]
    
    \node[above] () at (0, 2.25) {$R$};
    \node[below] () at (0,-1.5) {$R$};
    \draw[] (0, -1.5) -- (0, 2.25);
    \draw[] (1, -0.5) -- (1, 1.25);
    
    \node[fill=blue, regular polygon, regular polygon sides=3, minimum size=0.35cm, inner sep=0pt, rotate=180] (u1d) at (0, 0) {};
    \node[fill=blue, regular polygon, regular polygon sides=3, minimum size=0.35cm, inner sep=0pt] (u2d) at (1, 0) {};
    \node[left] () at (u1d) {$u^{\dagger}$};
    
    \node[fill=blue, regular polygon, regular polygon sides=3, minimum size=0.35cm, inner sep=0pt] (u1) at (0, 0.75) {};
    \node[fill=blue, regular polygon, regular polygon sides=3, minimum size=0.35cm, inner sep=0pt, rotate=180] (u2) at (1, 0.75) {};
    \node[left] () at (u1) {$u\,$};
    
    \draw[] (u1d) -- (u2d);
    \draw[] (u1) -- (u2);
    \node[above] () at (1, 1.25) {$\langle (\varphi_I')^*|)$};
    \node[below] () at (1, -0.5) {$|(\varphi_I')^*\rangle$};
    \end{tikzpicture}
     = 
    \begin{tikzpicture}[scale=0.8, baseline={([yshift=-.5ex]current bounding box.center)}]
    
    \node[below] () at (0,-2) {$R$};
    \node[below] () at (1, -1.25) {$|(\varphi_I')^*\rangle$};
    \node[fill=blue, regular polygon, regular polygon sides=3, minimum size=0.35cm, inner sep=0pt, rotate=180] (u1d) at (0, 0) {};
    \node[fill=blue, regular polygon, regular polygon sides=3, minimum size=0.35cm, inner sep=0pt, rotate=180] (u2d) at (1, 0) {};
    \draw[] (u1d) -- (u2d);
    \node[left] () at (u1d) {$u^{\dagger}$};
    
    \draw[] (0, -2) -- (0, 1.25);
    \draw[] plot [smooth, tension=0.8] coordinates {(1,0.125) (1.125, 0.375) (1.5, 0.25) (1.25, -0.25) (1.125, -0.625) (1.05, -1) (1,-1.25)};
    \node[fill=white, draw=white] () at (1.2, -0.6) {};
    \draw[] plot [smooth, tension=0.8] coordinates {(1, 0.125) (1.05, -0.5) (1.5, -0.875) (1.875, -0.375) (1.75, 0.25) (1.25, 0.625) (1.05, 1) (1, 1.25)};
    
    \begin{scope}[yshift=2.5cm]
    \node[above] () at (0, 2) {$R$};
    \node[above] () at (1, 1.25) {$\langle (\varphi_I')^*|)$};
    
    \node[fill=blue, regular polygon, regular polygon sides=3, minimum size=0.35cm, inner sep=0pt] (u1) at (0, 0) {};
    \node[fill=blue, regular polygon, regular polygon sides=3, minimum size=0.35cm, inner sep=0pt] (u2) at (1, 0) {};
    \draw[] (u1) -- (u2);
    \node[left] () at (u1) {$u\,$};
    
    \draw[] (0, -1.25) -- (0, 2);
    \draw[] plot [smooth, tension=0.8] coordinates {(1,0.125) (1.125, 0.375) (1.5, 0.25) (1.25, -0.25) (1.125, -0.625) (1.05, -1) (1,-1.25)};
    \node[fill=white, draw=white] () at (1.2, -0.6) {};
    \draw[] plot [smooth, tension=0.8] coordinates {(1, 0.125) (1.05, -0.5) (1.5, -0.875) (1.875, -0.375) (1.75, 0.25) (1.25, 0.625) (1.05, 1) (1, 1.25)};
    \end{scope}
    \end{tikzpicture}
    =
    \begin{tikzpicture}[scale=0.8, baseline={([yshift=-.5ex]current bounding box.center)}]
    \node[below] () at (0,-1.25) {$R$};
    \draw[] (1,-0.5) -- ++ (0.75, 0) -- ++ (0, 1.75);
    \draw[] (1,0.5) -- (1,-0.5);
    \node[above] () at (1, 0.5) {$\langle\varphi_I'|$};
    
    \node[fill=blue, regular polygon, regular polygon sides=3, minimum size=0.35cm, inner sep=0pt, rotate=180] (u1d) at (0, 0) {};
    \node[fill=blue, regular polygon, regular polygon sides=3, minimum size=0.35cm, inner sep=0pt, rotate=180] (u2d) at (1, 0) {};
    \draw[] (u1d) -- (u2d);
    \node[left] () at (u1d) {$u^{\dagger}$};
    
    \draw[] (0, -1.25) -- (0, 1.25);

    \begin{scope}[yshift=2.5cm]    
    \node[above] () at (0, 1.25) {$R$};
    \draw[] (1,0.5) -- (1,-0.5);
    \node[below] () at (1, -0.5) {$| \varphi_I'\rangle$};
    \node[fill=blue, regular polygon, regular polygon sides=3, minimum size=0.35cm, inner sep=0pt] (u1) at (0, 0) {};
    \node[fill=blue, regular polygon, regular polygon sides=3, minimum size=0.35cm, inner sep=0pt] (u2) at (1, 0) {};
    \draw[] (u1) -- (u2);
    \node[left] () at (u1) {$u\,$};

    \draw[] (0, -1.25) -- (0, 1.25);
    \draw[] (1,0.5) -- ++ (0.75, 0) -- ++ (0, -1.75);
    \node[right] () at (1.75, -1.25) {$I$};
    \end{scope}
    \end{tikzpicture}.
\end{aligned}
\end{equation}
This can be formally viewed as an operator acting on $R$, with a norm bounded by 
\begin{equation}
    \|\text{Tr}_I(u^{\dagger} (I_R \otimes |\varphi_I'\rangle\langle \varphi_I'|) u) \| \leq |I|.
\end{equation}
An analogous bound can be obtained for the bottom half of in Eq.~\eqref{eq:pushing3}. After taking a trace over $R$, and recalling that $d=|B| |R|$, we thus arrive at the following bound
\begin{equation}
    |(\langle \varphi| \otimes \langle \varphi'|)\Delta (|\varphi'\rangle \otimes |\varphi\rangle)| \leq \frac{|I|^2}{d^2-1}
\end{equation}
for any $|\varphi\rangle$ and $|\varphi'\rangle$ which are product states over $I$ and $\ell$.

One may ask how our conclusion changes if we allow states such that $I$ and $\ell$ are entangled. In this case, without loss of generality we can assume that $|\varphi\rangle$ and $|\varphi'\rangle$ admit the Schmidt decomposition
\begin{equation}
    \begin{aligned}
        |\varphi\rangle &= \sum_k \sqrt{p_k}|\varphi_k\rangle, \qquad &|\varphi_k\rangle := |\varphi_{I,k}\rangle \otimes |\varphi_{\ell,k}\rangle, \\
        |\varphi'\rangle &= \sum_k \sqrt{p_k'} |\varphi_k'\rangle, \qquad &|\varphi_k'\rangle := |\varphi_{I,k}'\rangle \otimes |\varphi_{\ell,k}'\rangle,
    \end{aligned}
\end{equation}
where $\{|\varphi_{I,k}\rangle\}, \{|\varphi_{I,k}'\rangle\} \subset \mathcal{H}_I$ and $\{|\varphi_{\ell,k}\rangle \}, \{|\varphi_{\ell,k}'\rangle \} \subset \mathcal{H}_{\ell}$ are orthonormal basis sets, with the index $k$ ranging from $0$ to $|I| -1$, assuming $|I| \leq |\ell|$. Thus, the first term can be bounded by
\begin{equation}
\begin{aligned}
    |\left( \langle \varphi| \otimes \langle \varphi'| \right)| \Delta (|\varphi'\rangle \otimes |\varphi\rangle) &\leq \left(\sum_{k_1, k_2, k_3, k_4} \sqrt{p_{k_1} p_{k_2} p_{k_3} p_{k_4}}\right) \max_{k, k'} (|\langle \varphi_k| \otimes \langle \varphi_{k'}'|)\Delta (|\varphi_{k'}'\rangle\otimes |\varphi_k\rangle|) \\
    &\leq |I|^2 \max_{k, k'} (|\langle \varphi_k| \otimes \langle \varphi_{k'}'|)\Delta (|\varphi_{k'}'\rangle\otimes |\varphi_k\rangle|).
\end{aligned}
\end{equation}
We can thus bound the variance for general states $|\varphi\rangle, |\varphi'\rangle \in \mathcal{H}_{I}\otimes \mathcal{H}_{\ell}$ in terms of the worst-case variance for product states, up to an extra factor of $|I|^2$. Accounting for this factor, we obtain
\begin{equation}
     |(\langle \varphi| \otimes \langle \varphi'|)\Delta (|\varphi'\rangle \otimes |\varphi\rangle)| \leq \frac{|I|^4}{d^2-1}.
\end{equation}

To summarize, the fluctuations of $V(U')^{\dagger }V(U')$ about its mean value can be bounded as
\begin{equation}
\boxed{    \int dU' \left| \langle \varphi| V(U')^{\dagger }V(U')|\varphi'\rangle-\langle \varphi|\varphi'\rangle  \right|^2  \leq 
    \begin{cases}
    \frac{|I|^2-1}{d^2-1} & \textrm{ if } |\varphi\rangle, |\varphi'\rangle\in \mathcal{H}_I\otimes \mathcal{H}_{\ell} \textrm{ are product states,} \\
    \frac{|I|^4-1}{d^2-1} & \textrm{more generally,}
    \end{cases}}
\end{equation}
where $d = |BR|$.
In both cases, as long as the initial black hole is macroscopic and the infaller is small in comparison, the fluctuation is small.

\section{Pseudorandomness}
\label{sec:pseudorandomness}
\subsection{Pseudorandom unitary transformations}
\label{subsec:pseudo-unitary}

The analysis in Section~\ref{sec:average} establishes that, for Haar-random unitary $U'$, the black hole $S$-matrix $V(U')$ is very nearly unitary with high probability. However, a typical random unitary has computational complexity which scales exponentially with the initial entropy of the black hole, while for a realistic black hole we expect $U'$ to have polynomial computational complexity. Can we conclude as in Section~\ref{sec:average} that deviations of $V(U')$ from unitarity are very small even if $U'$ has polynomial complexity? Here we answer this question in the affirmative under reasonable assumptions.

The key underlying assumption is the pseudorandomness of the unitary $U'$~\cite{Kim2020}. In the rest of this Section, we introduce this concept and argue that this assumption is plausible for a realistic black hole. We then show in Section~\ref{sec:unitarity_pseudorandomness} that the pseudorandomness assumption implies that $V(U')$ is very close to unitary. Moreover, we show in Section~\ref{sec:complexity} that the complexity of $V(U')$ scales polynomially in the initial black hole entropy, despite the postselection inherent in the holographic map. 

Let us first define pseudorandomness in a general setup. Consider the physical process
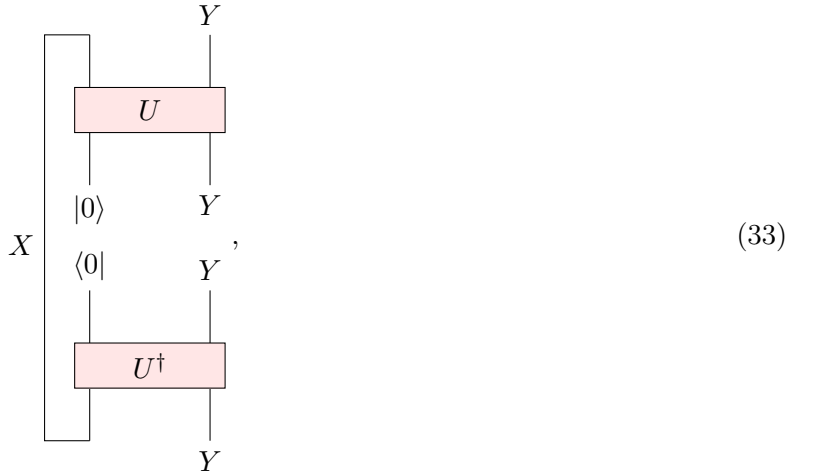
\begin{equation}
    \begin{tikzpicture}[scale=0.8, baseline={([yshift=-.5ex]current bounding box.center)}]
    \draw[] (0,0) -- (0,1);
    \draw[] (2,0) -- (2,1);
    \draw[] (0, 1.5) -- (0, 2.5);
    \draw[] (2, 1.5) -- (2, 2.5);
    
    \draw[fill=red!10!white] (-0.25, 0.875) -- (2.25, 0.875) -- (2.25, 1.625) -- (-0.25, 1.625) -- cycle;
    
    \node[] at (1,1.25) {$U$};
    \node[below] at (0,0) {$|0\rangle$};
    \node[below] at (2,0) {$Y$};
    \node[above] () at (2, 2.5) {$Y$};
    \draw[] (0, 2.5) -- (-0.75, 2.5) -- (-0.75, -4.25) -- (0, -4.25);
    \node[left] () at (-0.75, -1) {$X$};
    \begin{scope}[yshift=-1.75cm, yscale=-1]
    \draw[] (0,0) -- (0,1);
    \draw[] (2,0) -- (2,1);
    \draw[] (0, 1.5) -- (0, 2.5);
    \draw[] (2, 1.5) -- (2, 2.5);
    
    \draw[fill=red!10!white] (-0.25, 0.875) -- (2.25, 0.875) -- (2.25, 1.625) -- (-0.25, 1.625) -- cycle;
    
    \node[] at (1,1.25) {$U^{\dagger}$};
    \node[above] at (0,0) {$\langle 0|$};
    \node[above] at (2,0) {$Y$};
    \node[below] () at (2, 2.5) {$Y$};
    \end{scope}
    \end{tikzpicture},
\end{equation}
for some unitary $U: \mathcal{H}_X \otimes \mathcal{H}_Y \to \mathcal{H}_X \otimes \mathcal{H}_Y$, where $|0\rangle$ is shorthand for $|0^{\log |X|}\rangle$. In this process, a unitary is applied to $XY$, where $X$ is initialized in some fixed pure state, and then $X$ is discarded. This defines a quantum channel acting on $Y$, denoted as $\mathcal{E}^{(U)}_Y$. The action of this channel on an input pure state $|\phi\rangle\langle \phi|$ is represented by

\begin{equation}
    \begin{tikzpicture}[scale=0.8, baseline={([yshift=-.5ex]current bounding box.center)}]
    \draw[] (0,0) -- (0,1);
    \draw[] (2,0) -- (2,1);
    \draw[] (0, 1.5) -- (0, 2.5);
    \draw[] (2, 1.5) -- (2, 2.5);
    
    \draw[fill=red!10!white] (-0.25, 0.875) -- (2.25, 0.875) -- (2.25, 1.625) -- (-0.25, 1.625) -- cycle;
    
    \node[] at (1,1.25) {$U$};
    \node[below] at (0,0) {$|0\rangle$};
    \node[below] at (2,0) {$|\phi\rangle$};
    \node[above] () at (2, 2.5) {$Y$};
    \draw[] (0, 2.5) -- (-0.75, 2.5) -- (-0.75, -4.25) -- (0, -4.25);
    \node[left] () at (-0.75, -1) {$X$};
    \begin{scope}[yshift=-1.75cm, yscale=-1]
    \draw[] (0,0) -- (0,1);
    \draw[] (2,0) -- (2,1);
    \draw[] (0, 1.5) -- (0, 2.5);
    \draw[] (2, 1.5) -- (2, 2.5);
    
    \draw[fill=red!10!white] (-0.25, 0.875) -- (2.25, 0.875) -- (2.25, 1.625) -- (-0.25, 1.625) -- cycle;
    
    \node[] at (1,1.25) {$U^{\dagger}$};
    \node[above] at (0,0) {$\langle 0|$};
    \node[above] at (2,0) {$\langle\phi|$};
    \node[below] () at (2, 2.5) {$Y$};
    \end{scope}
    \end{tikzpicture},
    \label{eq:pseudo-phi-insert}
\end{equation}
where the ``ket'' side of the output density operator is at the top of the diagram and the ``bra'' side is at the bottom.

We ask whether there is a fixed unitary $U$, with complexity polynomial in $\log|XY|$, such that it is virtually impossible to distinguish $\mathcal{E}^{(U)}_Y$ from a channel defined by a Haar-uniform average of $\mathcal{E}^{({\widetilde{V}})}_Y$ over ${\widetilde{V}}$. If $|X|\gg |Y|$, this can be achieved by choosing $U$ to be an efficient scrambling unitary~\cite{Hayden2007}. On the other hand, if $|X| \ll |Y|$, then for any fixed $U$ $\mathcal{E}^{(U)}_Y$ is distinguishable from the Haar averaged channel, at least information theoretically. (In Appendix~\ref{appendix:two-design_nogo}, we derive a quantitative bound which establishes this fact rigorously.) What may be possible, however, is to ensure the indistinguishability of the two channels against computationally bounded adversaries. Specifically, we propose that there is a unitary $U$ of complexity polynomial in $\log |XY|$, such that the following approximate identity holds:
\begin{equation}
    \begin{tikzpicture}[scale=0.8, baseline={([yshift=-.5ex]current bounding box.center)}]
    \draw[] (0,0) -- (0,1);
    \draw[] (2,0) -- (2,1);
    \draw[] (0, 1.5) -- (0, 2.5);
    \draw[] (2, 1.5) -- (2, 3.25);
    \draw[] (3, 0) -- (3, 3.75);
    \draw[] (2, 3.25) -- ++(-3.5, 0) --++ (0, -4.125);
    \draw[fill=red!10!white] (-0.25, 0.875) -- (2.25, 0.875) -- (2.25, 1.625) -- (-0.25, 1.625) -- cycle;
    \draw[fill=blue!10!white] (1.75, 2.5) -- (3.25, 2.5) -- (3.25, 3) -- (1.75, 3) --cycle;
    \node[] at (2.5, 2.75) {$W$};
    \node[] at (1,1.25) {$U$};
    \node[below] at (0,0) {$|0\rangle$};
    \node[] () at (2.5, -0.25) {$\phi_2$};
    \draw[] (0, 2.5) -- (-0.75, 2.5) -- (-0.75, -4.25) -- (0, -4.25);
    \draw[] (1.75, 0) -- (3.25, 0) -- (2.5, -0.75) -- cycle;
    \draw[] (2.5, 3.75) -- ++ (1, 0) -- ++ (-0.5, 0.75) -- cycle;
    \node[] () at (3, 4.0) {$\phi_1$};
    
    \node[left] () at (-0.75, -0.75) {$X$};
    \node[left] () at (2, 2) {$Y$};
    \node[left] () at (2, -3.75) {$Y$};
    \node[right] () at (3, 2) {$Z$};
    \node[right] () at (3, -3.75) {$Z$};
    \node[left] () at (-1.5, -0.75) {$Y'$};
    \node[right] () at (3, 3.375) {$Z'$};
    \node[right] () at (3, -5.125) {$Z'$};
    
    \begin{scope}[yshift=-1.75cm, yscale=-1]
    \draw[] (0,0) -- (0,1);
    \draw[] (2,0) -- (2,1);
    \draw[] (0, 1.5) -- (0, 2.5);
    \draw[] (2, 1.5) -- (2, 3.25);
    \draw[] (3, 0) -- (3, 3.75);
    \draw[] (2, 3.25) -- ++(-3.5, 0) --++ (0, -4.125);
    
    \draw[fill=red!10!white] (-0.25, 0.875) -- (2.25, 0.875) -- (2.25, 1.625) -- (-0.25, 1.625) -- cycle;
    \draw[fill=blue!10!white] (1.75, 2.5) -- (3.25, 2.5) -- (3.25, 3) -- (1.75, 3) --cycle;
    \node[] at (2.5, 2.75) {$W^{\dagger}$};
    
    \node[] at (1,1.25) {$U^{\dagger}$};
    \node[above] at (0,0) {$\langle 0|$};
    \node[] () at (2.5, -0.25) {$\phi_2$};
    \draw[] (1.75, 0) -- (3.25, 0) -- (2.5, -0.75) -- cycle;
    \draw[] (2.5, 3.75) -- ++ (1, 0) -- ++ (-0.5, 0.75) -- cycle;
    \node[] () at (3, 4) {$\phi_1$};
    \end{scope}
    \end{tikzpicture}
    \approx 
    \int d\widetilde{V}
    \left(
    \begin{tikzpicture}[scale=0.8, baseline={([yshift=-.5ex]current bounding box.center)}]
    \draw[] (0,0) -- (0,1.5);
    \draw[] (2,0) -- (2,1);
    \draw[] (0, 1.5) -- (0, 2.5);
    \draw[] (2, 1.5) -- (2, 3.25);
    \draw[] (3, 0) -- (3, 3.75);
    \draw[] (2, 3.25) -- ++(-3.5, 0) --++ (0, -4.125);
    \draw[fill=red!10!white] (-0.25, 0.875) -- (2.25, 0.875) -- (2.25, 1.625) -- (-0.25, 1.625) -- cycle;
    \draw[fill=blue!10!white] (1.75, 2.5) -- (3.25, 2.5) -- (3.25, 3) -- (1.75, 3) --cycle;
    \node[] at (2.5, 2.75) {$W$};
    \node[] at (1.125,1.25) {$\widetilde{V}$};
    \node[below] at (0,0) {$|0\rangle$};
    \node[] () at (2.5, -0.25) {$\phi_2$};
    \draw[] (0, 2.5) -- (-0.75, 2.5) -- (-0.75, -4.25) -- (0, -4.25);
    \draw[] (1.75, 0) -- (3.25, 0) -- (2.5, -0.75) -- cycle;
    \draw[] (2.5, 3.75) -- ++ (1, 0) -- ++ (-0.5, 0.75) -- cycle;
    \node[] () at (3, 4) {$\phi_1$};
    \node[left] () at (-0.75, -0.75) {$X$};
    \node[left] () at (2, 2) {$Y$};
    \node[left] () at (2, -3.75) {$Y$};
    \node[right] () at (3, 2) {$Z$};
    \node[right] () at (3, -3.75) {$Z$};
    \node[left] () at (-1.5, -0.75) {$Y'$};
    \node[right] () at (3, 3.375) {$Z'$};
    \node[right] () at (3, -5.125) {$Z'$};
    
    \begin{scope}[yshift=-1.75cm, yscale=-1]
    \draw[] (0,0) -- (0,1.5);
    \draw[] (2,0) -- (2,1);
    \draw[] (0, 1.5) -- (0, 2.5);
    \draw[] (2, 1.5) -- (2, 3.25);
    \draw[] (3, 0) -- (3, 3.75);
    \draw[] (2, 3.25) -- ++(-3.5, 0) --++ (0, -4.125);
    \draw[fill=red!10!white] (-0.25, 0.875) -- (2.25, 0.875) -- (2.25, 1.625) -- (-0.25, 1.625) -- cycle;
    \draw[fill=blue!10!white] (1.75, 2.5) -- (3.25, 2.5) -- (3.25, 3) -- (1.75, 3) --cycle;
    \node[] at (2.5, 2.75) {$W^{\dagger}$};
    
    \node[] at (1.125,1.25) {$\widetilde{V}^{\dagger}$};
    \node[above] at (0,0) {$\langle 0|$};
    \node[] () at (2.5, -0.25) {$\phi_2$};
    \draw[] (1.75, 0) -- (3.25, 0) -- (2.5, -0.75) -- cycle;
    \draw[] (2.5, 3.75) -- ++ (1, 0) -- ++ (-0.5, 0.75) -- cycle;
    \node[] () at (3, 4) {$\phi_1$};
    \end{scope}
    \end{tikzpicture}
    \right),
    \label{eq:pseudorandom_general}
\end{equation}
with an approximation error superpolynomially small in $\log |X|$. Here we assume that $W$, $|\phi_1\rangle$, and $|\phi_2\rangle$ all have complexity polynomial in $\log |X|$ but are otherwise arbitrary. In this process, we have introduced a third subsystem $Z$, and $|\phi_2\rangle$ is an input pure state for system $YZ$. The ``adversary,'' after applying $W$ to $YZ$, performs a projective measurement on a subsystem of $YZ$, obtaining the outcome $|\phi_1\rangle$. (We denote the measured subsystem by $Z'$, and the complementary subsystem by $Y'$; hence  $\mathcal{H}_{Y'}\otimes \mathcal{H}_{Z'}$ serves as an alternative tensor product decompmosition of $\mathcal{H}_Y \otimes \mathcal{H}_Z$.)
The diagram in Eq.~\eqref{eq:pseudorandom_general} represents the probability of obtaining this particular measurement outcome.

The existence of such a unitary $U$ would follow from the assumption that \emph{pseudorandom unitaries} exist~\cite{Ji2018}. At a high level, this assumption means that there is a family of efficient unitary transformations such that an average over the family cannot be distinguished by computationally bounded observers from an average over Haar measure. More concretely, let $\{U_{\text{PR}, x} \}$ be a family of unitaries acting on $\mathcal{H}_Y$, indexed by a $\log |X|$-bit string $x$. 
\begin{definition}
\label{definition:pru}
A family $\{U_{\text{PR},x} \}$ is pseudorandom if the following conditions are met.
\begin{enumerate}
    \item There is a unitary $U_{\text{PR}}:\mathcal{H}_X \otimes \mathcal{H}_Y \to \mathcal{H}_X \otimes \mathcal{H}_Y$ with complexity polynomial in $\log |XY|$ such that
    \begin{equation}
        U_{\text{PR}}|x\rangle_X \otimes |\psi\rangle_Y = |x\rangle_X \otimes U_{\text{PR}, x}|\psi\rangle_Y \label{eq:pru_ingredient1}
    \end{equation}
    for any $|\psi\rangle$.
    \item Consider a circuit consisting of a polynomial number of copies of $U_{\text{PR}, x}$ and a polynomial number (in $\log |X|$) of additional gates, followed by a measurement of a single qubit. If we average over $x$ uniformly, the measurement statistics are indistinguishable from the same circuit in which $U_{\text{PR}, x}$ is replaced by a Haar-random unitary acting on $Y$, with an error decaying faster than any polynomial in $\log |X|$.
\end{enumerate}
\end{definition}

Candidate constructions of pseudorandom unitaries were proposed in~\cite{Ji2018} and are briefly reviewed in Appendix~\ref{appendix:pru_candidate}. Though the existence of pseudorandom unitaries seems quite plausible, at present there is no proof premised on other widely accepted complexity theory assumptions. Nevertheless, we will assume the existence of pseudorandom unitaries in what follows. In the definition, we assume an approximation error that is superpolynomially small in $\log|X|$. In Section~\ref{sec:unitarity_pseudorandomness} and Section~\ref{sec:complexity} we consider consequences of a stronger assumption, that the approximation error is exponentially small instead, and draw sharper conclusions about the unitarity and computational complexity of the black hole $S$-matrix in that case. Our qualitative conclusions continue to apply, however, under the weaker assumption of a superpolynomially small error. 

Assuming a pseudorandom unitary exists, we can construct $U$ satisfying Eq.~\eqref{eq:pseudorandom_general} as follows. Suppose there is a pseudorandom unitary $\{U_{\text{PR}, x} \}$ such that the distinguishability error (see the second condition in Definition~\ref{definition:pru}) is superpolynomially small in $\log |X|$. Choose $U$ as 
\begin{equation}
    U = U_{\text{PR}}(\text{UNIFORM}_X \otimes I_Y),
    \label{eq:pru_construction}
\end{equation}
where $U_{\text{PR}}$ is defined in Eq.~\eqref{eq:pru_ingredient1} and $\text{UNIFORM}_X$ is a unitary that maps $|0\ldots 0\rangle_X$ to the uniform superposition state. Since the unitaries appearing in Eq.~\eqref{eq:pru_construction} have complexity polynomial in $\log |XY|$,\footnote{The preparation of the uniform superposition state can be realized by a tensor product of $\log_2 |X|$ Hadamard gates, and as such, uses at most $\log_2 |X|$ gates.} the first condition in Definition~\ref{definition:pru} is satisfied. Moreover, the second condition implies, upon tracing out $X$, the following identity 
\begin{equation}
    \begin{tikzpicture}[scale=0.8, baseline={([yshift=-.5ex]current bounding box.center)}]
    \draw[] (0,0) -- (0,1);
    \draw[] (2,0) -- (2,1);
    \draw[] (0, 1.5) -- (0, 2.5);
    \draw[] (2, 1.5) -- (2, 3.25);
    \draw[] (3, 0) -- (3, 3.75);
    \draw[] (2, 3.25) -- ++(-3.5, 0) --++ (0, -4.125);
    \draw[fill=red!10!white] (-0.25, 0.875) -- (2.25, 0.875) -- (2.25, 1.625) -- (-0.25, 1.625) -- cycle;
    \draw[fill=blue!10!white] (1.75, 2.5) -- (3.25, 2.5) -- (3.25, 3) -- (1.75, 3) --cycle;
    \node[] at (2.5, 2.75) {$W$};
    \node[] at (1,1.25) {$U$};
    \node[below] at (0,0) {$|0\rangle$};
    \node[] () at (2.5, -0.25) {$\phi_2$};
    \draw[] (0, 2.5) -- (-0.75, 2.5) -- (-0.75, -4.25) -- (0, -4.25);
    \draw[] (1.75, 0) -- (3.25, 0) -- (2.5, -0.75) -- cycle;
    \draw[] (2.5, 3.75) -- ++ (1, 0) -- ++ (-0.5, 0.75) -- cycle;
    \node[] () at (3, 4) {$\phi_1$};
    
    \node[left] () at (-0.75, -0.75) {$X$};
    \node[left] () at (2, 2) {$Y$};
    \node[left] () at (2, -3.75) {$Y$};
    \node[right] () at (3, 2) {$Z$};
    \node[right] () at (3, -3.75) {$Z$};
    \node[left] () at (-1.5, -0.75) {$Y'$};
    \node[right] () at (3, 3.375) {$Z'$};
    \node[right] () at (3, -5.125) {$Z'$};
    
    \begin{scope}[yshift=-1.75cm, yscale=-1]
    \draw[] (0,0) -- (0,1);
    \draw[] (2,0) -- (2,1);
    \draw[] (0, 1.5) -- (0, 2.5);
    \draw[] (2, 1.5) -- (2, 3.25);
    \draw[] (3, 0) -- (3, 3.75);
    \draw[] (2, 3.25) -- ++(-3.5, 0) --++ (0, -4.125);
    
    \draw[fill=red!10!white] (-0.25, 0.875) -- (2.25, 0.875) -- (2.25, 1.625) -- (-0.25, 1.625) -- cycle;
    \draw[fill=blue!10!white] (1.75, 2.5) -- (3.25, 2.5) -- (3.25, 3) -- (1.75, 3) --cycle;
    \node[] at (2.5, 2.75) {$W^{\dagger}$};
    
    \node[] at (1,1.25) {$U^{\dagger}$};
    \node[above] at (0,0) {$\langle 0|$};
    \node[] () at (2.5, -0.25) {$\phi_2$};
    \draw[] (1.75, 0) -- (3.25, 0) -- (2.5, -0.75) -- cycle;
    \draw[] (2.5, 3.75) -- ++ (1, 0) -- ++ (-0.5, 0.75) -- cycle;
    \node[] () at (3, 4) {$\phi_1$};
    \end{scope}
    \end{tikzpicture}
    \approx 
    \int d\widetilde{V}
    \left(
    \begin{tikzpicture}[scale=0.8, baseline={([yshift=-.5ex]current bounding box.center)}]
    \draw[] (0,0) -- (0,1.5);
    \draw[] (2,0) -- (2,1);
    \draw[] (0, 1.5) -- (0, 2.5);
    \draw[] (2, 1.5) -- (2, 3.25);
    \draw[] (3, 0) -- (3, 3.75);
    \draw[] (2, 3.25) -- ++(-3.5, 0) --++ (0, -4.125);
    \draw[fill=red!10!white] (1.5, 0.875) -- (2.5, 0.875) -- (2.5, 1.625) -- (1.5, 1.625) -- cycle;
    \draw[fill=blue!10!white] (1.75, 2.5) -- (3.25, 2.5) -- (3.25, 3) -- (1.75, 3) --cycle;
    \node[] at (2.5, 2.75) {$W$};
    \node[] at (2,1.25) {$\widetilde{V}$};
    \node[below] at (0,0) {$|0\rangle$};
    \node[] () at (2.5, -0.25) {$\phi_2$};
    \draw[] (0, 2.5) -- (-0.75, 2.5) -- (-0.75, -4.25) -- (0, -4.25);
    \draw[] (1.75, 0) -- (3.25, 0) -- (2.5, -0.75) -- cycle;
    \draw[] (2.5, 3.75) -- ++ (1, 0) -- ++ (-0.5, 0.75) -- cycle;
    \node[] () at (3, 4) {$\phi_1$};
    
    \node[left] () at (-0.75, -0.75) {$X$};
    \node[left] () at (2, 2) {$Y$};
    \node[left] () at (2, -3.75) {$Y$};
    \node[right] () at (3, 2) {$Z$};
    \node[right] () at (3, -3.75) {$Z$};
    \node[left] () at (-1.5, -0.75) {$Y'$};
    \node[right] () at (3, 3.375) {$Z'$};
    \node[right] () at (3, -5.125) {$Z'$};
    
    \begin{scope}[yshift=-1.75cm, yscale=-1]
    \draw[] (0,0) -- (0,1.5);
    \draw[] (2,0) -- (2,1);
    \draw[] (0, 1.5) -- (0, 2.5);
    \draw[] (2, 1.5) -- (2, 3.25);
    \draw[] (3, 0) -- (3, 3.75);
    \draw[] (2, 3.25) -- ++(-3.5, 0) --++ (0, -4.125);
    \draw[fill=red!10!white] (1.5, 0.875) -- (2.5, 0.875) -- (2.5, 1.625) -- (1.5, 1.625) -- cycle;
    \draw[fill=blue!10!white] (1.75, 2.5) -- (3.25, 2.5) -- (3.25, 3) -- (1.75, 3) --cycle;
    \node[] at (2.5, 2.75) {$W^{\dagger}$};
    
    \node[] at (2,1.25) {$\widetilde{V}^{\dagger}$};
    \node[above] at (0,0) {$\langle 0|$};
    \node[] () at (2.5, -0.25) {$\phi_2$};
    \draw[] (1.75, 0) -- (3.25, 0) -- (2.5, -0.75) -- cycle;
    \draw[] (2.5, 3.75) -- ++ (1, 0) -- ++ (-0.5, 0.75) -- cycle;
    \node[] () at (3, 4) {$\phi_1$};
    \end{scope}
    \end{tikzpicture}
    \right),
    \label{eq:pru_diagram}
\end{equation}
with an approximation error superpolynomially small in $\log |X|$. Since conjugation by a Haar-random unitary is a depolarizing channel, we can moreover conclude that 
\begin{equation}
    \int d\widetilde{V} 
    \left(
    \begin{tikzpicture}[scale=0.8, baseline={([yshift=-.5ex]current bounding box.center)}]
    \draw[] (0,0) -- (0,1.5);
    \draw[] (2,0) -- (2,1);
    \draw[] (0, 1.5) -- (0, 2.5);
    \draw[] (2, 1.5) -- (2, 3.25);
    \draw[] (3, 0) -- (3, 3.75);
    \draw[] (2, 3.25) -- ++(-3.5, 0) --++ (0, -4.125);
    \draw[fill=red!10!white] (1.5, 0.875) -- (2.5, 0.875) -- (2.5, 1.625) -- (1.5, 1.625) -- cycle;
    \draw[fill=blue!10!white] (1.75, 2.5) -- (3.25, 2.5) -- (3.25, 3) -- (1.75, 3) --cycle;
    \node[] at (2.5, 2.75) {$W$};
    \node[] at (2,1.25) {$\widetilde{V}$};
    \node[below] at (0,0) {$|0\rangle$};
    \node[] () at (2.5, -0.25) {$\phi_2$};
    \draw[] (0, 2.5) -- (-0.75, 2.5) -- (-0.75, -4.25) -- (0, -4.25);
    \draw[] (1.75, 0) -- (3.25, 0) -- (2.5, -0.75) -- cycle;
    \draw[] (2.5, 3.75) -- ++ (1, 0) -- ++ (-0.5, 0.75) -- cycle;
    \node[] () at (3, 4) {$\phi_1$};
    \node[left] () at (-0.75, -0.75) {$X$};
    \node[left] () at (2, 2) {$Y$};
    \node[left] () at (2, -3.75) {$Y$};
    \node[right] () at (3, 2) {$Z$};
    \node[right] () at (3, -3.75) {$Z$};
    \node[left] () at (-1.5, -0.75) {$Y'$};
    \node[right] () at (3, 3.375) {$Z'$};
    \node[right] () at (3, -5.125) {$Z'$};
    
    \begin{scope}[yshift=-1.75cm, yscale=-1]
    \draw[] (0,0) -- (0,1.5);
    \draw[] (2,0) -- (2,1);
    \draw[] (0, 1.5) -- (0, 2.5);
    \draw[] (2, 1.5) -- (2, 3.25);
    \draw[] (3, 0) -- (3, 3.75);
    \draw[] (2, 3.25) -- ++(-3.5, 0) --++ (0, -4.125);
    \draw[fill=red!10!white] (1.5, 0.875) -- (2.5, 0.875) -- (2.5, 1.625) -- (1.5, 1.625) -- cycle;
    \draw[fill=blue!10!white] (1.75, 2.5) -- (3.25, 2.5) -- (3.25, 3) -- (1.75, 3) --cycle;
    \node[] at (2.5, 2.75) {$W^{\dagger}$};
    
    \node[] at (2,1.25) {$\widetilde{V}^{\dagger}$};
    \node[above] at (0,0) {$\langle 0|$};
    \node[] () at (2.5, -0.25) {$\phi_2$};
    \draw[] (1.75, 0) -- (3.25, 0) -- (2.5, -0.75) -- cycle;
    \draw[] (2.5, 3.75) -- ++ (1, 0) -- ++ (-0.5, 0.75) -- cycle;
    \node[] () at (3, 4) {$\phi_1$};
    \end{scope}
    \end{tikzpicture}
    \right)
    =    
    \int d\widetilde{V}
    \left(
    \begin{tikzpicture}[scale=0.8, baseline={([yshift=-.5ex]current bounding box.center)}]
    \draw[] (0,0) -- (0,1.5);
    \draw[] (2,0) -- (2,1);
    \draw[] (0, 1.5) -- (0, 2.5);
    \draw[] (2, 1.5) -- (2, 3.25);
    \draw[] (3, 0) -- (3, 3.75);
    \draw[] (2, 3.25) -- ++(-3.5, 0) --++ (0, -4.125);
    \draw[fill=red!10!white] (-0.25, 0.875) -- (2.25, 0.875) -- (2.25, 1.625) -- (-0.25, 1.625) -- cycle;
    \draw[fill=blue!10!white] (1.75, 2.5) -- (3.25, 2.5) -- (3.25, 3) -- (1.75, 3) --cycle;
    \node[] at (2.5, 2.75) {$W$};
    \node[] at (1.125,1.25) {$\widetilde{V}$};
    \node[below] at (0,0) {$|0\rangle$};
    \node[] () at (2.5, -0.25) {$\phi_2$};
    \draw[] (0, 2.5) -- (-0.75, 2.5) -- (-0.75, -4.25) -- (0, -4.25);
    \draw[] (1.75, 0) -- (3.25, 0) -- (2.5, -0.75) -- cycle;
    \draw[] (2.5, 3.75) -- ++ (1, 0) -- ++ (-0.5, 0.75) -- cycle;
    \node[] () at (3, 4) {$\phi_1$};
    \node[left] () at (-0.75, -0.75) {$X$};
    \node[left] () at (2, 2) {$Y$};
    \node[left] () at (2, -3.75) {$Y$};
    \node[right] () at (3, 2) {$Z$};
    \node[right] () at (3, -3.75) {$Z$};
    \node[left] () at (-1.5, -0.75) {$Y'$};
    \node[right] () at (3, 3.375) {$Z'$};
    \node[right] () at (3, -5.125) {$Z'$};
    
    \begin{scope}[yshift=-1.75cm, yscale=-1]
    \draw[] (0,0) -- (0,1.5);
    \draw[] (2,0) -- (2,1);
    \draw[] (0, 1.5) -- (0, 2.5);
    \draw[] (2, 1.5) -- (2, 3.25);
    \draw[] (3, 0) -- (3, 3.75);
    \draw[] (2, 3.25) -- ++(-3.5, 0) --++ (0, -4.125);
    \draw[fill=red!10!white] (-0.25, 0.875) -- (2.25, 0.875) -- (2.25, 1.625) -- (-0.25, 1.625) -- cycle;
    \draw[fill=blue!10!white] (1.75, 2.5) -- (3.25, 2.5) -- (3.25, 3) -- (1.75, 3) --cycle;
    \node[] at (2.5, 2.75) {$W^{\dagger}$};
    
    \node[] at (1.125,1.25) {$\widetilde{V}^{\dagger}$};
    \node[above] at (0,0) {$\langle 0|$};
    \node[] () at (2.5, -0.25) {$\phi_2$};
    \draw[] (1.75, 0) -- (3.25, 0) -- (2.5, -0.75) -- cycle;
    \draw[] (2.5, 3.75) -- ++ (1, 0) -- ++ (-0.5, 0.75) -- cycle;
    \node[] () at (3, 4) {$\phi_1$};
    \end{scope}
    \end{tikzpicture}
    \right).
\end{equation}
Thus, assuming the existence of pseudorandom unitaries, we can conclude that there is a unitary $U$ of complexity polynomial in $\log |XY|$ such that Eq.~\eqref{eq:pseudorandom_general} holds.

In Eq.~\eqref{eq:pseudorandom_general}, the initial state $|0\rangle$ is a pure state of subsystem $X$, and we trace out subsystem $X$ after the action of the unitary transformation $U$ or $\tilde V$. What we will need to consider in our analysis of a partially evaporated black hole is a somewhat more general situation, in which we trace out a system $X'$ which does not necessarily have the same dimension as $X$. In this more general setting, we wish to argue that a fixed unitary $U$ exists which satisfies the identity
\begin{equation}
    \begin{tikzpicture}[scale=0.8, baseline={([yshift=-.5ex]current bounding box.center)}]
    \draw[] (0,0) -- (0,1);
    \draw[] (2,0) -- (2,1);
    \draw[] (0, 1.5) -- (0, 2.5);
    \draw[] (2, 1.5) -- (2, 3.25);
    \draw[] (3, 0) -- (3, 3.75);
    \draw[] (2, 3.25) -- ++(-3.5, 0) --++ (0, -4.125);
    \draw[fill=red!10!white] (-0.25, 0.875) -- (2.25, 0.875) -- (2.25, 1.625) -- (-0.25, 1.625) -- cycle;
    \draw[fill=blue!10!white] (1.75, 2.5) -- (3.25, 2.5) -- (3.25, 3) -- (1.75, 3) --cycle;
    \node[] at (2.5, 2.75) {$W$};
    \node[] at (1,1.25) {$U$};
    \node[below] at (0,0) {$|0\rangle$};
    \node[] () at (2.5, -0.25) {$\phi_2$};
    \draw[] (0, 2.5) -- (-0.75, 2.5) -- (-0.75, -4.25) -- (0, -4.25);
    \draw[] (1.75, 0) -- (3.25, 0) -- (2.5, -0.75) -- cycle;
    \draw[] (2.5, 3.75) -- ++ (1, 0) -- ++ (-0.5, 0.75) -- cycle;
    \node[] () at (3, 4.0) {$\phi_1$};
    
    \node[left] () at (-0.65, -0.75) {$X'$};
    \node[left] () at (2, 2) {$Y'$};
    \node[left] () at (2, -3.75) {$Y'$};
    \node[right] () at (3, 2) {$Z$};
    \node[right] () at (3, -3.75) {$Z$};
    \node[left] () at (-1.5, -0.75) {$Y''$};
    \node[right] () at (3, 3.375) {$Z'$};
    \node[right] () at (3, -5.125) {$Z'$};
    \node[right] () at (0,0.5) {$X$};
    \node[left] () at (2,0.5) {$Y$};
    
    \begin{scope}[yshift=-1.75cm, yscale=-1]
    \draw[] (0,0) -- (0,1);
    \draw[] (2,0) -- (2,1);
    \draw[] (0, 1.5) -- (0, 2.5);
    \draw[] (2, 1.5) -- (2, 3.25);
    \draw[] (3, 0) -- (3, 3.75);
    \draw[] (2, 3.25) -- ++(-3.5, 0) --++ (0, -4.125);
    \node[right] () at (0,0.5) {$X$};
    \node[left] () at (2,0.5) {$Y$};
    
    \draw[fill=red!10!white] (-0.25, 0.875) -- (2.25, 0.875) -- (2.25, 1.625) -- (-0.25, 1.625) -- cycle;
    \draw[fill=blue!10!white] (1.75, 2.5) -- (3.25, 2.5) -- (3.25, 3) -- (1.75, 3) --cycle;
    \node[] at (2.5, 2.75) {$W^{\dagger}$};
    
    \node[] at (1,1.25) {$U^{\dagger}$};
    \node[above] at (0,0) {$\langle 0|$};
    \node[] () at (2.5, -0.25) {$\phi_2$};
    \draw[] (1.75, 0) -- (3.25, 0) -- (2.5, -0.75) -- cycle;
    \draw[] (2.5, 3.75) -- ++ (1, 0) -- ++ (-0.5, 0.75) -- cycle;
    \node[] () at (3, 4) {$\phi_1$};
    \end{scope}
    \end{tikzpicture}
    \approx 
    \int d\widetilde{V}
    \left(
    \begin{tikzpicture}[scale=0.8, baseline={([yshift=-.5ex]current bounding box.center)}]
    \draw[] (0,0) -- (0,1.5);
    \draw[] (2,0) -- (2,1);
    \draw[] (0, 1.5) -- (0, 2.5);
    \draw[] (2, 1.5) -- (2, 3.25);
    \draw[] (3, 0) -- (3, 3.75);
    \draw[] (2, 3.25) -- ++(-3.5, 0) --++ (0, -4.125);
    \draw[fill=red!10!white] (-0.25, 0.875) -- (2.25, 0.875) -- (2.25, 1.625) -- (-0.25, 1.625) -- cycle;
    \draw[fill=blue!10!white] (1.75, 2.5) -- (3.25, 2.5) -- (3.25, 3) -- (1.75, 3) --cycle;
    \node[] at (2.5, 2.75) {$W$};
    \node[] at (1.125,1.25) {$\widetilde{V}$};
    \node[below] at (0,0) {$|0\rangle$};
    \node[] () at (2.5, -0.25) {$\phi_2$};
    \draw[] (0, 2.5) -- (-0.75, 2.5) -- (-0.75, -4.25) -- (0, -4.25);
    \draw[] (1.75, 0) -- (3.25, 0) -- (2.5, -0.75) -- cycle;
    \draw[] (2.5, 3.75) -- ++ (1, 0) -- ++ (-0.5, 0.75) -- cycle;
    \node[] () at (3, 4) {$\phi_1$};
    \node[left] () at (-0.65, -0.75) {$X'$};
    \node[left] () at (2, 2) {$Y'$};
    \node[left] () at (2, -3.75) {$Y'$};
    \node[right] () at (3, 2) {$Z$};
    \node[right] () at (3, -3.75) {$Z$};
    \node[left] () at (-1.5, -0.75) {$Y''$};
    \node[right] () at (3, 3.375) {$Z'$};
    \node[right] () at (3, -5.125) {$Z'$};
    \node[right] () at (0,0.5) {$X$};
    \node[left] () at (2,0.5) {$Y$};
    
    \begin{scope}[yshift=-1.75cm, yscale=-1]
    \node[right] () at (0,0.5) {$X$};
    \node[left] () at (2,0.5) {$Y$};
    \draw[] (0,0) -- (0,1.5);
    \draw[] (2,0) -- (2,1);
    \draw[] (0, 1.5) -- (0, 2.5);
    \draw[] (2, 1.5) -- (2, 3.25);
    \draw[] (3, 0) -- (3, 3.75);
    \draw[] (2, 3.25) -- ++(-3.5, 0) --++ (0, -4.125);
    \draw[fill=red!10!white] (-0.25, 0.875) -- (2.25, 0.875) -- (2.25, 1.625) -- (-0.25, 1.625) -- cycle;
    \draw[fill=blue!10!white] (1.75, 2.5) -- (3.25, 2.5) -- (3.25, 3) -- (1.75, 3) --cycle;
    \node[] at (2.5, 2.75) {$W^{\dagger}$};
    
    \node[] at (1.125,1.25) {$\widetilde{V}^{\dagger}$};
    \node[above] at (0,0) {$\langle 0|$};
    \node[] () at (2.5, -0.25) {$\phi_2$};
    \draw[] (1.75, 0) -- (3.25, 0) -- (2.5, -0.75) -- cycle;
    \draw[] (2.5, 3.75) -- ++ (1, 0) -- ++ (-0.5, 0.75) -- cycle;
    \node[] () at (3, 4) {$\phi_1$};
    \end{scope}
    \end{tikzpicture}
    \right)
    \label{eq:pseudorandom_more_general}
\end{equation}
up to a small error. In the notation we are now using, $U$ acts on $\mathcal{H}_X\otimes \mathcal{H}_Y$, which has an alternative tensor decomposition as $\mathcal{H}_{X'}\otimes \mathcal{H}_{Y'}$, and $W$ acts on $\mathcal{H}_{Y'}\otimes \mathcal{H}_Z$, which has an alternative tensor decomposition as $\mathcal{H}_{Y''}\otimes \mathcal{H}_{Z'}$. This more general identity is needed in the black hole setting because in the fundamental description we decompose the Hilbert space into the black hole system $B$ and the radiation system $R$; $B$ is inaccessible and therefore traced out, and the dimension of $B$ need not be the same as the dimension of the fixed input state that appears in the definition of the holographic map.

To argue in favor of Eq.~\eqref{eq:pseudorandom_more_general}, and to characterize the approximation error, we distinguish two cases. If $\dim(\mathcal{H}_{X'})$ is larger or equal to $\dim(\mathcal{H}_{X})$, the existence of a unitary that satisfies Eq.~\eqref{eq:pseudorandom_more_general} follows straightforwardly from the one that satisfies Eq.~\eqref{eq:pseudorandom_general}. One can simply choose the unitary in Eq.~\eqref{eq:pseudorandom_more_general} to be the one appearing in Eq.~\eqref{eq:pseudorandom_general}, followed by a swapping of qubits so that the bitstring $x$ that determines $U_{\text{PR}, x}$ is discarded by the partial trace of $X'$. If instead $\dim(\mathcal{H}_{X'})$ is smaller than $\dim(\mathcal{H}_X)$, we can choose the ensemble of unitaries $\{U_{\text{PR}, x} \}$ labeled by a $\log |X'|$-bit string. These bits can be then swapped so that the partial trace of $X'$ discards the label information. Therefore, even in this generalized setup there is a unitary that satisfies Eq.~\eqref{eq:pseudorandom_more_general}. The approximation error is superpolynomially small in $\min(\log |X|, \log |X'|)$.

Will the unitary that describes the formation and evaporation of the black hole also satisfy Eq.~\eqref{eq:pseudorandom_more_general}? Without knowing the complete theory of quantum gravity, we cannot provide a definitive answer to this question. However, since a black hole is one of the best information scramblers in nature, we may expect that, if achieving Eq.~\eqref{eq:pseudorandom_more_general} is possible in principle, a black hole will be able to do it.\footnote{A stronger justification of this assumption may be made by showing that more realistic toy models, e.g., a typical instance of the random circuit model, satisfies the pseudorandomness condition with high probability. However, it is currently unclear how one can prove such a statement. Existing approaches to construct pseudorandom quantum states rely on one-way functions which are hard to invert on a quantum computer~\cite{Ji2018}, and it is not yet clear how such constructions can be used to construct pseudorandom unitaries. Establishing such a connection is an area of ongoing research in quantum complexity theory.} From now on, we shall assume that the unitary $U'$ that describes the formation and the evaporation of the black hole is pseudorandom in the sense of Eq.~\eqref{eq:pseudorandom_more_general}.\footnote{Note that we are assuming that the \emph{unitary} that describes the black hole is pseudorandom, as opposed to the \emph{state} of the radiation being pseudorandom, as assumed in Ref.~\cite{Kim2020}. The former is a stronger assumption than the latter; see Ref.~\cite{Ji2018} for discussion of this distinction.} 

We will use Eq.~\eqref{eq:pseudorandom_more_general} to investigate the black hole $S$-matrix, and in particular to verify its unitarity. For this purpose it is convenient to rewrite the Eq.~\eqref{eq:pseudorandom_more_general} in the alternative but equivalent form
\begin{equation}
    \begin{tikzpicture}[scale=0.8, baseline={([yshift=-.5ex]current bounding box.center)}]
    \draw[] (0,0) -- (0,4.75);
    \draw[] (2,0) -- (2,4.75);
    \draw[] (0, 1.5) -- (0, 2.5);
    \draw[] (2, 1.5) -- (2, 3.25);
    \draw[] (3, 0) -- (3, 3.75);
    \draw[fill=red!10!white] (-0.25, 0.875) -- (2.25, 0.875) -- (2.25, 1.625) -- (-0.25, 1.625) -- cycle;
    \draw[fill=blue!10!white] (1.75, 2.5) -- (3.25, 2.5) -- (3.25, 3) -- (1.75, 3) --cycle;
    \node[] at (2.5, 2.75) {$W$};
    \node[] at (1,1.25) {$U$};
    \node[below] at (0,0) {$|0\rangle$};
    \node[] () at (2.5, -0.25) {$\phi_2$};
    \draw[] (1.75, 0) -- (3.25, 0) -- (2.5, -0.75) -- cycle;
    \draw[] (2.5, 3.75) -- ++ (1, 0) -- ++ (-0.5, 0.75) -- cycle;
    \node[] () at (3, 4) {$\phi_1$};
    
    \node[left] () at (0, 4.625) {$X'$};
    \node[left] () at (2, 4.625) {$Y''$};
    \node[left] () at (2, 7.125) {$Y'$};
    \node[left] () at (2, 2.125) {$Y'$};
    \node[right] () at (3, 1.25) {$Z$};
    \node[right] () at (3, 3.375) {$Z'$};
    \node[left] () at (0, 0.5) {$X$};
    \node[left] () at (2, 0.5) {$Y$};

    \begin{scope}[yshift=9.25cm, yscale=-1]
    \node[right] () at (3, 1.25) {$Z$};
    \node[right] () at (3, 3.375) {$Z'$};
    \node[left] () at (0, 0.5) {$X$};
    \node[left] () at (2, 0.5) {$Y$};
    \draw[] (0,0) -- (0,4.75);
    \draw[] (2,0) -- (2,4.75);
    \draw[] (0, 1.5) -- (0, 2.5);
    \draw[] (2, 1.5) -- (2, 3.25);
    \draw[] (3, 0) -- (3, 3.75);
    
    \draw[fill=red!10!white] (-0.25, 0.875) -- (2.25, 0.875) -- (2.25, 1.625) -- (-0.25, 1.625) -- cycle;
    \draw[fill=blue!10!white] (1.75, 2.5) -- (3.25, 2.5) -- (3.25, 3) -- (1.75, 3) --cycle;
    \node[] at (2.5, 2.75) {$W^{\dagger}$};
    
    \node[] at (1,1.25) {$U^{\dagger}$};
    \node[above] at (0,0) {$\langle 0|$};
    \node[] () at (2.5, -0.25) {$\phi_2$};
    \draw[] (1.75, 0) -- (3.25, 0) -- (2.5, -0.75) -- cycle;
    \draw[] (2.5, 3.75) -- ++ (1, 0) -- ++ (-0.5, 0.75) -- cycle;
    \node[] () at (3, 4) {$\phi_1$};
    \end{scope}
    \end{tikzpicture}
    \approx
    \int d\widetilde{V} \left( 
    \begin{tikzpicture}[scale=0.8, baseline={([yshift=-.5ex]current bounding box.center)}]
    \draw[] (0,0) -- (0,4.75);
    \draw[] (2,0) -- (2,4.75);
    \draw[] (0, 1.5) -- (0, 2.5);
    \draw[] (2, 1.5) -- (2, 3.25);
    \draw[] (3, 0) -- (3, 3.75);
    \draw[fill=red!10!white] (-0.25, 0.875) -- (2.25, 0.875) -- (2.25, 1.625) -- (-0.25, 1.625) -- cycle;
    \draw[fill=blue!10!white] (1.75, 2.5) -- (3.25, 2.5) -- (3.25, 3) -- (1.75, 3) --cycle;
    \node[] at (2.5, 2.75) {$W$};
    \node[] at (1,1.25) {$\widetilde{V}$};
    \node[below] at (0,0) {$|0\rangle$};
    \node[] () at (2.5, -0.25) {$\phi_2$};
    \draw[] (1.75, 0) -- (3.25, 0) -- (2.5, -0.75) -- cycle;
    \draw[] (2.5, 3.75) -- ++ (1, 0) -- ++ (-0.5, 0.75) -- cycle;
    \node[] () at (3, 4) {$\phi_1$};
    
    \node[left] () at (0, 4.625) {$X'$};
    \node[left] () at (2, 4.625) {$Y''$};
    \node[left] () at (2, 7.125) {$Y'$};
    \node[left] () at (2, 2.125) {$Y'$};
    \node[right] () at (3, 1.25) {$Z$};
    \node[right] () at (3, 3.375) {$Z'$};
    \node[left] () at (0, 0.5) {$X$};
    \node[left] () at (2, 0.5) {$Y$};
    
    \begin{scope}[yshift=9.25cm, yscale=-1]
    \node[right] () at (3, 1.25) {$Z$};
    \node[right] () at (3, 3.375) {$Z'$};
    \node[left] () at (0, 0.5) {$X$};
    \node[left] () at (2, 0.5) {$Y$};
    \draw[] (0,0) -- (0,4.75);
    \draw[] (2,0) -- (2,4.75);
    \draw[] (0, 1.5) -- (0, 2.5);
    \draw[] (2, 1.5) -- (2, 3.25);
    \draw[] (3, 0) -- (3, 3.75);
    
    \draw[fill=red!10!white] (-0.25, 0.875) -- (2.25, 0.875) -- (2.25, 1.625) -- (-0.25, 1.625) -- cycle;
    \draw[fill=blue!10!white] (1.75, 2.5) -- (3.25, 2.5) -- (3.25, 3) -- (1.75, 3) --cycle;
    \node[] at (2.5, 2.75) {$W^{\dagger}$};
    
    \node[] at (1,1.25) {$\widetilde{V}^{\dagger}$};
    \node[above] at (0,0) {$\langle 0|$};
    \node[] () at (2.5, -0.25) {$\phi_2$};
    \draw[] (1.75, 0) -- (3.25, 0) -- (2.5, -0.75) -- cycle;
    \draw[] (2.5, 3.75) -- ++ (1, 0) -- ++ (-0.5, 0.75) -- cycle;
    \node[] () at (3, 4) {$\phi_1$};
    \end{scope}
    \end{tikzpicture} \right).
    \label{eq:alternative-pseudo}
\end{equation}
In this form, the pseudorandomness of $U$ has a different interpretation than before. Now we consider a process in which the initial pure state is $|0\rangle_X\otimes |\phi_2\rangle_{YZ}$, to which is applied the unitary transformation $\left(I_{X'}\otimes W_{Y'Z}\right)\left(U_{XY}\otimes I_Z\right)$. Then system $Z'$ is projected onto the pure state $|\phi_1\rangle$, the inverse unitary transformation $\left(U_{XY}^\dagger\otimes I_Z\right)\left(I_{X'}\otimes W_{Y'Z}^\dagger\right)$ is applied, and a measurement on $XYZ$ yields the outcome $|0\rangle_X\otimes|\phi_2\rangle_{YZ}$. The diagram represents the probability that this projection succeeds and this measurement outcome is obtained. Here pseudorandomness of $U$ means that if $W$, $|\phi_1\rangle$, and $|\phi_2\rangle$ have complexity polynomial in $\log |X|$, then replacing $U$ by a Haar average alters this probability by only a superpolynomially small amount.

\subsection{Unitarity of the black hole $S$-matrix}
\label{sec:unitarity_pseudorandomness}

In this section we show that if $U'$ is pseudorandom, the black hole $S$-matrix $V(U')$ is approximately unitary. In the next section, we study the computational complexity of $V(U')$. In both cases our analysis hinges on the identity
\begin{equation}
|r|
\begin{tikzpicture}[scale=0.8, baseline={([yshift=-.5ex]current bounding box.center)}]
    \draw[] (0,0) -- (0,1);
    \draw[] (2,0) -- (2,1);
    \draw[] (4,0) -- (4,1);
    \draw[] (6,0) -- (6,1);
    \draw[] (4,0) -- (5, -0.5) -- (6,0);
    \draw[] (0, 1.5) -- (0, 2.5);
    \draw[] (2, 1.5) -- (2, 2.5);
    \draw[] (2, 2.5) -- (3, 3) -- (4, 2.5) -- (4,1);
    \draw[] (3,0) -- (3,1);
    
    \draw[fill=red!10!white] (-0.25, 1) -- (3.25, 1) -- (3.25, 1.75) -- (-0.25, 1.75) -- cycle;
    
     \node[fill=blue, regular polygon, regular polygon sides=3, minimum size=0.35cm, inner sep=0pt] (u1) at (3, 0.25) {};
    \node[fill=blue, regular polygon, regular polygon sides=3, minimum size=0.35cm, inner sep=0pt] (u2) at (6, 0.25) {};
    \node[fill=red, regular polygon, regular polygon sides=3, minimum size=0.35cm, inner sep=0pt] (v1) at (3, 0.7) {};
    \node[fill=red, regular polygon, regular polygon sides=3, minimum size=0.35cm, inner sep=0pt] (v2) at (4, 0.7) {};
    
    \node[right] at (u2) {$u$};
    \node[right] at (v2) {$v$};
    
    \draw[blue, thick] (u1) -- (u2);
    \draw[red, thick] (v1) -- (v2);
    
    
    \node[above] at (0, 2.5) {$B$};
    \node[above] at (3, 3) {$\langle \text{MAX}|_{r'r}$};
    \node[] at (1.5,1.375) {$U'$};
    \node[below] at (0,0) {$|\psi\rangle_{\ell}$};
    \node[below] at (2,0) {$|\psi_0\rangle_f$};
    \node[below] at (3,0) {$I$};
    \node[below] at (4,0) {$r$};
    \node[below] at (6,0) {$R$};
    \node[left] at (2, 2.25) {$r'$};
    \node[below] at (5,-0.5) {$|\text{MAX}\rangle_{rR}$};
    \end{tikzpicture}
    =
    |I|
    \begin{tikzpicture}[scale=0.8, baseline={([yshift=-.5ex]current bounding box.center)}]
    \draw[] (0,0) -- (0,1);
    \draw[] (2,0) -- (2,1);
    \draw[] (0, 1.5) -- (0, 2.75);
    \draw[] (2, 1.5) -- (2, 2.75);
    \draw[] (3,0) -- (3,1);
    \node[above] () at (0, 2.75) {$B$};
    \node[above] () at (2, 2.75) {$R$};
    
    \draw[fill=red!10!white] (-0.25, 1) -- (3.25, 1) -- (3.25, 1.75) -- (-0.25, 1.75) -- cycle;

    \node[] at (1.5,1.375) {$U'$};
    \node[below] at (0,0) {$|\psi\rangle_{\ell}$};
    \node[below] at (2,0) {$| \psi_0\rangle_f$};
    
    \draw[] (3, 0) -- ++ (0.75, -0.5) -- ++ (0.75, 0.5);
    \draw[] (6, -0.5) -- ++ (0, 3.25);
    \draw[] (4.5, 0) -- ++ (0, 2.75); 
    \draw[] (4.5, 2.75) -- ++ (0.75, 0.5) -- ++ (0.75, -0.5);
    
    \node[below] () at (4.125,-0.5) {$|\text{MAX}\rangle_{I_1 I_2}$};
    \node[below] () at (6.125, -0.5) {$I$};
    \node[left] () at (3, 0.5) {$I_1$};
    \node[right] () at (4.5, 0.5) {$I_2$};

    \node[above] () at (5.3, 3.25) {$\langle \text{MAX}|_{I_2 I}$};
    
    \node[fill=blue, regular polygon, regular polygon sides=3, minimum size=0.35cm, inner sep=0pt] (u1) at (6, 2.375) {};
    \node[fill=blue, regular polygon, regular polygon sides=3, minimum size=0.35cm, inner sep=0pt] (u2) at (2, 2.375) {};
    \node[fill=red, regular polygon, regular polygon sides=3, minimum size=0.35cm, inner sep=0pt, rotate=180] (v1) at (4.5, 2) {};
    \node[fill=red, regular polygon, regular polygon sides=3, minimum size=0.35cm, inner sep=0pt, rotate=180] (v2) at (2, 2) {};
    
    \node[left] at (u2) {$u\phantom{l}$};
    \node[left] at (v2) {$v\phantom{l}$};
    
    \draw[blue, thick] (u1) -- (u2);
    \draw[red, thick] (v1) --  (v2);
    \end{tikzpicture},
    \label{eq:I-postselect}
\end{equation}
 where the left-hand and right-hand sides of the identity are both equivalent to the black hole $S$-matrix  
\begin{equation}
V(U')=
        \begin{tikzpicture}[scale=0.8, baseline={([yshift=-.5ex]current bounding box.center)}]
    \draw[] (0,0) -- (0,1);
    \draw[] (2,0) -- (2,1);
    \draw[] (0, 1.5) -- (0, 2.75);
    \draw[] (2, 1.5) -- (2, 2.75);
    \draw[] (3,0) -- (3,1);
    \node[above] () at (0, 2.75) {$B$};
    \node[above] () at (2, 2.75) {$R$};
    
    \draw[fill=red!10!white] (-0.25, 1) -- (3.25, 1) -- (3.25, 1.75) -- (-0.25, 1.75) -- cycle;
    
     \node[fill=blue, regular polygon, regular polygon sides=3, minimum size=0.35cm, inner sep=0pt] (u1) at (3, 0.25) {};
    \node[fill=blue, regular polygon, regular polygon sides=3, minimum size=0.35cm, inner sep=0pt] (u2) at (2, 2.5) {};
    \node[fill=red, regular polygon, regular polygon sides=3, minimum size=0.35cm, inner sep=0pt] (v1) at (3, 0.7) {};
    \node[fill=red, regular polygon, regular polygon sides=3, minimum size=0.35cm, inner sep=0pt, rotate=180] (v2) at (2, 2) {};
    
    \node[left] at (u2) {$u\phantom{l}$};
    \node[left] at (v2) {$v\phantom{l}$};
    
    \draw[blue, thick] (u1) -- ++ (0.75, 0) -- ++ (0,2.25) -- (u2);
    \draw[red, thick] (v1) -- ++ (0.5, 0) -- ++ (0, 1.3) --  (v2);
    
    \node[] at (1.5,1.375) {$U'$};
    \node[below] at (0,0) {$|\psi\rangle_{\ell}$};
    \node[below] at (2,0) {$|\psi_0\rangle_f$};
    \node[below] at (3,0) {$I$};
    \end{tikzpicture}.
    \label{eq:V-of-U'}
\end{equation}
On the right-hand side of Eq.~\eqref{eq:I-postselect} we have introduced auxiliary subsystems $\mathcal{H}_{I_1}$ and $\mathcal{H}_{I_2}$ which have the same dimension as  $\mathcal{H}_I$. 
What is particularly notable is that $V(U')$, originally realized in a unitary circuit accompanied by postselection of a measurement outcome occurring with probability $1/|r|^2$, can be alternatively realized in a unitary circuit accompanied by postselection of a measurement outcome occurring with the much larger probability $1/|I|^2$.

Now we would like to show that $V(U')$ is very nearly unitary if $U'$ is pseudorandom and the operations $u$ and $v$ performed by the infaller have low complexity. To do so, we will use the identity Eq.~\eqref{eq:I-postselect} and the pseudorandomness property Eq.~\eqref{eq:alternative-pseudo}. We will also assume that the infaller's initial state is uncorrelated with the black hole's microstate, so we may take the initial state of $\ell f I$ in Eq.~\eqref{eq:V-of-U'} to be a product state $|\psi_\ell \otimes |\psi_0\rangle_f\otimes |\varphi\rangle_I$. Furthermore, we assume that this initial state has low complexity. Using Eq.~\eqref{eq:alternative-pseudo}, we can verify that the diagonal matrix element
\begin{equation}
   \langle \psi|_\ell \langle \psi_0|_f \langle \varphi|_I \left[ V(U')^\dagger V(U') \right] |\psi\rangle_\ell |\psi_0\rangle_f |\varphi\rangle_I
\end{equation}
deviates from 1 by an amount that 
is superpolynomially small in $\log |B|$ for any choice of the input product state. Because this holds for any product state, it follows that off-diagonal matrix elements between product states also satisfy 
\begin{equation}
   \langle \psi'|_\ell \langle \psi_0|_f \langle \varphi'|_I \left[ V(U')^\dagger V(U') \right] |\psi\rangle_\ell |\psi_0\rangle_f |\varphi\rangle_I
   \approx  \langle \psi'|\psi\rangle \langle \varphi'|\varphi\rangle
\end{equation}
with a similar approximation error. We may limit our attention to product states if we assume that the initial state of the infaller is prepared outside the black hole, and so is uncorrelated with the left-moving system $\ell$, which is behind the black hole horizon.

We evaluate the diagonal matrix element as follows: 
\begin{equation}
    \begin{tikzpicture}[scale=0.8, baseline={([yshift=-.5ex]current bounding box.center)}]
    \draw[] (0,0) -- (0,1);
    \draw[] (2,0) -- (2,1);
    \draw[] (0, 1.5) -- (0, 4.25);
    \draw[] (2, 1.5) -- (2, 4.25);
    \draw[] (3,0) -- (3,1);
    \node[left] () at (0, 4.25) {$B$};
    \node[left] () at (2, 4.25) {$R$};
    
    \draw[fill=red!10!white] (-0.25, 1) -- (3.25, 1) -- (3.25, 1.75) -- (-0.25, 1.75) -- cycle;

    \node[] at (1.5,1.375) {$U'$};
    \node[below] at (0,0) {$|\psi\rangle_{\ell}$};
    \node[below] at (2,0) {$| \psi_0\rangle_f$};
    
    \draw[] (3, 0) -- ++ (0.75, -0.5) -- ++ (0.75, 0.5);
    \draw[] (6, -0.5) -- ++ (0, 3.25);
    \draw[] (4.5, 0) -- ++ (0, 2.75); 
    \draw[] (4.5, 2.75) -- ++ (0.75, 0.5) -- ++ (0.75, -0.5);

    \draw[dashed] (2.75, 0.25) -- (6.75, 0.25) -- (6.75, -1.25) -- (2.75, -1.25) -- cycle;
    \node[below] () at (4.125,-0.5) {$|\text{MAX}\rangle_{I_1 I_2}$};
    \node[below] () at (6.125, -0.5) {$|\varphi\rangle_{I}$};

    \draw[dotted, thick] (3.875, 2.625) -- (6.625, 2.625) -- (6.625, 4.125) -- (3.875, 4.125) -- cycle;
    \node[above] () at (5.3, 3.25) {$\langle \text{MAX}|_{I_2 I}$};
    
    \node[fill=blue, regular polygon, regular polygon sides=3, minimum size=0.35cm, inner sep=0pt] (u1) at (6, 2.375) {};
    \node[fill=blue, regular polygon, regular polygon sides=3, minimum size=0.35cm, inner sep=0pt] (u2) at (2, 2.375) {};
    \node[fill=red, regular polygon, regular polygon sides=3, minimum size=0.35cm, inner sep=0pt, rotate=180] (v1) at (4.5, 2) {};
    \node[fill=red, regular polygon, regular polygon sides=3, minimum size=0.35cm, inner sep=0pt, rotate=180] (v2) at (2, 2) {};
    
    \node[left] at (u2) {$u\phantom{l}$};
    \node[left] at (v2) {$v\phantom{l}$};
    
    \draw[blue, thick] (u1) -- (u2);
    \draw[red, thick] (v1) --  (v2);
    \begin{scope}[yshift = 8.5cm, yscale=-1]
    
    \draw[] (0,0) -- (0,1);
    \draw[] (2,0) -- (2,1);
    \draw[] (0, 1.5) -- (0, 4.25);
    \draw[] (2, 1.5) -- (2, 4.25);
    \draw[] (3,0) -- (3,1);
    
    \draw[fill=red!10!white] (-0.25, 1) -- (3.25, 1) -- (3.25, 1.75) -- (-0.25, 1.75) -- cycle;

    \node[] at (1.5,1.375) {${U'}^{\dagger}$};
    \node[above] at (0,0) {$\langle\psi|_{\ell}$};
    \node[above] at (2,0) {$\langle \psi_0|_f$};
    
    \draw[] (3, 0) -- ++ (0.75, -0.5) -- ++ (0.75, 0.5);
    \draw[] (6, -0.5) -- ++ (0, 3.25);
    \draw[] (4.5, 0) -- ++ (0, 2.75); 
    \draw[] (4.5, 2.75) -- ++ (0.75, 0.5) -- ++ (0.75, -0.5);
    
   \draw[dashed] (2.75, 0.25) -- (6.75, 0.25) -- (6.75, -1.25) -- (2.75, -1.25) -- cycle;
    \node[above] () at (4.125,-0.5) {$\langle\text{MAX}|_{I_1 I_2}$};
    \node[above] () at (6.125, -0.5) {$\langle \varphi|_{I}$};
    
    \draw[dotted, thick] (3.875, 2.625) -- (6.625, 2.625) -- (6.625, 4.125) -- (3.875, 4.125) -- cycle;
    \node[below] () at (5.3, 3.25) {$| \text{MAX}\rangle_{I_2 I}$};
    \node[fill=blue, regular polygon, regular polygon sides=3, minimum size=0.35cm, inner sep=0pt, rotate=180] (u1) at (6, 2.375) {};
    \node[fill=blue, regular polygon, regular polygon sides=3, minimum size=0.35cm, inner sep=0pt, rotate=180] (u2) at (2, 2.375) {};
    \node[fill=red, regular polygon, regular polygon sides=3, minimum size=0.35cm, inner sep=0pt] (v1) at (4.5, 2) {};
    \node[fill=red, regular polygon, regular polygon sides=3, minimum size=0.35cm, inner sep=0pt] (v2) at (2, 2) {};
    
    \node[left] at (u2) {$u\phantom{l}$};
    \node[left] at (v2) {$v\phantom{l}$};
    
    \draw[blue, thick] (u1) -- (u2);
    \draw[red, thick] (v1) --  (v2);
    
    \end{scope}
    \end{tikzpicture}
    \approx
    \int d \widetilde{V}
    \left(    
    \begin{tikzpicture}[scale=0.8, baseline={([yshift=-.5ex]current bounding box.center)}]
    \draw[] (0,0) -- (0,1);
    \draw[] (2,0) -- (2,1);
    \draw[] (0, 1.5) -- (0, 4.25);
    \draw[] (2, 1.5) -- (2, 4.25);
    \draw[] (3,0) -- (3,1);
    \node[left] () at (0, 4.25) {$B$};
    \node[left] () at (2, 4.25) {$R$};
    
    \draw[fill=red!10!white] (-0.25, 1) -- (3.25, 1) -- (3.25, 1.75) -- (-0.25, 1.75) -- cycle;

    \node[] at (1.5,1.375) {$\widetilde{V}$};
    \node[below] at (0,0) {$|\psi\rangle_{\ell}$};
    \node[below] at (2,0) {$| \psi_0\rangle_f$};
    
    \draw[] (3, 0) -- ++ (0.75, -0.5) -- ++ (0.75, 0.5);
    \draw[] (6, -0.5) -- ++ (0, 3.25);
    \draw[] (4.5, 0) -- ++ (0, 2.75); 
    \draw[] (4.5, 2.75) -- ++ (0.75, 0.5) -- ++ (0.75, -0.5);

    \draw[dashed] (2.75, 0.25) -- (6.75, 0.25) -- (6.75, -1.25) -- (2.75, -1.25) -- cycle;
    \node[below] () at (4.125,-0.5) {$|\text{MAX}\rangle_{I_1 I_2}$};
    \node[below] () at (6.125, -0.5) {$|\varphi\rangle_{I}$};

    \draw[dotted, thick] (3.875, 2.625) -- (6.625, 2.625) -- (6.625, 4.125) -- (3.875, 4.125) -- cycle;
    \node[above] () at (5.3, 3.25) {$\langle \text{MAX}|_{I_2 I}$};
    
    \node[fill=blue, regular polygon, regular polygon sides=3, minimum size=0.35cm, inner sep=0pt] (u1) at (6, 2.375) {};
    \node[fill=blue, regular polygon, regular polygon sides=3, minimum size=0.35cm, inner sep=0pt] (u2) at (2, 2.375) {};
    \node[fill=red, regular polygon, regular polygon sides=3, minimum size=0.35cm, inner sep=0pt, rotate=180] (v1) at (4.5, 2) {};
    \node[fill=red, regular polygon, regular polygon sides=3, minimum size=0.35cm, inner sep=0pt, rotate=180] (v2) at (2, 2) {};
    
    \node[left] at (u2) {$u\phantom{l}$};
    \node[left] at (v2) {$v\phantom{l}$};
    
    \draw[blue, thick] (u1) -- (u2);
    \draw[red, thick] (v1) --  (v2);
    \begin{scope}[yshift = 8.5cm, yscale=-1]
    
    \draw[] (0,0) -- (0,1);
    \draw[] (2,0) -- (2,1);
    \draw[] (0, 1.5) -- (0, 4.25);
    \draw[] (2, 1.5) -- (2, 4.25);
    \draw[] (3,0) -- (3,1);
    
    \draw[fill=red!10!white] (-0.25, 1) -- (3.25, 1) -- (3.25, 1.75) -- (-0.25, 1.75) -- cycle;

    \node[] at (1.5,1.375) {$\widetilde{V}^{\dagger}$};
    \node[above] at (0,0) {$\langle\psi|_{\ell}$};
    \node[above] at (2,0) {$\langle \psi_0|_f$};
    
    \draw[] (3, 0) -- ++ (0.75, -0.5) -- ++ (0.75, 0.5);
    \draw[] (6, -0.5) -- ++ (0, 3.25);
    \draw[] (4.5, 0) -- ++ (0, 2.75); 
    \draw[] (4.5, 2.75) -- ++ (0.75, 0.5) -- ++ (0.75, -0.5);
    
   \draw[dashed] (2.75, 0.25) -- (6.75, 0.25) -- (6.75, -1.25) -- (2.75, -1.25) -- cycle;
    \node[above] () at (4.125,-0.5) {$\langle\text{MAX}|_{I_1 I_2}$};
    \node[above] () at (6.125, -0.5) {$\langle \varphi|_{I}$};
    
    \draw[dotted, thick] (3.875, 2.625) -- (6.625, 2.625) -- (6.625, 4.125) -- (3.875, 4.125) -- cycle;
    \node[below] () at (5.3, 3.25) {$| \text{MAX}\rangle_{I_2 I}$};
    \node[fill=blue, regular polygon, regular polygon sides=3, minimum size=0.35cm, inner sep=0pt, rotate=180] (u1) at (6, 2.375) {};
    \node[fill=blue, regular polygon, regular polygon sides=3, minimum size=0.35cm, inner sep=0pt, rotate=180] (u2) at (2, 2.375) {};
    \node[fill=red, regular polygon, regular polygon sides=3, minimum size=0.35cm, inner sep=0pt] (v1) at (4.5, 2) {};
    \node[fill=red, regular polygon, regular polygon sides=3, minimum size=0.35cm, inner sep=0pt] (v2) at (2, 2) {};
    
    \node[left] at (u2) {$u\phantom{l}$};
    \node[left] at (v2) {$v\phantom{l}$};
    
    \draw[blue, thick] (u1) -- (u2);
    \draw[red, thick] (v1) --  (v2);
    
    \end{scope}
    \end{tikzpicture}
    \right) = \frac{1}{|I|^2},
    \label{eq:pseudorandom_unitarity}
\end{equation}
First we use Eq.~\eqref{eq:I-postselect} to rewrite $\frac{1}{|I|}V(U')$, and we note that the entangled state $|\textrm{MAX}\rangle$ has low complexity. Therefore, the left-hand side has the same form as Eq.~\eqref{eq:alternative-pseudo}, where $X'=B$, $Y'=Y''=R$, and $Z=I_2I$. Comparing to Eq.~\eqref{eq:alternative-pseudo}, we identity $W$ with the product of the low-complexity unitaries $u$ and $v$, $|0\rangle$ with the low-complexity state $|\psi\rangle_\ell \otimes |\psi_0\rangle_f$, $|\phi_2\rangle$ with the low complexity state $|\textrm{MAX}\rangle_{I_1I_2}\otimes |\varphi\rangle_{I}$, and $|\phi_1\rangle$ with the low-complexity state $|\textrm{MAX}\rangle_{I_2I}$. Thus the first equality, in which we replace $U'$ by an integral over Haar measure (making an error which is superpolynomially small in $\log |B|$), follows from the pseudorandomness of $U'$. The second equality follows from Eq.~\eqref{eq:unitarity_average}, taking into account that the diagram represents a matrix element of $\frac{1}{|I|^2}V(U')^\dagger V(U')$. Thus, by straightening out the legs on the left-hand side of Eq.~\eqref{eq:pseudorandom_unitarity} and restoring the conventional normalization, we conclude that for any low-complexity states $|\psi\rangle_{\ell}$ and $|\varphi\rangle_I$,
\begin{equation}
    \label{eq:pseudorandomness_approximate_unitarity}
\left|
        \begin{tikzpicture}[scale=0.8, baseline={([yshift=-.5ex]current bounding box.center)}]
    \draw[] (0,0) -- (0,1);
    \draw[] (2,0) -- (2,1);
    \draw[] (0, 1.5) -- (0, 2.75);
    \draw[] (2, 1.5) -- (2, 2.75);
    \draw[] (3,0) -- (3,1);
    
    \draw[fill=red!10!white] (-0.25, 1) -- (3.25, 1) -- (3.25, 1.75) -- (-0.25, 1.75) -- cycle;
    
     \node[fill=blue, regular polygon, regular polygon sides=3, minimum size=0.35cm, inner sep=0pt] (u1) at (3, 0.25) {};
    \node[fill=blue, regular polygon, regular polygon sides=3, minimum size=0.35cm, inner sep=0pt] (u2) at (2, 2.5) {};
    \node[fill=red, regular polygon, regular polygon sides=3, minimum size=0.35cm, inner sep=0pt] (v1) at (3, 0.7) {};
    \node[fill=red, regular polygon, regular polygon sides=3, minimum size=0.35cm, inner sep=0pt, rotate=180] (v2) at (2, 2) {};
    
    \node[left] at (u2) {$u\phantom{l}$};
    \node[left] at (v2) {$v\phantom{l}$};
    
    \draw[blue, thick] (u1) -- ++ (0.75, 0) -- ++ (0,2.25) -- (u2);
    \draw[red, thick] (v1) -- ++ (0.5, 0) -- ++ (0, 1.3) --  (v2);
    
    \node[] at (1.5,1.375) {$U'$};
    \node[below] at (0,0) {$|\psi\rangle_{\ell}$};
    \node[below] at (2,0) {$|\psi_0\rangle_f$};
    \node[below] at (3,0) {$|\varphi\rangle_I$};
    \begin{scope}[yscale=-1, yshift=-5.5cm]
    
    \draw[] (0,0) -- (0,1);
    \draw[] (2,0) -- (2,1);
    \draw[] (0, 1.5) -- (0, 2.75);
    \draw[] (2, 1.5) -- (2, 2.75);
    \draw[] (3,0) -- (3,1);
    
    \draw[fill=red!10!white] (-0.25, 1) -- (3.25, 1) -- (3.25, 1.75) -- (-0.25, 1.75) -- cycle;
    
    \node[fill=blue, regular polygon, regular polygon sides=3, minimum size=0.35cm, inner sep=0pt, rotate=180] (u1) at (3, 0.25) {};
    \node[fill=blue, regular polygon, regular polygon sides=3, minimum size=0.35cm, inner sep=0pt, rotate=180] (u2) at (2, 2.5) {};
    \node[fill=red, regular polygon, regular polygon sides=3, minimum size=0.35cm, inner sep=0pt, rotate=180] (v1) at (3, 0.7) {};
    \node[fill=red, regular polygon, regular polygon sides=3, minimum size=0.35cm, inner sep=0pt] (v2) at (2, 2) {};
    
    \node[left] at (u2) {$u^{\dagger}$};
    \node[left] at (v2) {$v^{\dagger}$};
    
    \draw[blue, thick] (u1) -- ++ (0.75, 0) -- ++ (0,2.25) -- (u2);
    \draw[red, thick] (v1) -- ++ (0.5, 0) -- ++ (0, 1.3) --  (v2);
    
    \node[] at (1.5,1.375) {${U'}^{\dagger}$};
    \node[above] at (0,0) {$\langle\psi|_{\ell}$};
    \node[above] at (2,0) {$\langle\psi_0|_f$};
    \node[above] at (3,0) {$\langle\varphi|_I$};
    \end{scope}
    \end{tikzpicture}
    -1 \right| \leq |I|^2 \epsilon(\log |B|),
\end{equation}
where $\epsilon(x)$ denotes a function that decays superpolynomially with $x$. Note that $\log |I|$ quantifies the number of qubits that constitute the infaller. We conclude that, as long as the black hole has low complexity and the right-hand side of Eq.~\eqref{eq:pseudorandomness_approximate_unitarity} is small, then the deviation of $V(U')$ from unitarity is small. In particular, if we assume $\epsilon(x)\leq e^{-\alpha x}$ for some constant $\alpha >0$, then $V(U')$ is unitary up to corrections which are exponentially small in $\log |B|$ provided that the size $\log |I|$ of the infaller is small compared to the size $\log |B|$ of the black hole.

This argument applies if the Hilbert space of the infaller has dimension $|I|= e^{\beta \log |B|}$ as long as $\beta$ is strictly less than $\alpha/2$ and the initial state of the infaller is ``simple'' (i.e., has low complexity). In fact, we can relax the low-complexity requirement for the infaller's initial state, and still conclude that unitarity holds to high accuracy if $|I|$ is small enough. First we note that $I$ has a complete orthonormal basis of simple states $\{|\varphi_i\rangle: i = 1,\ldots, |I|\}$, which we shall choose to be the computational basis states, and we can expand an arbitrary state $|\widetilde{\varphi}\rangle_I = \sum_{i=1}^{|I|} c_i |\varphi_i\rangle$ in this basis. Because each of the basis states satisfies Eq.~\eqref{eq:pseudorandomness_approximate_unitarity}, we find
\begin{equation}
\left|
        \begin{tikzpicture}[scale=0.8, baseline={([yshift=-.5ex]current bounding box.center)}]
    \draw[] (0,0) -- (0,1);
    \draw[] (2,0) -- (2,1);
    \draw[] (0, 1.5) -- (0, 2.75);
    \draw[] (2, 1.5) -- (2, 2.75);
    \draw[] (3,0) -- (3,1);
    
    \draw[fill=red!10!white] (-0.25, 1) -- (3.25, 1) -- (3.25, 1.75) -- (-0.25, 1.75) -- cycle;
    
     \node[fill=blue, regular polygon, regular polygon sides=3, minimum size=0.35cm, inner sep=0pt] (u1) at (3, 0.25) {};
    \node[fill=blue, regular polygon, regular polygon sides=3, minimum size=0.35cm, inner sep=0pt] (u2) at (2, 2.5) {};
    \node[fill=red, regular polygon, regular polygon sides=3, minimum size=0.35cm, inner sep=0pt] (v1) at (3, 0.7) {};
    \node[fill=red, regular polygon, regular polygon sides=3, minimum size=0.35cm, inner sep=0pt, rotate=180] (v2) at (2, 2) {};
    
    \node[left] at (u2) {$u\phantom{l}$};
    \node[left] at (v2) {$v\phantom{l}$};
    
    \draw[blue, thick] (u1) -- ++ (0.75, 0) -- ++ (0,2.25) -- (u2);
    \draw[red, thick] (v1) -- ++ (0.5, 0) -- ++ (0, 1.3) --  (v2);
    
    \node[] at (1.5,1.375) {$U'$};
    \node[below] at (0,0) {$|\psi\rangle_{\ell}$};
    \node[below] at (2,0) {$|\psi_0\rangle_f$};
    \node[below] at (3,0) {$|\widetilde{\varphi}\rangle_I$};
    \begin{scope}[yscale=-1, yshift=-5.5cm]
    
    \draw[] (0,0) -- (0,1);
    \draw[] (2,0) -- (2,1);
    \draw[] (0, 1.5) -- (0, 2.75);
    \draw[] (2, 1.5) -- (2, 2.75);
    \draw[] (3,0) -- (3,1);
    
    \draw[fill=red!10!white] (-0.25, 1) -- (3.25, 1) -- (3.25, 1.75) -- (-0.25, 1.75) -- cycle;
    
    \node[fill=blue, regular polygon, regular polygon sides=3, minimum size=0.35cm, inner sep=0pt, rotate=180] (u1) at (3, 0.25) {};
    \node[fill=blue, regular polygon, regular polygon sides=3, minimum size=0.35cm, inner sep=0pt, rotate=180] (u2) at (2, 2.5) {};
    \node[fill=red, regular polygon, regular polygon sides=3, minimum size=0.35cm, inner sep=0pt, rotate=180] (v1) at (3, 0.7) {};
    \node[fill=red, regular polygon, regular polygon sides=3, minimum size=0.35cm, inner sep=0pt] (v2) at (2, 2) {};
    
    \node[left] at (u2) {$u^{\dagger}$};
    \node[left] at (v2) {$v^{\dagger}$};
    
    \draw[blue, thick] (u1) -- ++ (0.75, 0) -- ++ (0,2.25) -- (u2);
    \draw[red, thick] (v1) -- ++ (0.5, 0) -- ++ (0, 1.3) --  (v2);
    
    \node[] at (1.5,1.375) {${U'}^{\dagger}$};
    \node[above] at (0,0) {$\langle\psi|_{\ell}$};
    \node[above] at (2,0) {$\langle\psi_0|_f$};
    \node[above] at (3,0) {$\langle\widetilde{\varphi}|_I$};
    \end{scope}
    \end{tikzpicture}
    -1 \right| \leq |I|^2 e^{-\alpha \log |B|} + 
    \left|
    \sum_{i\neq j} c_i c_j^*
            \begin{tikzpicture}[scale=0.8, baseline={([yshift=-.5ex]current bounding box.center)}]
    \draw[] (0,0) -- (0,1);
    \draw[] (2,0) -- (2,1);
    \draw[] (0, 1.5) -- (0, 2.75);
    \draw[] (2, 1.5) -- (2, 2.75);
    \draw[] (3,0) -- (3,1);
    
    \draw[fill=red!10!white] (-0.25, 1) -- (3.25, 1) -- (3.25, 1.75) -- (-0.25, 1.75) -- cycle;
    
     \node[fill=blue, regular polygon, regular polygon sides=3, minimum size=0.35cm, inner sep=0pt] (u1) at (3, 0.25) {};
    \node[fill=blue, regular polygon, regular polygon sides=3, minimum size=0.35cm, inner sep=0pt] (u2) at (2, 2.5) {};
    \node[fill=red, regular polygon, regular polygon sides=3, minimum size=0.35cm, inner sep=0pt] (v1) at (3, 0.7) {};
    \node[fill=red, regular polygon, regular polygon sides=3, minimum size=0.35cm, inner sep=0pt, rotate=180] (v2) at (2, 2) {};
    
    \node[left] at (u2) {$u\phantom{l}$};
    \node[left] at (v2) {$v\phantom{l}$};
    
    \draw[blue, thick] (u1) -- ++ (0.75, 0) -- ++ (0,2.25) -- (u2);
    \draw[red, thick] (v1) -- ++ (0.5, 0) -- ++ (0, 1.3) --  (v2);
    
    \node[] at (1.5,1.375) {$U'$};
    \node[below] at (0,0) {$|\psi\rangle_{\ell}$};
    \node[below] at (2,0) {$|\psi_0\rangle_f$};
    \node[below] at (3,0) {$\,\,\,\,|\varphi_i\rangle_I$};
    \begin{scope}[yscale=-1, yshift=-5.5cm]
    
    \draw[] (0,0) -- (0,1);
    \draw[] (2,0) -- (2,1);
    \draw[] (0, 1.5) -- (0, 2.75);
    \draw[] (2, 1.5) -- (2, 2.75);
    \draw[] (3,0) -- (3,1);
    
    \draw[fill=red!10!white] (-0.25, 1) -- (3.25, 1) -- (3.25, 1.75) -- (-0.25, 1.75) -- cycle;
    
    \node[fill=blue, regular polygon, regular polygon sides=3, minimum size=0.35cm, inner sep=0pt, rotate=180] (u1) at (3, 0.25) {};
    \node[fill=blue, regular polygon, regular polygon sides=3, minimum size=0.35cm, inner sep=0pt, rotate=180] (u2) at (2, 2.5) {};
    \node[fill=red, regular polygon, regular polygon sides=3, minimum size=0.35cm, inner sep=0pt, rotate=180] (v1) at (3, 0.7) {};
    \node[fill=red, regular polygon, regular polygon sides=3, minimum size=0.35cm, inner sep=0pt] (v2) at (2, 2) {};
    
    \node[left] at (u2) {$u^{\dagger}$};
    \node[left] at (v2) {$v^{\dagger}$};
    
    \draw[blue, thick] (u1) -- ++ (0.75, 0) -- ++ (0,2.25) -- (u2);
    \draw[red, thick] (v1) -- ++ (0.5, 0) -- ++ (0, 1.3) --  (v2);
    
    \node[] at (1.5,1.375) {${U'}^{\dagger}$};
    \node[above] at (0,0) {$\langle\psi|_{\ell}$};
    \node[above] at (2,0) {$\langle\psi_0|_f$};
    \node[above] at (3,0) {$\,\,\,\,\langle\varphi_j|_I$};
    \end{scope}
    \end{tikzpicture}
    \right|.
    \label{eq:off-diagonal}
\end{equation}
The second term on the right-hand side of Eq.~\eqref{eq:off-diagonal} involves off-diagonal terms in the $|\varphi_i\rangle_I$ basis. To bound these terms, we apply Eq.~\eqref{eq:pseudorandomness_approximate_unitarity} to the states $|\varphi\rangle_I = \frac{1}{\sqrt{2}}(|\varphi_i\rangle_I \pm |\varphi_j\rangle_I)$ and use the triangle inequality to show that the real part of each off-diagonal term has absolute value bounded above by $|I|e^{-\alpha\log |B|}$. This works because a superposition of two computational basis states is simple to create. Similarly, we apply Eq.~\eqref{eq:pseudorandomness_approximate_unitarity} to the states $|\varphi\rangle_I = \frac{1}{\sqrt{2}}(|\varphi_i\rangle_I \pm i|\varphi_j\rangle_I)$ to bound the imaginary part of each off-diagonal term. Since for a normalized state vector $\sum_{i,j} c_ic_j^* \le |I|$, we conclude that
\begin{equation}
\label{eq:pseudorandomness_approximate_unitarity_general}
\left|
        \begin{tikzpicture}[scale=0.8, baseline={([yshift=-.5ex]current bounding box.center)}]
    \draw[] (0,0) -- (0,1);
    \draw[] (2,0) -- (2,1);
    \draw[] (0, 1.5) -- (0, 2.75);
    \draw[] (2, 1.5) -- (2, 2.75);
    \draw[] (3,0) -- (3,1);
    
    \draw[fill=red!10!white] (-0.25, 1) -- (3.25, 1) -- (3.25, 1.75) -- (-0.25, 1.75) -- cycle;
    
     \node[fill=blue, regular polygon, regular polygon sides=3, minimum size=0.35cm, inner sep=0pt] (u1) at (3, 0.25) {};
    \node[fill=blue, regular polygon, regular polygon sides=3, minimum size=0.35cm, inner sep=0pt] (u2) at (2, 2.5) {};
    \node[fill=red, regular polygon, regular polygon sides=3, minimum size=0.35cm, inner sep=0pt] (v1) at (3, 0.7) {};
    \node[fill=red, regular polygon, regular polygon sides=3, minimum size=0.35cm, inner sep=0pt, rotate=180] (v2) at (2, 2) {};
    
    \node[left] at (u2) {$u\phantom{l}$};
    \node[left] at (v2) {$v\phantom{l}$};
    
    \draw[blue, thick] (u1) -- ++ (0.75, 0) -- ++ (0,2.25) -- (u2);
    \draw[red, thick] (v1) -- ++ (0.5, 0) -- ++ (0, 1.3) --  (v2);
    
    \node[] at (1.5,1.375) {$U'$};
    \node[below] at (0,0) {$|\psi\rangle_{\ell}$};
    \node[below] at (2,0) {$|\psi_0\rangle_f$};
    \node[below] at (3,0) {$|\widetilde{\varphi}\rangle_I$};
    \begin{scope}[yscale=-1, yshift=-5.5cm]
    
    \draw[] (0,0) -- (0,1);
    \draw[] (2,0) -- (2,1);
    \draw[] (0, 1.5) -- (0, 2.75);
    \draw[] (2, 1.5) -- (2, 2.75);
    \draw[] (3,0) -- (3,1);
    
    \draw[fill=red!10!white] (-0.25, 1) -- (3.25, 1) -- (3.25, 1.75) -- (-0.25, 1.75) -- cycle;
    
    \node[fill=blue, regular polygon, regular polygon sides=3, minimum size=0.35cm, inner sep=0pt, rotate=180] (u1) at (3, 0.25) {};
    \node[fill=blue, regular polygon, regular polygon sides=3, minimum size=0.35cm, inner sep=0pt, rotate=180] (u2) at (2, 2.5) {};
    \node[fill=red, regular polygon, regular polygon sides=3, minimum size=0.35cm, inner sep=0pt, rotate=180] (v1) at (3, 0.7) {};
    \node[fill=red, regular polygon, regular polygon sides=3, minimum size=0.35cm, inner sep=0pt] (v2) at (2, 2) {};
    
    \node[left] at (u2) {$u^{\dagger}$};
    \node[left] at (v2) {$v^{\dagger}$};
    
    \draw[blue, thick] (u1) -- ++ (0.75, 0) -- ++ (0,2.25) -- (u2);
    \draw[red, thick] (v1) -- ++ (0.5, 0) -- ++ (0, 1.3) --  (v2);
    
    \node[] at (1.5,1.375) {${U'}^{\dagger}$};
    \node[above] at (0,0) {$\langle\psi|_{\ell}$};
    \node[above] at (2,0) {$\langle\psi_0|_f$};
    \node[above] at (3,0) {$\langle\widetilde{\varphi}|_I$};
    \end{scope}
    \end{tikzpicture}
    -1 \right| \leq |I|^2(1+|I|) e^{-\alpha \log |B|}.
\end{equation}
Thus we find that unitarity holds up to exponentially small corrections even if $|\widetilde\varphi\rangle_I$ does not have low complexity, but to reach that conclusion we required $|I| =e^{\beta\log|B|}$ for $\beta$ strictly less than $\alpha/3$, a somewhat stronger requirement on the dimension of the infaller's Hilbert space than we found for the case where the initial state of the infaller has low complexity ($\beta$ strictly less than $\alpha/2$). 

To summarize, Eq.~\eqref{eq:pseudorandomness_approximate_unitarity_general} implies that, as long as the infaller's Hilbert space is not too large, and assuming the infaller performs operations on the radiation with complexity polynomial in $\log |B|$, then the black hole $S$-matrix is unitary up to corrections which are exponentially small in $\log |B|$. Here $\log |B|$ is essentially the entropy of the black hole that the infaller encounters. We reached this conclusion by assuming that the black hole's scrambling unitary $U'$ is pseudorandom. Note that here $U' $ is one particular fixed unitary transformation, in contrast to the computation in Section~\ref{sec:average} which involved a Haar average over scrambling unitaries. 

Note that Eq.~\eqref{eq:off-diagonal} and Eq.~\eqref{eq:pseudorandomness_approximate_unitarity_general} were derived under the assumption that the approximation error $\epsilon(\log |B|)$ in Eq.~\eqref{eq:pseudorandomness_approximate_unitarity} is exponentially small. Under the weaker assumption that $\epsilon(\log |B|)$ is superpolynomially small, we conclude that $V(U')$ is approximately unitary with a superpolynomially small error provided that $|I|$ scales polynomially with $\log |B|$.

The holographic map we are considering includes postselection, and it is known that in some situations postselection can lead to closed time-like curves, violation of unitarity, and other pathologies~\cite{Lloyd2011,Lloyd2011a}. In our setting, pseudorandomness apparently obviates this problem, making postselection consistent with unitarity up to very small corrections. 

To conclude that the black hole $S$-matrix is very nearly unitary, we assumed that the infaller performs operations of low compexity. The conclusion may change if the infaller can perform operations with complexity exponential in the black hole entropy, resulting in violations of unitary that are detectable by an exterior observer who performs operations of polynomial complexity. We will discuss this issue in Section~\ref{sec:constraint_infaller}.

\subsection{Complexity of the black hole $S$-matrix}
\label{sec:complexity}

Eq.~\eqref{eq:pseudorandomness_approximate_unitarity} establishes that the a computationally bounded infaller cannot cause the black hole $S$-matrix to deviate significantly from unitarity. However, another issue deserves attention. The circuit Eq.~\eqref{eq:straightening}, which defines the linear map $V(U')$, involves postselection on a measurement outcome that occurs with probability $1/|r|^2$. This probability is exponentially small in the black hole entropy, raising the troubling possibility that the resulting $V(U')$ might have exponentially large computational complexity even if $U'$ does not. Indeed, it is known that quantum circuits of polynomial size, augmented by postselection, can solve arbitrary problems in the very large complexity class PP \cite{Aaronson2005}, which contains NP. Could a crafty observer who stays outside the black hole greatly amplify the computational power of the black hole $S$-matrix by sending suitably programmed robots into the black hole, thus overturning the quantum extended Church-Turing thesis \cite{Deutsch1985,Susskind2020}?

A significant enhancement of the black hole's computational power due to the infalling agent is conceivable. What we can say, though, is that this enhancement is governed, not by the very large Hilbert space dimension of the black hole, but rather by the much smaller Hilbert space dimension of the infaller. If we fix the dimension of the infaller, as well as the computational complexity of $u$ and $v$, then if $U'$ has computational complexity polynomial in the black hole $S$-matrix, so will $V(U')$. Furthermore, if $U'$ is pseudorandom and the infalling agent is small compared to the black hole we can make a stronger statement --- the computational complexity of the black hole $S$-matrix is polynomial in $\log |B|$ provided the interactions of the infaller with the radiation also have polynomial computational complexity, despite the postselection in the holographic map. 

As in our analysis of unitarity, our analysis of the complexity $V(U')$ again hinges on the identity Eq.~\eqref{eq:I-postselect}. What is crucial is that $V(U')$, originally realized in a unitary circuit accompanied by postselection on a measurement outcome occurring with probability $1/|r|^2$, can be alternatively realized in a unitary circuit accompanied by postselection on a measurement outcome occurring with the much larger probability $1/|I|^2$.

We now wish to show that the map
\begin{equation}
V(U')= |I|
    \begin{tikzpicture}[scale=0.8, baseline={([yshift=-.5ex]current bounding box.center)}]
    \draw[] (0,0) -- (0,1);
    \draw[] (2,0) -- (2,1);
    \draw[] (0, 1.5) -- (0, 2.75);
    \draw[] (2, 1.5) -- (2, 2.75);
    \draw[] (3,0) -- (3,1);
    \node[above] () at (0, 2.75) {$B$};
    \node[above] () at (2, 2.75) {$R$};
    
    \draw[fill=red!10!white] (-0.25, 1) -- (3.25, 1) -- (3.25, 1.75) -- (-0.25, 1.75) -- cycle;

    \node[] at (1.5,1.375) {$U'$};
    \node[below] at (0,0) {$|\psi\rangle_{\ell}$};
    \node[below] at (2,0) {$| \psi_0\rangle_f$};
    
    \draw[] (3, 0) -- ++ (0.75, -0.5) -- ++ (0.75, 0.5);
    \draw[] (6, -0.5) -- ++ (0, 3.25);
    \draw[] (4.5, 0) -- ++ (0, 2.75); 
    \draw[] (4.5, 2.75) -- ++ (0.75, 0.5) -- ++ (0.75, -0.5);

    \node[below] () at (4.125,-0.5) {$|\text{MAX}\rangle_{I_1 I_2}$};
    \node[below] () at (6.125, -0.5) {$I$};

    \node[above] () at (5.3, 3.25) {$\langle \text{MAX}|_{I_2 I}$};
    
    \node[fill=blue, regular polygon, regular polygon sides=3, minimum size=0.35cm, inner sep=0pt] (u1) at (6, 2.375) {};
    \node[fill=blue, regular polygon, regular polygon sides=3, minimum size=0.35cm, inner sep=0pt] (u2) at (2, 2.375) {};
    \node[fill=red, regular polygon, regular polygon sides=3, minimum size=0.35cm, inner sep=0pt, rotate=180] (v1) at (4.5, 2) {};
    \node[fill=red, regular polygon, regular polygon sides=3, minimum size=0.35cm, inner sep=0pt, rotate=180] (v2) at (2, 2) {};
    
    \node[left] at (u2) {$u\phantom{l}$};
    \node[left] at (v2) {$v\phantom{l}$};
    
    \draw[blue, thick] (u1) -- (u2);
    \draw[red, thick] (v1) --  (v2);
    \end{tikzpicture},
    \label{eq:circ_temp}
\end{equation}
defined by a nondeterministic circuit with postselection, can be accurately approximated deterministically. 
This is achieved via a technique in quantum computation known as the quantum singular value transformation (QSVT)~\cite{Gilyen2019,Martyn2021}. The basic setting for QSVT is access to a unitary $\mathcal{U}$ and two projectors, $\Pi$ and $\Pi'$, which together encode some rectangular matrix $A$:
\begin{equation}
    \Pi' \mathcal{U} \Pi = A.
\end{equation}
The main goal of QSVT is to construct another unitary $\mathcal{U}'$ that, together with the same projectors, encodes a new matrix $f(A)$, with different singular values than $A$. To be more specific, let $A= \mathsf{U} \mathsf{\Sigma} \mathsf{V}^{\dagger}$ be the singular value decomposition of $A$, where $\mathsf{U}$ and $\mathsf{V}$ are unitary matrices and $\mathsf{\Sigma}$ is a diagonal rectangular matrix with nonnegative entries.  The QSVT algorithm constructs $\mathcal{U}'$ such that 
\begin{equation}
    \Pi' \mathcal{U}' \Pi = f(A) := \mathsf{U} f(\mathsf{\Sigma})\mathsf{V}^{\dagger}
\end{equation}
for some polynomial function $f(x)$ where $f(\mathsf{\Sigma})$ is an entry-wise application of $f(x)$ to $\mathsf{\Sigma}$.

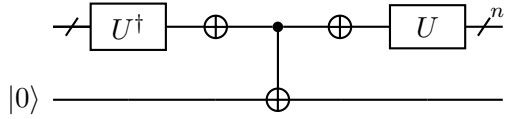
\begin{figure}[h]
    \centering
    \begin{quantikz}
    & \gate[wires=1][1cm]{U^{\dagger}}\qwbundle{} & \targ{} & \ctrl{1} & \targ{} &\gate[wires=1][1cm]{U} & \qwbundle{n} \\
    \lstick{$|0\rangle$}&\qw & \qw & \targ{} & \qw & \qw & \qw
    \end{quantikz}
    \caption{The circuit to implement Eq.~\eqref{eq:check_projector}, where $P = U{|0\ldots 0\rangle}{\langle 0\ldots 0|}U^{\dagger}$. Here $U$ is a general unitary matrix acting on $n$ qubits and the gate in the middle is a multi-qubit Toffoli gate controlled on $n$ qubits. The $NOT$ gates are applied to all the $n$ qubits.}
    \label{fig:complexity_checking}
\end{figure}

The unitary $\mathcal{U}'$ can be constructed from $\mathcal{U}$, its inverse, and a unitary process that ``checks'' if a state lies in the kernel of a projector. Specifically, given a projector $P$, the ``check'' can be realized by the map
\begin{equation}
\label{eq:check_projector}
        |\psi\rangle |0\rangle  \to 
        \begin{cases}
        |\psi \rangle |1\rangle &\text{ if } |\psi\rangle \in
        \text{ker}(P) \\
        |\psi \rangle |0\rangle &\text{ if } |\psi\rangle \in \text{ker}(P)^{\perp}
        \end{cases}
\end{equation}
where $\text{ker}(P)$ is the kernel of $P$ and $\text{ker}(P)^{\perp}$ is its orthogonal complement. If furthermore the projector $P$ can be expressed as  $P={|\varphi\rangle}{\langle\varphi|}$, where $|\varphi\rangle = U |0\ldots 0\rangle$, Eq.~\eqref{eq:check_projector} can be implemented by the circuit in Fig.~\ref{fig:complexity_checking}. Thus, in this case, the complexity of implementing Eq.~\eqref{eq:check_projector} is proportional to the complexity of preparing the state $|\varphi\rangle$, up to some additional gates whose cost is negligible in comparison. The QSVT makes use of such check maps for both the projector $\Pi$ and the projector $\Pi'$.

Importantly, the complexity of $\mathcal{U}'$ can be quantified in terms of the degree of the polynomial function $f(x)$. Let $\mathcal{C}$ be the sum of the complexity of $\mathcal{U}$ and the complexity of implementing Eq.~\eqref{eq:check_projector}. Then the complexity of $\mathcal{U}'$ is bounded by $O(\mathcal{C} d)$, where $d$ is the degree of the polynomial.

Ported to our setup, one can see that our goal of implementing $V(U')$ deterministically can be achieved using the QSVT approach. Note that $V(U')$ in fact can be constructed from a unitary process $\mathcal{U} := uv U'$, sandwiched between two projections, defined as
\begin{equation}
\begin{aligned}
    \Pi &= 
    {|\psi_0\rangle_f}{\langle\psi_0|} \otimes {|\text{MAX}\rangle_{I_1I_2}}{\langle\text{MAX}|}, \\
    \Pi' &= {|\text{MAX}\rangle_{I_2I}}{\langle \text{MAX}|}.
\end{aligned}
\end{equation}
Specifically, 
\begin{equation}
    \Pi' \mathcal{U} \Pi = \frac{1}{|I|} {|\text{MAX}\rangle_{I_2 I}} \cdot V(U') \cdot 
    {\langle \psi_0|_f}{\langle \text{MAX}|_{I_1I_2}}.
\end{equation}
Because both $\Pi$ and $\Pi'$ are projectors onto states of low complexity, the corresponding check maps have correspondingly low circuit complexity.

The singular values of the matrix $\Pi' \mathcal{U} \Pi$ are all $\approx 1/|I|$, up to an additive correction that decays exponentially in $\log |B|$; see Eq.~\eqref{eq:pseudorandomness_approximate_unitarity_general}. These small singular values occur because the projection onto the state $|\text{MAX}\rangle_{I_2 I}$ in Eq.~\eqref{eq:circ_temp} succeeds with probability $1/|I|^2$. Our goal is to replace $\mathcal{U}$ by a different efficient unitary transformation $\mathcal{U}'$ such that the singular values of $\Pi' \mathcal{U}' \Pi$ are all close to one. This would mean that if $\mathcal{U}'$ is performed instead of $\mathcal {U}$, the desired measurement outcome is obtained with probability close to one. Furthermore, we want the resulting map acting on $\ell  I$ to closely approximate $V(U')$.

These goals can be achieved using the QSVT where $f(x)$ is a polynomial approximation to the sign function:
\begin{equation}
f(x)\approx \, \text{sgn}(x-c)=
    \begin{cases}
    -1, & x<c, \\
    0, & x=c, \\
    1, & x>c.
    \end{cases}
\end{equation}
To see why this works, suppose we could apply the sign function, without any approximation, to the singular values of $\Pi'\mathcal{U}\Pi$, and that we choose the constant $c$ to be strictly smaller than any of these singular values. For example, we could choose $c$ to be $1/|I|$ minus a correction that decays more slowly than exponentially in $\log |B|$.\footnote{ Other choices of $c$ would suffice as well, for instance, $c=1/(2|I|)$, assuming $\log |B|$ is sufficiently large.} In this case, the QSVT would boost each singular value from a value close to $1/|I|$ to the value $1$. Furthermore, because $\Pi'\mathcal{U}\Pi$ and $\Pi'\mathcal{U}'\Pi$ are diagonalized by the same transformations $\mathsf{U}$ and $\mathsf{V}$ acting on left and right, and because the singular values of $\Pi\mathcal{U}\Pi'$ are very nearly all the same, the effect of the QSVT, to a very good approximation, is to multiply $\Pi\mathcal{U}\Pi'$ by the constant factor $|I|$; hence,
\begin{equation}
    \Pi\mathcal{U}'\Pi'\approx  |\textrm{MAX}\rangle_{I_2I}\cdot V(U') \cdot  
    \langle \psi_0|_{f} \langle \textrm{MAX}|_{I_1 I_2}.
\end{equation}
Therefore, by applying $\mathcal{U}'$ to the initial state
initial state $|\psi\rangle_{\ell}|\psi_0\rangle_f|\textrm{MAX}\rangle_{I_1I_2}|\varphi\rangle_I$, we obtain
\begin{equation}
    \mathcal{U'}\left(|\psi\rangle_{\ell}|\psi_0\rangle_f|\textrm{MAX}\rangle_{I_1I_2}|\varphi\rangle_I\right) \approx (V(U')|\psi_\ell\rangle|\varphi\rangle_I)|\textrm{MAX}\rangle_{I_2I}.
\end{equation}
Thus, we can deterministically apply the operator $V(U')$ to the initial state $|\psi_\ell\rangle|\varphi\rangle_I$.

In the above argument, we considered the QSVT in the idealized case where the sign function is applied exactly. However, it is impossible to represent the sign function exactly as a finite-degree polynomial; upon truncating to a finite degree, there is an approximation error that must be analyzed. It was shown in Ref.~\cite{Gilyen2019} that there is a polynomial of degree $O(\Delta^{-1} \log 1/\epsilon)$ which approximates the sign function up to an error of $\epsilon$, as long as $x$ does not belong to $[-\Delta/2, \Delta/2]$. Thus, in our setup, we can choose $\Delta = O(1/|I|)$, leading to the degree of $O(|I|)$. Therefore, up to a logarithmic factor in the inverse precision $\epsilon$, we can implement $V(U')$ deterministically, with the complexity $O(|I| \mathcal{C})$, where $\mathcal{C}$ is the combined complexity of $U', u, v,$ and the state preparation for $|\psi_0\rangle_f, |\text{MAX}\rangle_{I'I''}, |\text{MAX}\rangle_{I''I}$. Thus, the overall complexity of $V(U')$ remains polynomial in the initial black hole entropy, as long as the infaller's Hilbert space is sufficiently small.

We conclude that postselection can boost the computational complexity of $uvU'$ by at most an $O(|I|)$ factor. That is much smaller than a factor exponential in $\log |B|$, which one might have naively expected due to the postselection in the holographic map, but even $O(|I|)$ would be a big boost in computational power. If for example we send a programmed 1000-qubit quantum computer into a black hole, can we enhance the computational power of the black hole by a factor $O(|I|=2^{1000})$? The above argument does not rule out this possibility.

However, our complexity analysis has not yet invoked the pseudorandomness of $U'$. We show in Appendix \ref{appendix:bh_assisted_qc} that if $U'$ is pseudorandom, then a postselected quantum computation enabled by an infalling robot that interacts efficiently with radiation both outside and inside the black hole can be simulated by a conventional quantum circuit (without postselection) of size polynomial in $\log |B|$, where the error in the simulation is $|I|^2\epsilon(|B|)$, and $\epsilon(|B|)$ is the approximation error in the relation Eq.\eqref{eq:alternative-pseudo}. The simulation error is exponentially small in $\log |B|$ if $\epsilon(|B|)= e^{-\alpha\log|B|}$ and $|I|= e^{\beta\log |B|}$ where $\beta$ is strictly less than $\alpha/2$. Under these assumptions, the postselection in the holographic map does not result in superpolynomial computational power.

\section{Computationally powerful infaller}
\label{sec:constraint_infaller}

In Section~\ref{sec:average}, we showed that for Haar-random $U'$, and for any $u$ and $v$, $V(U')$ is approximately unitary with high probability. Furthermore, in Section~\ref{sec:unitarity_pseudorandomness} we showed that for any pseudorandom unitary $U'$, if $u$, $v$, and the initial state are polynomially complex, then $V(U')$ is approximately unitary. In this section, we consider what happens if, for an arbitrary fixed $U'$, the interactions of the infaller with the radiation (the unitary transformations $u$ and $v$) have unrestricted complexity. We will see that, in that case, $u$ and $v$ can be chosen so that $V(U')$ deviates substantially from unitarity. This non-unitarity could be detected by an exterior observer, as discussed in Section~\ref{sec:issues}.

To demonstrate the potential for non-unitarity, it will suffice to choose $u$ to be the identity --- the infaller will interact nontrivally only with the radiation modes inside the black hole. To construct the appropriate $v$, we use the observation that if $U'$ is a scrambling unitary, then for an old black hole such that $|R|\gg |B|$ one can recover the initial state of the infaller $I$ by performing a decoding operation on $R$ alone~\cite{Hayden2007}. That is, there is a unitary $v'$ (which might have superpolynomial computational complexity) that unscrambles the radiation and deposits the infalling state of $I$ in a quantum memory $I'$ which has the same Hilbert space dimension as $I$:

\begin{equation}
        \begin{tikzpicture}[scale=0.8, baseline={([yshift=-.5ex]current bounding box.center)}]
    \draw[] (0,0) -- (0,1);
    \draw[] (2,0) -- (2,1);
    \draw[] (0, 1.5) -- (0, 3);
    \draw[] (2, 1.5) -- (2, 3);
    \draw[] (3,0) -- (3,1);
    
    \draw[fill=red!10!white] (-0.25, 0.75) -- (3.25, 0.75) -- (3.25, 1.5) -- (-0.25, 1.5) -- cycle;
    
    \node[] at (1.5,1.125) {$U'$};
    \node[below] at (0,0) {$| \psi\rangle_{\ell}$};
    \node[below] at (2,0) {$| \psi_0\rangle_f$};
    \node[below] at (3,0) {$I$};
    
    \draw[] (3, 2.5) --++ (0, 0.5);
    \draw[fill=blue!10!white] (1.75,2.25) -- ++ (1.5,0 ) -- ++ (0, 0.5) -- ++ (-1.5, 0) --cycle;
    \node[] () at (2.5, 2.5) {$v'$};
    \node[above] () at (0,3) {$B$};
    \node[left] () at (2,1.875) {$R$};
    \node[above] () at (2,3) {$R'$};
    \node[above] () at (3,3) {$I'$};
    \end{tikzpicture}
    =
    \begin{tikzpicture}[scale=0.8, baseline={([yshift=-.5ex]current bounding box.center)}]
    \draw[] (0,0) -- (0,1);
    \draw[] (2,0) -- (2,1);
    \draw[] (0, 1.5) -- (0, 3);
    \draw[] (2, 1.5) -- (2, 3);
    \draw[] (3,0) -- (3,1);
    
    \draw[fill=red!10!white] (-0.25, 0.75) -- (2.75, 0.75) -- (2.75, 1.5) -- (-0.25, 1.5) -- cycle;

    \node[] at (1.25,1.25) {$U''$};
    \node[below] at (0,0) {$| \psi\rangle_{\ell}$};
    \node[below] at (2,0) {$| \psi_0\rangle_f$};
    \node[below] at (3,0) {$I$};
    
    \draw[] (3, 1) --++ (0, 2);
    \node[above] () at (0,3) {$B$};
    \node[above] () at (2,3) {$R'$};
    \node[above] () at (3,3) {$I'$};
    \end{tikzpicture}.
    \label{eq:unscrambler}
\end{equation}

Now we construct $v$ as the product of two unitary transformations: $v''$ acting on $Ir$, followed by $v'^T$ (the transpose of the unscrambler $v'$) acting on $r$ along. Here we choose $v''$ so that it couples $I$ to the subsystem of $r$ that is mapped to $I'$ by the holographic map and unscrambler:
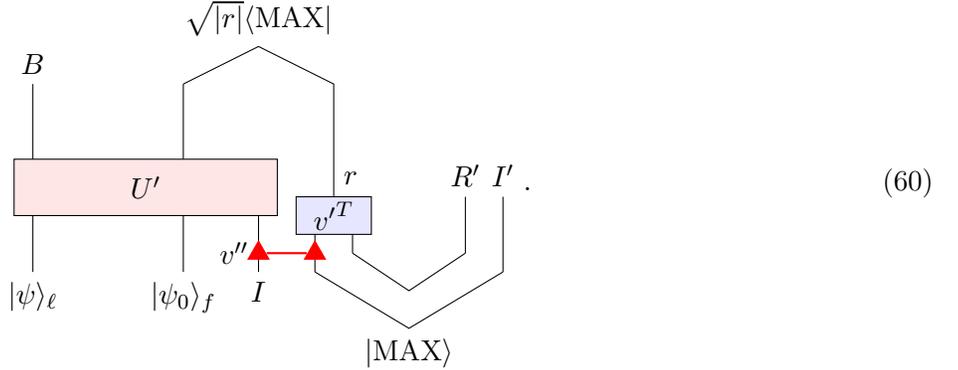
\begin{equation}
    \begin{tikzpicture}[baseline={([yshift=-.5ex]current bounding box.center)}]
    \draw[] (0,0) -- (0,1);
    \draw[] (2,0) -- (2,1);
    \draw[] (3.75,0) -- (3.75,1);
    \draw[] (4.25,0.25) -- (4.25,1);
    \draw[] (5.75,0.25) -- (5.75,1);
    \draw[] (6.25,0) -- (6.25,1);
    \draw[] (0, 1.5) -- (0, 2.5);
    \draw[] (2, 1.5) -- (2, 2.5);
    \draw[] (2, 2.5) -- (3, 3) -- (4, 2.5) -- (4,1);
    \draw[] (3,0) -- (3,1);

    \draw[] (3.75,0) -- (5, -0.75) -- (6.25,0);
    \draw[] (4.25, 0.25) -- (5, -0.25) -- (5.75, 0.25);
    
    \draw[fill=red!10!white] (-0.25, 0.75) -- (3.25, 0.75) -- (3.25, 1.5) -- (-0.25, 1.5) -- cycle;
    
    \node[fill=red, regular polygon, regular polygon sides=3, minimum size=0.35cm, inner sep=0pt] (v1) at (3, 0.25) {};
    \node[fill=red, regular polygon, regular polygon sides=3, minimum size=0.35cm, inner sep=0pt] (v2) at (3.75, 0.25) {};
    
    \node[left] at (v1) {$v''$};
    
    \draw[red, thick] (v1) -- (v2);
    
    \node[above] at (0, 2.5) {$B$};
    \node[above] at (3, 3) {$\sqrt{|r|} \langle \text{MAX}|$};
    \node[] at (1.5,1.125) {$U'$};
    \node[below] at (0,0) {$|\psi\rangle_{\ell}$};
    \node[below] at (2,0) {$|\psi_0\rangle_f$};
    \node[below] at (3,0) {$I$};
    \node[below] at (5,-0.75) {$|\text{MAX}\rangle$};
    
    \draw[fill=blue!10!white] (3.5, 0.5) -- ++ (1, 0) -- ++ (0, 0.5) -- ++ (-1, 0) -- cycle; 
     
    \node[] () at (4,0.75) {${v'}^T$};
    \node[above] at (5.75, 1) {$R'$};
    \node[above] at (6.25, 1) {$I'$};
    \node[right] at (4,1.25) {$r$};
\end{tikzpicture}.
\end{equation}
We see using Eq.~\eqref{eq:unscrambler} that $I$ is mapped to $I'$ by a process in which, in effect, $I$ interacts with its future self:
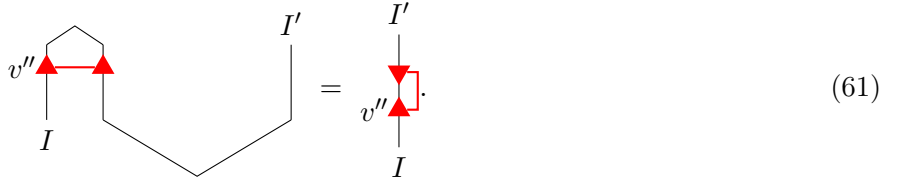
\begin{equation}
    \begin{tikzpicture}[baseline={([yshift=-.5ex]current bounding box.center)}]
    \draw[] (3.75,0) -- (3.75,1);
    \draw[] (6.25,0) -- (6.25,1);
 
    \draw[] (3,0) -- (3,1);
    
    \draw[] (3.75,0) -- (5, -0.75) -- (6.25,0);
    \draw[] (3,1) -- ++ (0.375, 0.25) -- ++ (0.375, -0.25);

    \node[fill=red, regular polygon, regular polygon sides=3, minimum size=0.35cm, inner sep=0pt] (v1) at (3, 0.7) {};
    \node[fill=red, regular polygon, regular polygon sides=3, minimum size=0.35cm, inner sep=0pt] (v2) at (3.75, 0.7) {};
    
    \node[left] at (v1) {$v''$};
    
    \draw[red, thick] (v1) -- (v2);
    
    \node[below] at (3,0) {$I$};
    \node[above] at (6.25, 1) {$I'$};
\end{tikzpicture}
=
\begin{tikzpicture}[baseline={([yshift=-.5ex]current bounding box.center)}]

\draw[] (0,0) -- (0,1.5);
\node[below] () at (0,0) {$I$};
\node[above] () at (0,1.5) {$I'$};
\node[fill=red, regular polygon, regular polygon sides=3, minimum size=0.35cm, inner sep=0pt] (v1) at (0, 0.5) {};
    \node[fill=red, regular polygon, regular polygon sides=3, minimum size=0.35cm, inner sep=0pt, rotate=180] (v2) at (0, 1.0) {};
    
    \node[left] at (v1) {$v''$};
    
    \draw[red, thick] (v1) -- ++ (0.25,0) -- ++ (0, 0.5) -- (v2);
\end{tikzpicture}.
\label{eq:postselect-IIprime}
\end{equation}
This map from $I$ to $I'$ is not necessarily unitary~\cite{gottesman2004comment}. To be concrete, let $\{|i\rangle, i=0, 1, \dots , |I|-1\}$ denote an orthonormal basis for $I$, and suppose the unitary transformation $v''$ is the ``controlled-sum gate''
\begin{equation}
    v'': |i\rangle\otimes |j\rangle \to |i\rangle\otimes | j+i ~(\text{mod } |I|)\rangle.
\end{equation}
Then after postselection and renormalization of the state, the map in Eq.~\eqref{eq:postselect-IIprime} takes an arbitrary input state to $|0\rangle$, flagrantly violating unitarity.

If the unitary acting on $Ir$ is $v=v'^T$ (i.e., if $v''$ is the identity), then as we have already emphasized the map from $I$ to $I'$ is the identity map. This means that we can program a (computationally powerful) robot so that after a scrambling time (the time needed for $U'$ to be a scrambling unitary) the information encoded in the infalling robot's memory emerges from the black hole in unscrambled form, encoded in just $O(\log |I|)$ qubits of the Hawking radiation.

We underscore again that the infaller induces detectable violations of unitarity, or arranges rapid emergence of quantum information in the Hawking radiation, by applying the unscrambling operation $v'^T$ to $r$ behind the horizon. In fact we expect this unscrambler to have computational complexity which is exponential in $\log |B|$~\cite{Harlow2013,Aaronson2005,Kim2020,brown2020python}. The proper time for the infaller to reach the black hole singularity after crossing the event horizon scales like the square root of $\log |B|$, far too short a time to apply so complex an operation. A physical process occurring in such a short time is expected to have polynomial complexity, and as we saw in Section~\ref{sec:unitarity_pseudorandomness}, the unitarity of the black hole $S$-matrix is well protected against such computationally bounded infallers. 

\section{Conclusion}
\label{sec:conclusion}

The ``central dogma'' of black hole physics \cite{almheiri2021entropy} asserts that a black hole when viewed from outside is a highly chaotic quantum system with finite Hilbert space dimension $|B|$, where $\log |B|$ is the black hole's Bekenstein-Hawking entropy, and that the joint evolution of the black hole and its surroundings is exactly unitary. In the language of Akers et al.~\cite{Akers2022}, this description of a black hole when viewed from outside is the ``fundamental picture.'' 
If we accept the central dogma, we face the challenge of reconciling this fundamental picture with an ``effective picture,'' which describes the black hole as viewed by observers who fall into the black hole. Black hole complementarity~\cite{Susskind1993} posits that, away from regions of large spacetime curvature, the effective picture is accurately characterized by local effective quantum field theory, and that a well-defined dictionary (a ``holographic map'') relates physics in the effective picture to a corresponding description of the same physics in the fundamental picture. 

Akers et al.~\cite{Akers2022} proposed such a holographic map in a toy model of black hole physics, and they argued persuasively that this model captures some essential features of realistic black holes. Our goal in this work has been to subject the holographic map of \cite{Akers2022} to further scrutiny. Our conclusion is that, under appropriate assumptions, the proposed holographic map passes some nontrivial consistency tests beyond those investigated in Ref.~\cite{Akers2022}.

For an old black hole that has already radiated away more than half of its initial entropy, the Hilbert space of the black hole interior in the effective picture has a much larger dimension than the black hole Hilbert space in the fundamental picture. Therefore, the holographic map from the effective picture to the fundamental picture is necessarily non-isometric, annihilating many of the states in the effective picture. Ref.~\cite{Akers2022} argued that local effective field theory can nevertheless accurately describe the interior as seen by observers who perform operations of bounded computational complexity, thus providing a satisfying explanation of how the two pictures can fit together. 

However, the non-isometric map is realized by unitary evolution accompanied by measurement and postselection; this raises concerns, because it is known that postselection can result in pathological effects such as flagrant violations of unitarity \cite{gottesman2004comment,Lloyd2011,Lloyd2011a,Lloyd2014}. In the proposal of Ref.~\cite{Akers2022}, postselection is a feature of the holographic map rather than a process occurring in spacetime. Nevertheless, we should examine whether the consequences of postselection might disrupt the putative compatibility of the effective and fundamental pictures under more general conditions than those considered in \cite{Akers2022}.

In particular, we have considered the effects of infalling agents that interact with radiation modes both outside and inside the black hole. As we have emphasized, it is possible in principle for such an infalling agent to induce flagrant violations of unitary that can be detected by an observer who stays outside the black hole. We have found, though, that the violations of unitarity are extremely small with high probability if the unitary transformation $U'$ describing the black hole's internal unitary dynamics is a typical transformation sampled from the Haar distribution. In the more realistic case where $U'$ is an efficient but pseudorandom unitary, we found that deviations from unitarity are highly suppressed if the infaller's Hilbert space dimension is not too large. We also investigated whether, in the presence of an infaller, the postselection inherent to the holographic map results in substantial amplification of the black hole's computational power. We concluded that this amplification is modest if the infaller's Hilbert space is small. For our analysis of both the unitarity and the complexity of the black hole $S$-matrix $V(U')$ we rely on the same essential trick --- we can replace the description of $V(U')$ in Ref.~\cite{Akers2022}, which involves postselection in the very large Hilbert space associated with the black hole interior, by an equivalent description involving postselection in a much smaller Hilbert space determined by the size of the infalling agent. 

From a quantum information perspective, it is intriguing to see that there are settings where postselection is compatible with (very small deviations from) unitarity and does not result in excessive computational power. But what guiding principles fix the context in which postselection is physically acceptable? Further exploration of the connections between exhausitive search and superluminal signalling ~\cite{bao2016grover} may be helpful for addressing this question. 

Of course, what we have analyzed is merely a toy model of black hole physics; we have not attempted to fold in essential features such as diffeomorphism invariance, weakly-coupled propagating gravitons, etc. Finding that well-motivated assumptions imply very small violations of unitarity in the toy model boosts our confidence that a more complete model can reconcile a semiclassical treatment in the effective picture with exact unitarity in the fundamental picture, in accord with the standard dogma. But that remains to be seen.

\section*{Acknowledgments}
We thank Chris Akers and Daniel Harlow for valuable discussions. JP acknowledges funding provided by the Institute for Quantum Information and Matter, an NSF Physics Frontiers Center (NSF Grant {PHY}-{1733907}), the Simons Foundation It from Qubit Collaboration, the DOE QuantISED program ({DE}-{SC0018407}), and the Air Force Office of Scientific Research ({FA9550}-{19}-{1}-{0360}).

\bibliographystyle{myhamsplain2}
\bibliography{bib}

\providecommand{\bysame}{\leavevmode\hbox to3em{\hrulefill}\thinspace}
\begin{thebibliography}{10}

\bibitem{Akers2022}
Chris Akers, Netta Engelhardt, Daniel Harlow, Geoff Penington, and Shreya
  Vardhan, \emph{The black hole interior from non-isometric codes and
  complexity}, July 2022, \href{http://arxiv.org/abs/2207.06536}{2207.06536}.

\bibitem{Ryu2006}
Shinsei Ryu and Tadashi Takayanagi, \emph{Holographic derivation of
  entanglement entropy from the anti--de sitter space/conformal field theory
  correspondence}, Phys. Rev. Lett. \textbf{96} (2006), 181602.

\bibitem{Faulkner2013}
Thomas Faulkner, Aitor Lewkowycz, and Juan Maldacena, \emph{Quantum corrections
  to holographic entanglement entropy}, Journal of High Energy Physics
  \textbf{2013} (2013), no.~11, 74.

\bibitem{Engelhardt2015}
Netta Engelhardt and Aron~C. Wall, \emph{Quantum extremal surfaces: Holographic
  entanglement entropy beyond the classical regime}, Journal of High Energy
  Physics \textbf{2015} (2015), no.~1, 73.

\bibitem{Horowitz2004}
Gary~T. Horowitz and Juan Maldacena, \emph{The black hole final state}, Journal
  of High Energy Physics \textbf{2004} (2004), no.~02, 008--008.

\bibitem{Lloyd2014}
Seth Lloyd and John Preskill, \emph{Unitarity of black hole evaporation in
  final-state projection models}, Journal of High Energy Physics \textbf{2014}
  (2014), no.~8, 126.

\bibitem{Aaronson2005}
Scott Aaronson, \emph{Quantum computing, postselection, and probabilistic
  polynomial-time}, Proceedings of the Royal Society A: Mathematical, Physical
  and Engineering Sciences \textbf{461} (2005), no.~2063, 3473--3482.

\bibitem{Deutsch1985}
D.~{Deutsch}, \emph{{Quantum theory, the Church-Turing principle and the
  universal quantum computer}}, Proceedings of the Royal Society of London
  Series A \textbf{400} (1985), no.~1818, 97--117.

\bibitem{Susskind2020}
Leonard Susskind, \emph{Horizons protect church-turing}, 2020,
  \href{http://arxiv.org/abs/arXiv:2003.01807}{arXiv:2003.01807}.

\bibitem{Ji2018}
Zhengfeng Ji, Yi-Kai Liu, and Fang Song, \emph{Pseudorandom quantum states},
  Cryptology ePrint Archive, Paper 2018/544, 2018,
  \url{https://eprint.iacr.org/2018/544}.

\bibitem{Kim2020}
Isaac Kim, Eugene Tang, and John Preskill, \emph{The ghost in the radiation:
  Robust encodings of the black hole interior}, Journal of High Energy Physics
  \textbf{2020} (2020), no.~6, 31.

\bibitem{almheiri2021entropy}
Ahmed Almheiri, Thomas Hartman, Juan Maldacena, Edgar Shaghoulian, and
  Amirhossein Tajdini, \emph{The entropy of hawking radiation}, Reviews of
  Modern Physics \textbf{93} (2021), no.~3, 035002.

\bibitem{gottesman2004comment}
Daniel Gottesman and John Preskill, \emph{Comment on ``the black hole final
  state''}, Journal of High Energy Physics \textbf{2004} (2004), no.~03, 026.

\bibitem{Harlow2013}
Daniel Harlow and Patrick Hayden, \emph{Quantum computation vs. firewalls},
  Journal of High Energy Physics \textbf{2013} (2013), no.~6, 85.

\bibitem{Hayden2007}
Patrick Hayden and John Preskill, \emph{Black holes as mirrors: Quantum
  information in random subsystems}, Journal of High Energy Physics
  \textbf{2007} (2007), no.~09, 120--120.

\bibitem{Collins2003}
Beno{\^i}t Collins, \emph{Moments and cumulants of polynomial random variables
  on unitarygroups, the {{Itzykson-Zuber}} integral, and free probability},
  International Mathematics Research Notices \textbf{2003} (2003), no.~17,
  953--982.

\bibitem{Lloyd2011}
Seth Lloyd, Lorenzo Maccone, Raul Garcia-Patron, Vittorio Giovannetti, Yutaka
  Shikano, Stefano Pirandola, Lee~A. Rozema, Ardavan Darabi, Yasaman Soudagar,
  Lynden~K. Shalm, and Aephraim~M. Steinberg, \emph{Closed timelike curves via
  postselection: Theory and experimental test of consistency}, Phys. Rev. Lett.
  \textbf{106} (2011), 040403.

\bibitem{Lloyd2011a}
Seth Lloyd, Lorenzo Maccone, Raul Garcia-Patron, Vittorio Giovannetti, and
  Yutaka Shikano, \emph{Quantum mechanics of time travel through post-selected
  teleportation}, Phys. Rev. D \textbf{84} (2011), 025007.

\bibitem{Gilyen2019}
Andr\'{a}s Gily\'{e}n, Yuan Su, Guang~Hao Low, and Nathan Wiebe, \emph{Quantum
  singular value transformation and beyond: Exponential improvements for
  quantum matrix arithmetics}, Proceedings of the 51st Annual ACM SIGACT
  Symposium on Theory of Computing (New York, NY, USA), STOC 2019,
  p.~193–204, Association for Computing Machinery, 2019.

\bibitem{Martyn2021}
John~M. Martyn, Zane~M. Rossi, Andrew~K. Tan, and Isaac~L. Chuang, \emph{Grand
  unification of quantum algorithms}, PRX Quantum \textbf{2} (2021), 040203.

\bibitem{brown2020python}
Adam~R Brown, Hrant Gharibyan, Geoff Penington, and Leonard Susskind, \emph{The
  python’s lunch: geometric obstructions to decoding hawking radiation},
  Journal of High Energy Physics \textbf{2020} (2020), no.~8, 1--53.

\bibitem{Susskind1993}
Leonard Susskind, L\'arus Thorlacius, and John Uglum, \emph{The stretched
  horizon and black hole complementarity}, Phys. Rev. D \textbf{48} (1993),
  3743--3761.

\bibitem{bao2016grover}
Ning Bao, Adam Bouland, and Stephen~P Jordan, \emph{Grover search and the
  no-signaling principle}, Physical review letters \textbf{117} (2016), no.~12,
  120501.

\bibitem{Fannes1973}
M.~Fannes, \emph{A continuity property of the entropy density for spin lattice
  systems}, Communications in Mathematical Physics \textbf{31} (1973), no.~4,
  291--294.

\bibitem{Audenaert2007}
Koenraad M~R Audenaert, \emph{A sharp continuity estimate for the von neumann
  entropy}, Journal of Physics A: Mathematical and Theoretical \textbf{40}
  (2007), no.~28, 8127.

\end{thebibliography}

\appendix

\section{Distinguishing a low-rank channel from a depolarizing channel}
\label{appendix:two-design_nogo}
Here we show that any quantum channel that can be purified using a small Hilbert space is information-theoretically distinguishable from the depolarizing channel. Consider a physical process acting on $\mathcal{H}_Y$, in which one (i) prepares a fixed pure state on $X$, (ii) applies a unitary $U$ acting on $\mathcal{H}_X \otimes \mathcal{H}_Y$, and (iii) traces out $X$. This defines a quantum channel on $Y$, denoted as $\mathcal{E}_Y$.

We claim that, assuming $|X|\ll |Y|$, $\mathcal{E}_Y$ is statistically distinguishable from the depolarizing channel $\mathcal{D}_Y$, defined as follows:
\begin{equation}
    \mathcal{D}_Y(\rho) = \frac{I_Y}{|Y|}
\end{equation}
for any density matrix $\rho$, where $I_Y$ is the identity operator on $\mathcal{H}_Y$. Two quantum states can be distinguished by some measurement with $O(1)$ success probability if the two states are distance $O(1)$ apart in the 1-norm. Therefore, two channels are information-theoretically distinguishable if there is an input state such that the two corresponding outputs are distance $O(1)$ apart in the 1-norm.

If the channel $\mathcal{E}_Y$ acts on the input pure state $|\psi\rangle\langle\psi|$, then the entropy of the output state is upper bounded by $\log |X|$. On the other hand, if we apply $\mathcal{D}_Y$ to $|\psi\rangle\langle\psi|$, we get a state with an entropy of $\log |Y|$. 
At this point we can invoke the continuity of entropy due to Fannes and Audenaert~\cite{Fannes1973,Audenaert2007}:
\begin{equation}
    |S(\rho) - S(\sigma)| \leq T\log (d-1) + H(\{T, 1-T\});
\end{equation}
here $T=\frac{1}{2}\|\rho - \sigma\|_1$, $d$ is the Hilbert space dimension, and $H(\{p, 1-p \}) = -p\log p - (1-p) \log (1-p)$ is the binary entropy. Choosing $\rho = \mathcal{E}_Y(|\psi\rangle\langle\psi|)$ and $\sigma = \mathcal{D}_Y(|\psi\rangle\langle\psi|)$, we obtain the following bound:
\begin{equation}
    |\log |Y| - \log |X|| \leq T \log (|Y|-1) + H(\{T, 1-T \}).
\end{equation}
The binary entropy is no larger than 1;
therefore, if $|X| \ll |Y|$ we conclude that $T$ must be close to 1 if $\log|Y|$ is large. Thus, with high probability one can distinguish $\mathcal{E}_Y$ from $\mathcal{D}_Y$, by applying both channels to the same pure state.

This argument shows that a measurement exists that distinguishes the two channels, but provides no guarantee concerning the computational complexity of that measurement. As discussed in Section \ref{sec:pseudorandomness}, under plausible assumptions, one can choose the unitary $U$ acting on $XY$ such that no efficient measurement on $Y$ can distinguish $\mathcal{E}_Y$ from $\mathcal{D}_Y$ with nonnegligible success probability.

\section{Partially transposed unitary}
\label{appendix:partial_transpose}

Here we bound the operator norm of a partially transposed unitary transformation. Consider a unitary $U$ acting on a bipartite Hilbert space $\mathcal{H}_A \otimes \mathcal{H}_B$, and let $U^{T_B}$ denotes its partial transpose.  Below we provide a tight bound on $\left\|U^{T_B}\right\|$, which will be useful in Appendix \ref{appendix:time_ordering}.

Recall that the operator norm of an operator $O$ acting on $\mathcal{H}_A \otimes \mathcal{H}_B$ is defined as 
\begin{equation}
    \| O \| = \sup_{|\psi\rangle} \sqrt{\langle \psi| O^{\dagger} O|\psi\rangle},
\end{equation}
where the supremum is taken over normalized vectors in $\mathcal{H}_A \otimes \mathcal{H}_B$. The partially transposed unitary $U^{T_B}$ can be diagrammatically expressed as 
\begin{equation}
    U^{T_B} = 
    \begin{tikzpicture}[baseline={([yshift=-.5ex]current bounding box.center)}]
    \draw[] (0,0) -- (2,0) -- (2,1) -- (0,1) -- cycle;
    \node[] () at (1, 0.5) {$U$};
    \draw[] (0.5, 0) -- ++ (0, -0.5);
    \draw[] (1.5, 0) -- ++ (0, -0.25) -- ++ (0.75, 0) -- ++ (0.5, 1.75);
    \draw[] (0.5, 1) -- ++ (0, 0.5);
    \node[draw=white, fill=white, circle, inner sep = 0pt, minimum size=0.2cm] () at (2.45, 0.5) {};
    \draw[] (1.5, 1) -- ++ (0, 0.25) -- ++ (0.75, 0) -- ++ (0.5, -1.75);
    \node[below] () at (0.5, -0.5) {$A$};
    \node[above] () at (0.5, 1.5) {$A$};
    \node[below] () at (2.75, -0.5) {$B$};
    \node[above] () at (2.75, 1.5) {$B$};
    \end{tikzpicture},
\end{equation}
with the input to $U^{T_B}$ on the bottom and the output on the top. The norm of this operator is equal to the norm of the operator 
\begin{equation}
    \overline{U}^{T_B} =  d_B \langle \Phi|_{BB_1}U|\Phi\rangle_{BB_2},
\end{equation}
which maps $AB_1$ to $AB_2$ and can be expressed diagrammatically as
\begin{equation}
    \overline{U}^{T_B} = d_B\,\,\, 
    \begin{tikzpicture}[baseline={([yshift=-.5ex]current bounding box.center)}]
    \draw[] (0,0) -- (2,0) -- (2,1) -- (0,1) -- cycle;
    \node[] () at (1, 0.5) {$U$};
    \draw[] (0.5, 0) -- ++ (0, -0.5);
    \draw[] (1.5, 0) -- ++ (0, -0.5);
    \draw[] (1.5, -0.5) -- ++ (0.5, -0.25) -- ++ (0.5, 0.25);

    \draw[] (2.5, -0.5) -- ++ (0, 0.75) -- ++ (1, 0.5) -- ++ (0, 0.75);
    \draw[] (0.5, 1) -- ++ (0, 0.5);
    \node[draw=white, fill=white, circle, inner sep = 0pt, minimum size=0.25cm] () at (3, 0.5) {};
    \draw[] (1.5, 1) -- ++ (0, 0.5);
    \draw[] (1.5, 1.5) -- ++ (0.5, 0.25) -- ++ (0.5, -0.25);
    \draw[] (2.5, 1.5) -- ++ (0, -0.75) -- ++ (1, -0.5) -- ++ (0, -0.75);

    \node[below] () at (0.5, -0.5) {$A$};
    \node[above] () at (0.5, 1.5) {$A$};
    \node[below] () at (3.5, -0.5) {$B_1$};
    \node[above] () at (3.5, 1.5) {$B_2$};
    \node[below] () at (2.325, -0.75) {$|\Phi\rangle_{BB_2}$};
    \node[above] () at (2.2875, 1.75) {$\langle\Phi|_{BB_1}$};

    \end{tikzpicture}
\end{equation}
Here $\mathcal{H}_{B_1}$ and $\mathcal{H}_{B_2}$ are auxiliary Hilbert spaces such that $\dim (\mathcal{H}_{B_1}) = \dim (\mathcal{H}_{B_2}) = \dim (\mathcal{H}_{B})$, and $|\Phi\rangle_{BB_1} = \frac{1}{\sqrt{d_B}}\sum_{k=0}^{d_B} |k\rangle_{B}|k\rangle_{B_1}$ and $|\Phi\rangle_{BB_2} = \frac{1}{\sqrt{d_B}} \sum_{k=0}^{d_B} |k\rangle_{B}|k\rangle_{B_2}$ are maximally entangled states. The norm squared of $\overline{U}^{T_B}$ is
\begin{equation}
    \begin{aligned}
         \|\overline{U}^{T_B} \|^2 &=  \sup_{|\psi\rangle} \langle \psi|_{AB_1}
         \overline{U}^{T_B\dagger} \overline{U}^{T_B} |\psi\rangle_{AB_1}
         = d_B^2 \sup_{|\psi\rangle} \langle \psi|_{AB_1}
         \left(\langle \Phi|_{BB_2} U^\dagger|\Phi\rangle_{BB_1} \right)
        \left(\langle \Phi|_{BB_1}U|\Phi\rangle_{BB_2}\right) |\psi\rangle_{AB_1}\\
         &=d_B^2\sup_{|\psi\rangle} \left(\langle \psi|_{AB_1} \langle \Phi_{BB_2}|\right)
         \left(U^\dagger |\Phi\rangle_{BB_1}\langle \Phi| U\right)
         \left(|\psi\rangle_{AB_1}|\Phi\rangle_{BB_2}\right)\\
    &\leq d_B^2 \|\left(U^\dagger |\Phi\rangle_{BB_1}\langle \Phi| U\right) \| \\
    &= d_B^2.
    \end{aligned}
\end{equation}
 Since the operator norm of $U^{T_B}$ is equal to that of $\overline{U}^{T_B}$, we thus conclude
\begin{equation}
    \|U^{T_B} \| \leq d_B.\label{eq:partial_transpose_bound}
\end{equation}

Eq.~\eqref{eq:partial_transpose_bound} is in fact tight. Setting the dimensions of $\mathcal{H}_A$ and $\mathcal{H}_B$ to be equal, consider a swap operator between $A$ and $B$, denoted as $\text{SWAP}_{A\leftrightarrow B}$. Its partial transpose becomes 
\begin{equation}
    \text{SWAP}_{A\leftrightarrow B}^{T_B} = d_B |\Phi\rangle_{AB} \langle \Phi|,
\end{equation}
where $|\Phi\rangle_{AB}$ is a maximally entangled state of $AB$. Thus, the operator norm of the partially transposed swap is $d_B$, saturating the bound in Eq.~\eqref{eq:partial_transpose_bound}.

\section{Reversed time ordering}
\label{appendix:time_ordering}

In Section \ref{sec:average} we assumed that the infaller $I$ interacts first with the radiation system $R$ outside the black hole, and then with the system $r$ inside the black hole. This is the natural order of operations enforced by the black hole's causal structure. Nevertheless, it is of interest to see whether the conclusion would be different if this ordering were reversed. 

In this appendix, we consider the hypothetical setup in which the infaller interacts with the interior mode prior to interacting with the exterior mode. In this scenario, the black hole $S$-matrix becomes 
\begin{equation}
\begin{tikzpicture}[baseline={([yshift=-.5ex]current bounding box.center)}]
    \draw[] (0,0) -- (0,1);
    \draw[] (2,0) -- (2,1);
    \draw[] (4,0) -- (4,1);
    \draw[] (6,0) -- (6,1);
    \draw[] (4,0) -- (5, -0.5) -- (6,0);
    \draw[] (0, 1.5) -- (0, 2.5);
    \draw[] (2, 1.5) -- (2, 2.5);
    \draw[] (2, 2.5) -- (3, 3) -- (4, 2.5) -- (4,1);
    \draw[] (3,0) -- (3,1);
    
    \draw[fill=red!10!white] (-0.25, 1) -- (3.25, 1) -- (3.25, 1.75) -- (-0.25, 1.75) -- cycle;
    
     \node[fill=blue, regular polygon, regular polygon sides=3, minimum size=0.35cm, inner sep=0pt] (u1) at (3, 0.7) {};
    \node[fill=blue, regular polygon, regular polygon sides=3, minimum size=0.35cm, inner sep=0pt] (u2) at (6, 0.7) {};
    \node[fill=red, regular polygon, regular polygon sides=3, minimum size=0.35cm, inner sep=0pt] (v1) at (3, 0.25) {};
    \node[fill=red, regular polygon, regular polygon sides=3, minimum size=0.35cm, inner sep=0pt] (v2) at (4, 0.25) {};
    
    \node[right] at (u2) {$u$};
    \node[right] at (v2) {$v$};
    
    \draw[blue, thick] (u1) -- (u2);
    \draw[red, thick] (v1) -- (v2);

    \node[above] at (0, 2.5) {$B$};
    \node[above] at (3, 3) {$\sqrt{|r|} \langle \text{MAX}|$};
    \node[] at (1.5,1.375) {$U'$};
    \node[below] at (0,0) {$\ell$};
    \node[below] at (2,0) {$|\psi_0\rangle_f$};
    \node[below] at (3,0) {$I$};
    \node[below] at (4,0) {$r$};
    \node[below] at (6,0) {$R$};
    \node[below] at (5,-0.5) {$|\text{MAX}\rangle$};
    
    \node[] at (7, 1.25) {$=$};
    
    \begin{scope}[xshift=8cm]
    \draw[] (0,0) -- (0,1);
    \draw[] (2,0) -- (2,1);
    \draw[] (0, 1.5) -- (0, 2.75);
    \draw[] (2, 1.5) -- (2, 2.75);
    \draw[] (3,0) -- (3,1);
    
    \draw[fill=red!10!white] (-0.25, 1) -- (3.25, 1) -- (3.25, 1.75) -- (-0.25, 1.75) -- cycle;
    
     \node[fill=blue, regular polygon, regular polygon sides=3, minimum size=0.35cm, inner sep=0pt] (u1) at (3, 0.7) {};
    \node[fill=blue, regular polygon, regular polygon sides=3, minimum size=0.35cm, inner sep=0pt] (u2) at (2, 2.5) {};
    \node[fill=red, regular polygon, regular polygon sides=3, minimum size=0.35cm, inner sep=0pt] (v1) at (3, 0.25) {};
    \node[fill=red, regular polygon, regular polygon sides=3, minimum size=0.35cm, inner sep=0pt, rotate=180] (v2) at (2, 2) {};
    
    \node[left] at (u2) {$u$};
    \node[left] at (v2) {$v$};
    
    \draw[blue, thick] (u1) -- ++ (0.75, 0) -- ++ (0,1.8) -- (u2);
    \draw[red, thick] (v1) -- ++ (0.5, 0) -- ++ (0, 1.75) --  (v2);
    
    \node[above] at (2, 2.75) {$R$};
    \node[above] at (0, 2.75) {$B$};
    \node[] at (1.5,1.375) {$U'$};
    \node[below] at (0,0) {$\ell$};
    \node[below] at (2,0) {$|\psi_0\rangle_f$};
    \node[below] at (3,0) {$I$};
    \end{scope}.
    \end{tikzpicture}.
    \label{eq:straightening2}
\end{equation}
When we compute $V(U')^\dagger V(U')$ averaged over $U'$, we get
\begin{equation}
\begin{aligned}
\int dU'
    \left(
    \begin{tikzpicture}[scale=0.8, baseline={([yshift=-.5ex]current bounding box.center)}]
    \draw[] (0,0) -- (0,1);
    \draw[] (2,0) -- (2,1);
    \draw[] (0, 1.5) -- (0, 2.75);
    \draw[] (2, 1.5) -- (2, 2.75);
    \draw[] (3,0) -- (3,1);
    
    \draw[fill=red!10!white] (-0.25, 1) -- (3.25, 1) -- (3.25, 1.75) -- (-0.25, 1.75) -- cycle;
    
    \node[fill=blue, regular polygon, regular polygon sides=3, minimum size=0.35cm, inner sep=0pt] (u1) at (3, 0.7) {};
    \node[fill=blue, regular polygon, regular polygon sides=3, minimum size=0.35cm, inner sep=0pt] (u2) at (2, 2.5) {};
    \node[fill=red, regular polygon, regular polygon sides=3, minimum size=0.35cm, inner sep=0pt] (v1) at (3, 0.25) {};
    \node[fill=red, regular polygon, regular polygon sides=3, minimum size=0.35cm, inner sep=0pt, rotate=180] (v2) at (2, 2) {};
    
    \node[left] at (u2) {$u\phantom{l}$};
    \node[left] at (v2) {$v\phantom{l}$};
    
    \draw[blue, thick] (u1) -- ++ (0.75, 0) -- ++ (0,1.80) -- (u2);
    \draw[red, thick] (v1) -- ++ (0.5, 0) -- ++ (0, 1.75) --  (v2);
    
    \node[] at (1.5,1.375) {$U'$};
    \node[below] at (0,0) {$\ell$};
    \node[below] at (2,0) {$|\psi_0\rangle_f$};
    \node[below] at (3,0) {$I$};
    \begin{scope}[yscale=-1, yshift=-5.5cm]
    
    \draw[] (0,0) -- (0,1);
    \draw[] (2,0) -- (2,1);
    \draw[] (0, 1.5) -- (0, 2.75);
    \draw[] (2, 1.5) -- (2, 2.75);
    \draw[] (3,0) -- (3,1);
    
    \draw[fill=red!10!white] (-0.25, 1) -- (3.25, 1) -- (3.25, 1.75) -- (-0.25, 1.75) -- cycle;
    
    \node[fill=blue, regular polygon, regular polygon sides=3, minimum size=0.35cm, inner sep=0pt, rotate=180] (u1) at (3, 0.7) {};
    \node[fill=blue, regular polygon, regular polygon sides=3, minimum size=0.35cm, inner sep=0pt, rotate=180] (u2) at (2, 2.5) {};
    \node[fill=red, regular polygon, regular polygon sides=3, minimum size=0.35cm, inner sep=0pt, rotate=180] (v1) at (3, 0.25) {};
    \node[fill=red, regular polygon, regular polygon sides=3, minimum size=0.35cm, inner sep=0pt] (v2) at (2, 2) {};
    
    \node[left] at (u2) {$u^{\dagger}$};
    \node[left] at (v2) {$v^{\dagger}$};
    
    \draw[blue, thick] (u1) -- ++ (0.75, 0) -- ++ (0,1.8) -- (u2);
    \draw[red, thick] (v1) -- ++ (0.5, 0) -- ++ (0, 1.75) --  (v2);
    
    \node[] at (1.5,1.375) {${U'}^{\dagger}$};
    \node[above] at (0,0) {$\ell$};
    \node[above] at (2,0) {$|\psi_0\rangle_f$};
    \node[above] at (3,0) {$I$};
    \end{scope}
    \end{tikzpicture}
    \right)
    &= \frac{1}{|\ell| |I||f|}
    \begin{tikzpicture}[scale=0.8, baseline={([yshift=-.5ex]current bounding box.center)}]
    \draw[] (0,0) -- (0,1);
    \draw[] (2,0) -- (2,1);
    \draw[] (0, 1.5) -- (0, 2.75);
    \draw[] (2, 1.5) -- (2, 2.75);
    \draw[] (3,0) -- (3,1);

     \node[fill=blue, regular polygon, regular polygon sides=3, minimum size=0.35cm, inner sep=0pt] (u1) at (3, 0.7) {};
    \node[fill=blue, regular polygon, regular polygon sides=3, minimum size=0.35cm, inner sep=0pt] (u2) at (2, 2.5) {};
    \node[fill=red, regular polygon, regular polygon sides=3, minimum size=0.35cm, inner sep=0pt] (v1) at (3, 0.25) {};
    \node[fill=red, regular polygon, regular polygon sides=3, minimum size=0.35cm, inner sep=0pt, rotate=180] (v2) at (2, 2) {};
    
    \node[left] at (u2) {$u\phantom{l}$};
    \node[left] at (v2) {$v\phantom{l}$};
    
    \draw[blue, thick] (u1) -- ++ (0.75, 0) -- ++ (0,1.8) -- (u2);
    \draw[red, thick] (v1) -- ++ (0.5, 0) -- ++ (0, 1.75) --  (v2);
    
    \node[below] at (0,0) {$\ell$};
    \node[below] at (2,0) {$|\psi_0\rangle_f$};
    \node[below] at (3,0) {$I$};
    \begin{scope}[yscale=-1, yshift=-5.5cm]
    
    \draw[] (0,0) -- (0,1);
    \draw[] (2,0) -- (2,1);
    \draw[] (0, 1.5) -- (0, 2.75);
    \draw[] (2, 1.5) -- (2, 2.75);
    \draw[] (3,0) -- (3,1);
    
    \node[fill=blue, regular polygon, regular polygon sides=3, minimum size=0.35cm, inner sep=0pt, rotate=180] (u1) at (3, 0.7) {};
    \node[fill=blue, regular polygon, regular polygon sides=3, minimum size=0.35cm, inner sep=0pt, rotate=180] (u2) at (2, 2.5) {};
    \node[fill=red, regular polygon, regular polygon sides=3, minimum size=0.35cm, inner sep=0pt, rotate=180] (v1) at (3, 0.25) {};
    \node[fill=red, regular polygon, regular polygon sides=3, minimum size=0.35cm, inner sep=0pt] (v2) at (2, 2) {};
    
    \node[left] at (u2) {$u^{\dagger}$};
    \node[left] at (v2) {$v^{\dagger}$};
    
    \draw[blue, thick] (u1) -- ++ (0.75, 0) -- ++ (0,1.8) -- (u2);
    \draw[red, thick] (v1) -- ++ (0.5, 0) -- ++ (0, 1.75) --  (v2);
    
    \node[above] at (0,0) {$\ell$};
    \node[above] at (2,0) {$|\psi_0\rangle_f$};
    \node[above] at (3,0) {$I$};
    \end{scope}
    
    \draw[] (0, 1) --++ (-0.25, 0.25) --++ (0, 3) --++ (0.25, 0.25);    \draw[] (0, 1.5) --++ (0.5, 0) --++ (0, 2.5) --++ (-0.5, 0);
    
    \draw[] (2,1) --++ (-1.25, 0.25) --++ (0, 3) --++ (1.25, 0.25);
    \draw[] (2, 1.5) --++ (-0.75, 0) --++ (0, 2.5) --++ (0.75, 0);
    
    \draw[] (3,1) --++ (0, 3.5);
    \end{tikzpicture} \\
    &= 
    \frac{|B|}{|\ell| |I| |f|}
        \begin{tikzpicture}[scale=0.8, baseline={([yshift=-.5ex]current bounding box.center)}]
    \draw[] (2, 1.5) --++ (-0.75, 0) --++ (0, 2.5) --++ (0.75, 0) -- cycle;
    \draw[] (0,0) -- (0,1);
    \draw[] (0, 1.5) -- (0, 2.75);
    \draw[] (3,0) -- (3,1);

     \node[fill=blue, regular polygon, regular polygon sides=3, minimum size=0.35cm, inner sep=0pt] (u1) at (3, 0.7) {};
    \node[fill=blue, regular polygon, regular polygon sides=3, minimum size=0.35cm, inner sep=0pt] (u2) at (2, 2.5) {};
    \node[fill=red, regular polygon, regular polygon sides=3, minimum size=0.35cm, inner sep=0pt] (v1) at (3, 0.25) {};
    \node[fill=red, regular polygon, regular polygon sides=3, minimum size=0.35cm, inner sep=0pt, rotate=180] (v2) at (2, 2) {};
    
    \node[left] at (u2) {$u\phantom{l}$};
    \node[left] at (v2) {$v\phantom{l}$};
    
    \draw[blue, thick] (u1) -- ++ (0.75, 0) -- ++ (0,1.8) -- (u2);
    \draw[red, thick] (v1) -- ++ (0.5, 0) -- ++ (0, 1.75) --  (v2);
    
    \node[below] at (0,0) {$\ell$};
    \node[below] at (3,0) {$I$};
    \begin{scope}[yscale=-1, yshift=-5.5cm]
    
    \draw[] (0,0) -- (0,1);
    \draw[] (0, 1.5) -- (0, 2.75);
    \draw[] (3,0) -- (3,1);
    
    \node[fill=blue, regular polygon, regular polygon sides=3, minimum size=0.35cm, inner sep=0pt, rotate=180] (u1) at (3, 0.7) {};
    \node[fill=blue, regular polygon, regular polygon sides=3, minimum size=0.35cm, inner sep=0pt, rotate=180] (u2) at (2, 2.5) {};
    \node[fill=red, regular polygon, regular polygon sides=3, minimum size=0.35cm, inner sep=0pt, rotate=180] (v1) at (3, 0.25) {};
    \node[fill=red, regular polygon, regular polygon sides=3, minimum size=0.35cm, inner sep=0pt] (v2) at (2, 2) {};
    
    \node[left] at (u2) {$u^{\dagger}$};
    \node[left] at (v2) {$v^{\dagger}$};
    
    \draw[blue, thick] (u1) -- ++ (0.75, 0) -- ++ (0,1.8) -- (u2);
    \draw[red, thick] (v1) -- ++ (0.5, 0) -- ++ (0, 1.75) --  (v2);
    
    \node[above] at (0,0) {$\ell$};
    \node[above] at (3,0) {$I$};
    \end{scope}
    
    \draw[] (0, 1) --++ (0,3.5);   
    
    \draw[] (3,1) --++ (0, 3.5);
    \end{tikzpicture}\\
    &= I_{\ell} \otimes I_I,
\end{aligned}
\end{equation}

We can also compute the fluctuation:
\begin{equation}
    \int dU'
    \left(
    \begin{tikzpicture}[scale=0.8, baseline={([yshift=-.5ex]current bounding box.center)}]
    \draw[] (0,0) -- (0,1);
    \draw[] (2,0) -- (2,1);
    \draw[] (0, 1.5) -- (0, 2.75);
    \draw[] (2, 1.5) -- (2, 2.75);
    \draw[] (3,0) -- (3,1);
    
    \draw[fill=red!10!white] (-0.25, 1) -- (3.25, 1) -- (3.25, 1.75) -- (-0.25, 1.75) -- cycle;
    
    \node[fill=blue, regular polygon, regular polygon sides=3, minimum size=0.35cm, inner sep=0pt] (u1) at (3, 0.7) {};
    \node[fill=blue, regular polygon, regular polygon sides=3, minimum size=0.35cm, inner sep=0pt] (u2) at (2, 2.5) {};
    \node[fill=red, regular polygon, regular polygon sides=3, minimum size=0.35cm, inner sep=0pt] (v1) at (3, 0.25) {};
    \node[fill=red, regular polygon, regular polygon sides=3, minimum size=0.35cm, inner sep=0pt, rotate=180] (v2) at (2, 2) {};
    
    \node[left] at (u2) {$u\phantom{l}$};
    \node[left] at (v2) {$v\phantom{l}$};
    
    \draw[blue, thick] (u1) -- ++ (0.75, 0) -- ++ (0,1.8) -- (u2);
    \draw[red, thick] (v1) -- ++ (0.5, 0) -- ++ (0, 1.75) --  (v2);
    
    \node[] at (1.5,1.375) {$U'$};
    \node[below] at (0,0) {$\ell$};
    \node[below] at (2,0) {$|\psi_0\rangle_f$};
    \node[below] at (3,0) {$I$};
    \begin{scope}[yscale=-1, yshift=-5.5cm]
    
    \draw[] (0,0) -- (0,1);
    \draw[] (2,0) -- (2,1);
    \draw[] (0, 1.5) -- (0, 2.75);
    \draw[] (2, 1.5) -- (2, 2.75);
    \draw[] (3,0) -- (3,1);
    
    \draw[fill=red!10!white] (-0.25, 1) -- (3.25, 1) -- (3.25, 1.75) -- (-0.25, 1.75) -- cycle;
    
    \node[fill=blue, regular polygon, regular polygon sides=3, minimum size=0.35cm, inner sep=0pt, rotate=180] (u1) at (3, 0.7) {};
    \node[fill=blue, regular polygon, regular polygon sides=3, minimum size=0.35cm, inner sep=0pt, rotate=180] (u2) at (2, 2.5) {};
    \node[fill=red, regular polygon, regular polygon sides=3, minimum size=0.35cm, inner sep=0pt, rotate=180] (v1) at (3, 0.25) {};
    \node[fill=red, regular polygon, regular polygon sides=3, minimum size=0.35cm, inner sep=0pt] (v2) at (2, 2) {};
    
    \node[left] at (u2) {$u^{\dagger}$};
    \node[left] at (v2) {$v^{\dagger}$};
    
    \draw[blue, thick] (u1) -- ++ (0.75, 0) -- ++ (0,1.8) -- (u2);
    \draw[red, thick] (v1) -- ++ (0.5, 0) -- ++ (0, 1.75) --  (v2);
    
    \node[] at (1.5,1.375) {${U'}^{\dagger}$};
    \node[above] at (0,0) {$\ell$};
    \node[above] at (2,0) {$|\psi_0\rangle_f$};
    \node[above] at (3,0) {$I$};
    \end{scope}
    \end{tikzpicture}
    \right)^{\otimes 2} = \left(1- \frac{1}{(|\ell| |I| |f|)^2}\right)^{-1} (I_{\ell}\otimes I_R)^{\otimes 2} + \Delta',
\end{equation}
where
\begin{equation}
\label{eq:deltaprime}
    \Delta' = -\frac{|B|}{d(d^2-1)}\text{SWAP}_{\ell \leftrightarrow \ell'}\left(
    \begin{tikzpicture}[scale=0.8, baseline={([yshift=-.5ex]current bounding box.center)}]
    \draw[] (2, 1.5) -- (2, 2.75);
    \draw[] (3,0) -- (3,1);

    \node[fill=blue, regular polygon, regular polygon sides=3, minimum size=0.35cm, inner sep=0pt] (u1) at (3, 0.7) {};
    \node[fill=blue, regular polygon, regular polygon sides=3, minimum size=0.35cm, inner sep=0pt] (u2) at (2, 2.5) {};
    \node[fill=red, regular polygon, regular polygon sides=3, minimum size=0.35cm, inner sep=0pt] (v1) at (3, 0.25) {};
    \node[fill=red, regular polygon, regular polygon sides=3, minimum size=0.35cm, inner sep=0pt, rotate=180] (v2) at (2, 2) {};
    
    \node[left] at (u2) {$u\phantom{l}$};
    \node[left] at (v2) {$v\phantom{l}$};
    
    \node[circle, fill=black, inner sep=0pt, minimum size =0.2cm] (vertex1) at (3,1) {};
    \node[circle, fill=black, inner sep=0pt, minimum size =0.2cm] (vertex2) at (2,1.5) {};
    \node[above] at (vertex1) {$4$};
    \node[below] at (vertex2) {$3$};
    
    \draw[blue, thick] (u1) -- ++ (0.75, 0) -- ++ (0,1.8) -- (u2);
    \draw[red, thick] (v1) -- ++ (0.5, 0) -- ++ (0, 1.75) --  (v2);
    
    \node[below] at (3,0) {$I$};
    \begin{scope}[yscale=-1, yshift=-5.5cm]
    \draw[] (2, 1.5) -- (2, 2.75);
    \draw[] (3,0) -- (3,1);

    \node[fill=blue, regular polygon, regular polygon sides=3, minimum size=0.35cm, inner sep=0pt, rotate=180] (u1) at (3, 0.7) {};
    \node[fill=blue, regular polygon, regular polygon sides=3, minimum size=0.35cm, inner sep=0pt, rotate=180] (u2) at (2, 2.5) {};
    \node[fill=red, regular polygon, regular polygon sides=3, minimum size=0.35cm, inner sep=0pt, rotate=180] (v1) at (3, 0.25) {};
    \node[fill=red, regular polygon, regular polygon sides=3, minimum size=0.35cm, inner sep=0pt] (v2) at (2, 2) {};
    
    \node[left] at (u2) {$u^{\dagger}$};
    \node[left] at (v2) {$v^{\dagger}$};
    
    \node[circle, fill=black, inner sep=0pt, minimum size =0.2cm] (vertex1) at (3,1) {};
    \node[circle, fill=black, inner sep=0pt, minimum size =0.2cm] (vertex2) at (2,1.5) {};
    \node[below] at (vertex1) {$1$};
    \node[above] at (vertex2) {$2$};
    
    \draw[blue, thick] (u1) -- ++ (0.75, 0) -- ++ (0,1.8) -- (u2);
    \draw[red, thick] (v1) -- ++ (0.5, 0) -- ++ (0, 1.75) --  (v2);
    
    \node[above] at (3,0) {$I$};
    \end{scope}
    \begin{scope}[xshift=4cm]
    \draw[] (2, 1.5) -- (2, 2.75);
    \draw[] (3,0) -- (3,1);
  
     \node[fill=blue, regular polygon, regular polygon sides=3, minimum size=0.35cm, inner sep=0pt] (u1) at (3, 0.7) {};
    \node[fill=blue, regular polygon, regular polygon sides=3, minimum size=0.35cm, inner sep=0pt] (u2) at (2, 2.5) {};
    \node[fill=red, regular polygon, regular polygon sides=3, minimum size=0.35cm, inner sep=0pt] (v1) at (3, 0.25) {};
    \node[fill=red, regular polygon, regular polygon sides=3, minimum size=0.35cm, inner sep=0pt, rotate=180] (v2) at (2, 2) {};
    
    \node[left] at (u2) {$u\phantom{l}$};
    \node[left] at (v2) {$v\phantom{l}$};
    
    \node[circle, fill=black, inner sep=0pt, minimum size =0.2cm] (vertex1) at (3,1) {};
    \node[circle, fill=black, inner sep=0pt, minimum size =0.2cm] (vertex2) at (2,1.5) {};
    \node[above] at (vertex1) {$1$};
    \node[below] at (vertex2) {$2$};

    \draw[blue, thick] (u1) -- ++ (0.75, 0) -- ++ (0,1.8) -- (u2);
    \draw[red, thick] (v1) -- ++ (0.5, 0) -- ++ (0, 1.75) --  (v2);
    
    \node[below] at (3,0) {$I'$};
    \begin{scope}[yscale=-1, yshift=-5.5cm]
    
    \draw[] (2, 1.5) -- (2, 2.75);
    \draw[] (3,0) -- (3,1);

    \node[fill=blue, regular polygon, regular polygon sides=3, minimum size=0.35cm, inner sep=0pt, rotate=180] (u1) at (3, 0.7) {};
    \node[fill=blue, regular polygon, regular polygon sides=3, minimum size=0.35cm, inner sep=0pt, rotate=180] (u2) at (2, 2.5) {};
    \node[fill=red, regular polygon, regular polygon sides=3, minimum size=0.35cm, inner sep=0pt, rotate=180] (v1) at (3, 0.25) {};
    \node[fill=red, regular polygon, regular polygon sides=3, minimum size=0.35cm, inner sep=0pt] (v2) at (2, 2) {};
    
    \node[left] at (u2) {$u^{\dagger}$};
    \node[left] at (v2) {$v^{\dagger}$};
    
    \node[circle, fill=black, inner sep=0pt, minimum size =0.2cm] (vertex1) at (3,1) {};
    \node[circle, fill=black, inner sep=0pt, minimum size =0.2cm] (vertex2) at (2,1.5) {};
    \node[below] at (vertex1) {$4$};
    \node[above] at (vertex2) {$3$};

    \draw[blue, thick] (u1) -- ++ (0.75, 0) -- ++ (0,1.8) -- (u2);
    \draw[red, thick] (v1) -- ++ (0.5, 0) -- ++ (0, 1.75) --  (v2);
    
    \node[above] at (3,0) {$I'$};
    \end{scope}
    \end{scope}
    \end{tikzpicture}  
    \right).
\end{equation}

Similar to the discussion in Section~\ref{sec:average}, we can first consider product states $|\varphi\rangle = |\varphi_I\rangle \otimes |\varphi_{\ell}\rangle$ and $|\varphi'\rangle = |\varphi_I'\rangle \otimes |\varphi_{\ell}'\rangle$ over $\ell$ and $I$ and later extend the bound to general entangled states of $I\ell$. Using the definition of $\Delta'$ in Eq.~\eqref{eq:deltaprime}, we get
\begin{equation}
\begin{aligned}
    (\langle \varphi|\otimes \langle \varphi'|) \Delta' (|\varphi'\rangle \otimes |\varphi\rangle ) &= -\frac{|B|}{d(d^2-1)}
    \left(
    \begin{tikzpicture}[scale=0.8, baseline={([yshift=-.5ex]current bounding box.center)}]
    \draw[] (2, 1.5) -- (2, 2.75);
    \draw[] (3,0) -- (3,1);

    \node[fill=blue, regular polygon, regular polygon sides=3, minimum size=0.35cm, inner sep=0pt] (u1) at (3, 0.7) {};
    \node[fill=blue, regular polygon, regular polygon sides=3, minimum size=0.35cm, inner sep=0pt] (u2) at (2, 2.5) {};
    \node[fill=red, regular polygon, regular polygon sides=3, minimum size=0.35cm, inner sep=0pt] (v1) at (3, 0.25) {};
    \node[fill=red, regular polygon, regular polygon sides=3, minimum size=0.35cm, inner sep=0pt, rotate=180] (v2) at (2, 2) {};
    
    \node[left] at (u2) {$u\phantom{l}$};
    \node[left] at (v2) {$v\phantom{l}$};
    
    \node[circle, fill=black, inner sep=0pt, minimum size =0.2cm] (vertex1) at (3,1) {};
    \node[circle, fill=black, inner sep=0pt, minimum size =0.2cm] (vertex2) at (2,1.5) {};
    \node[above] at (vertex1) {$4$};
    \node[below] at (vertex2) {$3$};
    
    \draw[blue, thick] (u1) -- ++ (0.75, 0) -- ++ (0,1.8) -- (u2);
    \draw[red, thick] (v1) -- ++ (0.5, 0) -- ++ (0, 1.75) --  (v2);
    
    \node[below] at (3,0) {$|\varphi_I'\rangle$};
    \begin{scope}[yscale=-1, yshift=-5.5cm]
    \draw[] (2, 1.5) -- (2, 2.75);
    \draw[] (3,0) -- (3,1);

    \node[fill=blue, regular polygon, regular polygon sides=3, minimum size=0.35cm, inner sep=0pt, rotate=180] (u1) at (3, 0.7) {};
    \node[fill=blue, regular polygon, regular polygon sides=3, minimum size=0.35cm, inner sep=0pt, rotate=180] (u2) at (2, 2.5) {};
    \node[fill=red, regular polygon, regular polygon sides=3, minimum size=0.35cm, inner sep=0pt, rotate=180] (v1) at (3, 0.25) {};
    \node[fill=red, regular polygon, regular polygon sides=3, minimum size=0.35cm, inner sep=0pt] (v2) at (2, 2) {};
    
    \node[left] at (u2) {$u^{\dagger}$};
    \node[left] at (v2) {$v^{\dagger}$};
    
    \node[circle, fill=black, inner sep=0pt, minimum size =0.2cm] (vertex1) at (3,1) {};
    \node[circle, fill=black, inner sep=0pt, minimum size =0.2cm] (vertex2) at (2,1.5) {};
    \node[below] at (vertex1) {$1$};
    \node[above] at (vertex2) {$2$};
    
    \draw[blue, thick] (u1) -- ++ (0.75, 0) -- ++ (0,1.8) -- (u2);
    \draw[red, thick] (v1) -- ++ (0.5, 0) -- ++ (0, 1.75) --  (v2);
    
    \node[above] at (3,0) {$\langle \varphi_I|$};
    \end{scope}
    \begin{scope}[xshift=4cm]
    \draw[] (2, 1.5) -- (2, 2.75);
    \draw[] (3,0) -- (3,1);
  
     \node[fill=blue, regular polygon, regular polygon sides=3, minimum size=0.35cm, inner sep=0pt] (u1) at (3, 0.7) {};
    \node[fill=blue, regular polygon, regular polygon sides=3, minimum size=0.35cm, inner sep=0pt] (u2) at (2, 2.5) {};
    \node[fill=red, regular polygon, regular polygon sides=3, minimum size=0.35cm, inner sep=0pt] (v1) at (3, 0.25) {};
    \node[fill=red, regular polygon, regular polygon sides=3, minimum size=0.35cm, inner sep=0pt, rotate=180] (v2) at (2, 2) {};
    
    \node[left] at (u2) {$u\phantom{l}$};
    \node[left] at (v2) {$v\phantom{l}$};
    
    \node[circle, fill=black, inner sep=0pt, minimum size =0.2cm] (vertex1) at (3,1) {};
    \node[circle, fill=black, inner sep=0pt, minimum size =0.2cm] (vertex2) at (2,1.5) {};
    \node[above] at (vertex1) {$1$};
    \node[below] at (vertex2) {$2$};

    \draw[blue, thick] (u1) -- ++ (0.75, 0) -- ++ (0,1.8) -- (u2);
    \draw[red, thick] (v1) -- ++ (0.5, 0) -- ++ (0, 1.75) --  (v2);
    
    \node[below] at (3,0) {$|\varphi_I\rangle$};
    \begin{scope}[yscale=-1, yshift=-5.5cm]
    
    \draw[] (2, 1.5) -- (2, 2.75);
    \draw[] (3,0) -- (3,1);

    \node[fill=blue, regular polygon, regular polygon sides=3, minimum size=0.35cm, inner sep=0pt, rotate=180] (u1) at (3, 0.7) {};
    \node[fill=blue, regular polygon, regular polygon sides=3, minimum size=0.35cm, inner sep=0pt, rotate=180] (u2) at (2, 2.5) {};
    \node[fill=red, regular polygon, regular polygon sides=3, minimum size=0.35cm, inner sep=0pt, rotate=180] (v1) at (3, 0.25) {};
    \node[fill=red, regular polygon, regular polygon sides=3, minimum size=0.35cm, inner sep=0pt] (v2) at (2, 2) {};
    
    \node[left] at (u2) {$u^{\dagger}$};
    \node[left] at (v2) {$v^{\dagger}$};
    
    \node[circle, fill=black, inner sep=0pt, minimum size =0.2cm] (vertex1) at (3,1) {};
    \node[circle, fill=black, inner sep=0pt, minimum size =0.2cm] (vertex2) at (2,1.5) {};
    \node[below] at (vertex1) {$4$};
    \node[above] at (vertex2) {$3$};

    \draw[blue, thick] (u1) -- ++ (0.75, 0) -- ++ (0,1.8) -- (u2);
    \draw[red, thick] (v1) -- ++ (0.5, 0) -- ++ (0, 1.75) --  (v2);
    
    \node[above] at (3,0) {$\langle \varphi_I'|$};
    \end{scope}
    \end{scope}
    \end{tikzpicture}  
    \right) \\
    &= -\frac{|B|}{d(d^2-1)}
    \left(
    \begin{tikzpicture}[scale=0.8, baseline={([yshift=-.5ex]current bounding box.center)}]
    \draw[] (2, 1.5) -- (2, 3.75);
    \draw[] (3, 1.5) -- (3, 2.75);

    \node[fill=blue, regular polygon, regular polygon sides=3, minimum size=0.35cm, inner sep=0pt, rotate=180] (u1) at (3, 2) {};
    \node[fill=blue, regular polygon, regular polygon sides=3, minimum size=0.35cm, inner sep=0pt] (u2) at (2, 2.5) {};
    \node[fill=red, regular polygon, regular polygon sides=3, minimum size=0.35cm, inner sep=0pt, rotate=180] (v1) at (3, 2.5) {};
    \node[fill=red, regular polygon, regular polygon sides=3, minimum size=0.35cm, inner sep=0pt, rotate=180] (v2) at (2, 2) {};
    
    \node[left] at (u2) {$u\phantom{l}$};
    \node[left] at (v2) {$v\phantom{l}$};
    
    \node[circle, fill=black, inner sep=0pt, minimum size =0.2cm] (vertex1) at (3,1.5) {};
    \node[circle, fill=black, inner sep=0pt, minimum size =0.2cm] (vertex2) at (2,1.5) {};
    \node[below] at (vertex1) {$4$};
    \node[below] at (vertex2) {$3$};
    
    \draw[blue, thick] (u1) -- (u2);
    \draw[red, thick] (v1) --  (v2);
    
    \node[above] at (3,2.75) {$\langle (\varphi_I')^*|$};
    \begin{scope}[yscale=-1, yshift=-7.5cm]
    \draw[] (2, 1.5) -- (2, 3.75);
    \draw[] (3,1.5) -- (3,2.75);

    \node[fill=blue, regular polygon, regular polygon sides=3, minimum size=0.35cm, inner sep=0pt] (u1) at (3, 2) {};
    \node[fill=blue, regular polygon, regular polygon sides=3, minimum size=0.35cm, inner sep=0pt, rotate=180] (u2) at (2, 2.5) {};
    \node[fill=red, regular polygon, regular polygon sides=3, minimum size=0.35cm, inner sep=0pt] (v1) at (3, 2.5) {};
    \node[fill=red, regular polygon, regular polygon sides=3, minimum size=0.35cm, inner sep=0pt] (v2) at (2, 2) {};
    
    \node[left] at (u2) {$u^{\dagger}$};
    \node[left] at (v2) {$v^{\dagger}$};
    
    \node[circle, fill=black, inner sep=0pt, minimum size =0.2cm] (vertex1) at (3,1.5) {};
    \node[circle, fill=black, inner sep=0pt, minimum size =0.2cm] (vertex2) at (2,1.5) {};
    \node[above] at (vertex1) {$1$};
    \node[above] at (vertex2) {$2$};
    
    \draw[blue, thick] (u1)  -- (u2);
    \draw[red, thick] (v1) --  (v2);
    
    \node[below] at (3,2.75) {$|\varphi_I^*\rangle$};
    \end{scope}
    \begin{scope}[xshift=4cm]
    \draw[] (2, 1.5) -- (2, 3.75);
    \draw[] (3,1.5) -- (3,2.75);
  
     \node[fill=blue, regular polygon, regular polygon sides=3, minimum size=0.35cm, inner sep=0pt, rotate=180] (u1) at (3, 2) {};
    \node[fill=blue, regular polygon, regular polygon sides=3, minimum size=0.35cm, inner sep=0pt] (u2) at (2, 2.5) {};
    \node[fill=red, regular polygon, regular polygon sides=3, minimum size=0.35cm, inner sep=0pt, rotate=180] (v1) at (3, 2.5) {};
    \node[fill=red, regular polygon, regular polygon sides=3, minimum size=0.35cm, inner sep=0pt, rotate=180] (v2) at (2, 2) {};
    
    \node[left] at (u2) {$u\phantom{l}$};
    \node[left] at (v2) {$v\phantom{l}$};
    
    \node[circle, fill=black, inner sep=0pt, minimum size =0.2cm] (vertex1) at (3,1.5) {};
    \node[circle, fill=black, inner sep=0pt, minimum size =0.2cm] (vertex2) at (2,1.5) {};
    \node[below] at (vertex1) {$1$};
    \node[below] at (vertex2) {$2$};

    \draw[blue, thick] (u1)  -- (u2);
    \draw[red, thick] (v1)  --  (v2);
    
    \node[above] at (3,2.75) {$\langle \varphi_I^*|$};
    \begin{scope}[yscale=-1, yshift=-7.5cm]
    
    \draw[] (2, 1.5) -- (2, 3.75);
    \draw[] (3, 1.5) -- (3, 2.75);

    \node[fill=blue, regular polygon, regular polygon sides=3, minimum size=0.35cm, inner sep=0pt] (u1) at (3, 2) {};
    \node[fill=blue, regular polygon, regular polygon sides=3, minimum size=0.35cm, inner sep=0pt, rotate=180] (u2) at (2, 2.5) {};
    \node[fill=red, regular polygon, regular polygon sides=3, minimum size=0.35cm, inner sep=0pt] (v1) at (3, 2.5) {};
    \node[fill=red, regular polygon, regular polygon sides=3, minimum size=0.35cm, inner sep=0pt] (v2) at (2, 2) {};
    
    \node[left] at (u2) {$u^{\dagger}$};
    \node[left] at (v2) {$v^{\dagger}$};
    
    \node[circle, fill=black, inner sep=0pt, minimum size =0.2cm] (vertex1) at (3,1.5) {};
    \node[circle, fill=black, inner sep=0pt, minimum size =0.2cm] (vertex2) at (2,1.5) {};
    \node[above] at (vertex1) {$4$};
    \node[above] at (vertex2) {$3$};

    \draw[blue, thick] (u1)  -- (u2);
    \draw[red, thick] (v1) --  (v2);
    
    \node[below] at (3,2.75) {$|(\varphi_I')^*\rangle$};
    \end{scope}
    \end{scope}
    \end{tikzpicture}  
    \right) \\
    &= -\frac{|B|}{d(d^2-1)}
    \left(
    \begin{tikzpicture}[scale=0.8, baseline={([yshift=-.5ex]current bounding box.center)}]
    \draw[] (2, 1.5) -- (2, 3.75);
    \draw[] (3, 1.5) -- (3, 2.75);
    \draw[] (3, 2.75) -- ++ (0.5, 0.25) -- ++ (0.5, -0.25) -- (4, 1.5);
    
    \node[fill=blue, regular polygon, regular polygon sides=3, minimum size=0.35cm, inner sep=0pt] (u1) at (4, 2.5) {};
    \node[fill=blue, regular polygon, regular polygon sides=3, minimum size=0.35cm, inner sep=0pt] (u2) at (2, 2.5) {};
    \node[fill=red, regular polygon, regular polygon sides=3, minimum size=0.35cm, inner sep=0pt] (v1) at (4, 2) {};
    \node[fill=red, regular polygon, regular polygon sides=3, minimum size=0.35cm, inner sep=0pt, rotate=180] (v2) at (2, 2) {};
    
    \node[left] at (u2) {$u\phantom{l}$};
    \node[left] at (v2) {$v\phantom{l}$};
    
    \node[circle, fill=black, inner sep=0pt, minimum size =0.2cm] (vertex1) at (3,1.5) {};
    \node[circle, fill=black, inner sep=0pt, minimum size =0.2cm] (vertex2) at (2,1.5) {};
    \node[below] at (vertex1) {$4$};
    \node[below] at (vertex2) {$3$};
    
    \draw[blue, thick] (u1) -- (u2);
    \draw[red, thick] (v1) --  (v2);

    \node[below] at (4,1.5) {$|\varphi_I'\rangle$};
    \begin{scope}[yscale=-1, yshift=-7.5cm]
    \draw[] (2, 1.5) -- (2, 3.75);
    \draw[] (3,1.5) -- (3,2.75);
    \draw[] (3, 2.75) -- ++ (0.5, 0.25) -- ++ (0.5, -0.25) -- (4, 1.5);
    
    \node[fill=blue, regular polygon, regular polygon sides=3, minimum size=0.35cm, inner sep=0pt, rotate=180] (u1) at (4, 2.5) {};
    \node[fill=blue, regular polygon, regular polygon sides=3, minimum size=0.35cm, inner sep=0pt, rotate=180] (u2) at (2, 2.5) {};
    \node[fill=red, regular polygon, regular polygon sides=3, minimum size=0.35cm, inner sep=0pt, rotate=180] (v1) at (4, 2) {};
    \node[fill=red, regular polygon, regular polygon sides=3, minimum size=0.35cm, inner sep=0pt] (v2) at (2, 2) {};
    
    \node[left] at (u2) {$u^{\dagger}$};
    \node[left] at (v2) {$v^{\dagger}$};
    
    \node[circle, fill=black, inner sep=0pt, minimum size =0.2cm] (vertex1) at (3,1.5) {};
    \node[circle, fill=black, inner sep=0pt, minimum size =0.2cm] (vertex2) at (2,1.5) {};
    \node[above] at (vertex1) {$1$};
    \node[above] at (vertex2) {$2$};
    
    \draw[blue, thick] (u1)  -- (u2);
    \draw[red, thick] (v1) --  (v2);

    \node[above] at (4,1.5) {$\langle \varphi_I|$};
    \end{scope}
    \begin{scope}[xshift=4cm]
    \draw[] (2, 1.5) -- (2, 3.75);
    \draw[] (3,1.5) -- (3,2.75);
    \draw[] (3, 2.75) -- ++ (0.5, 0.25) -- ++ (0.5, -0.25) -- (4, 1.5);
  
     \node[fill=blue, regular polygon, regular polygon sides=3, minimum size=0.35cm, inner sep=0pt] (u1) at (4, 2.5) {};
    \node[fill=blue, regular polygon, regular polygon sides=3, minimum size=0.35cm, inner sep=0pt] (u2) at (2, 2.5) {};
    \node[fill=red, regular polygon, regular polygon sides=3, minimum size=0.35cm, inner sep=0pt] (v1) at (4, 2) {};
    \node[fill=red, regular polygon, regular polygon sides=3, minimum size=0.35cm, inner sep=0pt, rotate=180] (v2) at (2, 2) {};
    
    \node[left] at (u2) {$u\phantom{l}$};
    \node[left] at (v2) {$v\phantom{l}$};
    
    \node[circle, fill=black, inner sep=0pt, minimum size =0.2cm] (vertex1) at (3,1.5) {};
    \node[circle, fill=black, inner sep=0pt, minimum size =0.2cm] (vertex2) at (2,1.5) {};
    \node[below] at (vertex1) {$1$};
    \node[below] at (vertex2) {$2$};

    \draw[blue, thick] (u1)  -- (u2);
    \draw[red, thick] (v1)  --  (v2);
    
    \node[below] at (4,1.5) {$|\varphi_I\rangle$};
    \begin{scope}[yscale=-1, yshift=-7.5cm]
    
    \draw[] (2, 1.5) -- (2, 3.75);
    \draw[] (3, 1.5) -- (3, 2.75);
     \draw[] (3, 2.75) -- ++ (0.5, 0.25) -- ++ (0.5, -0.25) -- (4, 1.5);

    \node[fill=blue, regular polygon, regular polygon sides=3, minimum size=0.35cm, inner sep=0pt, rotate=180] (u1) at (4, 2.5) {};
    \node[fill=blue, regular polygon, regular polygon sides=3, minimum size=0.35cm, inner sep=0pt, rotate=180] (u2) at (2, 2.5) {};
    \node[fill=red, regular polygon, regular polygon sides=3, minimum size=0.35cm, inner sep=0pt, rotate=180] (v1) at (4, 2) {};
    \node[fill=red, regular polygon, regular polygon sides=3, minimum size=0.35cm, inner sep=0pt] (v2) at (2, 2) {};
    
    \node[left] at (u2) {$u^{\dagger}$};
    \node[left] at (v2) {$v^{\dagger}$};
    
    \node[circle, fill=black, inner sep=0pt, minimum size =0.2cm] (vertex1) at (3,1.5) {};
    \node[circle, fill=black, inner sep=0pt, minimum size =0.2cm] (vertex2) at (2,1.5) {};
    \node[above] at (vertex1) {$4$};
    \node[above] at (vertex2) {$3$};

    \draw[blue, thick] (u1)  -- (u2);
    \draw[red, thick] (v1) --  (v2);

    \node[above] at (4,1.5) {$\langle \varphi_I'|$};
    \end{scope}
    \end{scope}
    \end{tikzpicture}  
    \right)
    \end{aligned}
\end{equation}
where the second line is obtained by straightening the legs connecting the triangles and the third line is obtained by bending the black legs.

We can now reinterpret the diagram as
\begin{equation}
    (\langle \varphi|\otimes \langle \varphi'|) \Delta' (|\varphi'\rangle \otimes |\varphi\rangle ) = -\frac{|B|}{d(d^2-1)} \textrm{Tr}_{R}(M_{\varphi_I} M_{\varphi_I'}),
\end{equation}
where $M_{\varphi_i}$ is defined as 
\begin{equation}
    M_{\varphi_I}=
        \begin{tikzpicture}[scale=0.8, baseline={([yshift=-.5ex]current bounding box.center)}]
    \begin{scope}[yscale=-1, yshift=-1.5cm]
    \draw[] (2, 0.75) -- (2, 3.75);
    \draw[] (3,0.75) -- (3,2.75);
    \draw[] (3, 2.75) -- ++ (0.5, 0.25) -- ++ (0.5, -0.25) -- (4, 1.5);
    
    \node[fill=blue, regular polygon, regular polygon sides=3, minimum size=0.35cm, inner sep=0pt, rotate=180] (u1) at (4, 2.5) {};
    \node[fill=blue, regular polygon, regular polygon sides=3, minimum size=0.35cm, inner sep=0pt, rotate=180] (u2) at (2, 2.5) {};
    \node[fill=red, regular polygon, regular polygon sides=3, minimum size=0.35cm, inner sep=0pt, rotate=180] (v1) at (4, 2) {};
    \node[fill=red, regular polygon, regular polygon sides=3, minimum size=0.35cm, inner sep=0pt] (v2) at (2, 2) {};
    
    \node[left] at (u2) {$u^{\dagger}$};
    \node[left] at (v2) {$v^{\dagger}$};
    
    \node[circle, fill=black, inner sep=0pt, minimum size =0.2cm] (vertex1) at (3,0.75) {};
    \node[circle, fill=black, inner sep=0pt, minimum size =0.2cm] (vertex2) at (2,0.75) {};
    \node[left] at (vertex1) {$1$};
    \node[left] at (vertex2) {$2$};
    
    \draw[blue, thick] (u1)  -- (u2);
    \draw[red, thick] (v1) --  (v2);
    \node[below] () at (2,3.75) {$R$};

    \node[above] at (4,1.5) {$\langle \varphi_I|$};
    \end{scope}
    \begin{scope}
    \draw[] (2, 0.75) -- (2, 3.75);
    \draw[] (3,0.75) -- (3,2.75);
    \draw[] (3, 2.75) -- ++ (0.5, 0.25) -- ++ (0.5, -0.25) -- (4, 1.5);
  
     \node[fill=blue, regular polygon, regular polygon sides=3, minimum size=0.35cm, inner sep=0pt] (u1) at (4, 2.5) {};
    \node[fill=blue, regular polygon, regular polygon sides=3, minimum size=0.35cm, inner sep=0pt] (u2) at (2, 2.5) {};
    \node[fill=red, regular polygon, regular polygon sides=3, minimum size=0.35cm, inner sep=0pt] (v1) at (4, 2) {};
    \node[fill=red, regular polygon, regular polygon sides=3, minimum size=0.35cm, inner sep=0pt, rotate=180] (v2) at (2, 2) {};
    
    \node[left] at (u2) {$u\phantom{l}$};
    \node[left] at (v2) {$v\phantom{l}$};

    \draw[blue, thick] (u1)  -- (u2);
    \draw[red, thick] (v1)  --  (v2);
    
    \node[below] at (4,1.5) {$|\varphi_I\rangle$};
    
    \node[above] () at (2,3.75) {$R$};
    \end{scope}
    \end{tikzpicture}= \text{Tr}_I\left(u v^{T_I}(I_R\otimes |\varphi_I\rangle\langle \varphi_I|) \left(v^{T_I}\right)^{\dagger} u^{\dagger}\right).
\end{equation}
Recalling that $\|v^{T_I}\| \leq |I|$ (see Eq.\eqref{eq:partial_transpose_bound}), we obtain the following bound:
\begin{equation}
    \|M_{\varphi_I} \| \leq |I|^3.
\end{equation}
A similar expression for $M_{\varphi_I'}$ and its bound can be derived in the same way. We can thus obtain the following bound
\begin{equation}
    |(\langle \varphi|\otimes \langle \varphi'|) \Delta' (|\varphi'\rangle \otimes |\varphi\rangle )| \leq \frac{|I|^6}{d^2-1},
\end{equation}
assuming $|\varphi\rangle$ and $|\varphi'\rangle$ are product states over $I$ and $\ell$, using the fact that the partial trace on $R$ incurs a factor of $|R|$ and the fact that $d=|B| |R|$. Following the argument in Section~\ref{sec:average}, we see that for general entangled states between $I$ and $\ell$, we get
\begin{equation}
    |(\langle \varphi|\otimes \langle \varphi'|) \Delta' (|\varphi'\rangle \otimes |\varphi\rangle )| \leq \frac{|I|^8}{d^2-1}.
\end{equation}
Therefore, even with the reverse ordering we conclude that the fluctuation remains small in the limit that $|I| \ll d$.

\subsection{Complexity}
\label{sec:complexity_reverse_time_ordering}

Here we study the complexity of the circuit Eq.~\eqref{eq:straightening2} after making it deterministic. Note that the map $V(U')$ can is equivalent to a unitary circuit followed by postselection on some auxiliary systems $I_1, I_2, I_3, I_4$ and the infaller $I$:
\begin{equation}
|r|
\begin{tikzpicture}[scale=0.8, baseline={([yshift=-.5ex]current bounding box.center)}]
    \draw[] (0,0) -- (0,1);
    \draw[] (2,0) -- (2,1);
    \draw[] (4,0) -- (4,1);
    \draw[] (6,0) -- (6,1);
    \draw[] (4,0) -- (5, -0.5) -- (6,0);
    \draw[] (0, 1.5) -- (0, 2.5);
    \draw[] (2, 1.5) -- (2, 2.5);
    \draw[] (2, 2.5) -- (3, 3) -- (4, 2.5) -- (4,1);
    \draw[] (3,0) -- (3,1);
    
    \draw[fill=red!10!white] (-0.25, 1) -- (3.25, 1) -- (3.25, 1.75) -- (-0.25, 1.75) -- cycle;
    
     \node[fill=blue, regular polygon, regular polygon sides=3, minimum size=0.35cm, inner sep=0pt] (u1) at (3, 0.7) {};
    \node[fill=blue, regular polygon, regular polygon sides=3, minimum size=0.35cm, inner sep=0pt] (u2) at (6, 0.7) {};
    \node[fill=red, regular polygon, regular polygon sides=3, minimum size=0.35cm, inner sep=0pt] (v1) at (3, 0.25) {};
    \node[fill=red, regular polygon, regular polygon sides=3, minimum size=0.35cm, inner sep=0pt] (v2) at (4, 0.25) {};
    
    \node[right] at (u2) {$u$};
    \node[right] at (v2) {$v$};
    
    \draw[blue, thick] (u1) -- (u2);
    \draw[red, thick] (v1) -- (v2);

    \node[above] at (0, 2.5) {$B$};
    \node[above] at (3, 3) {$ \langle \text{MAX}|$};
    \node[] at (1.5,1.375) {$U'$};
    \node[below] at (0,0) {$|\psi\rangle_{\ell}$};
    \node[below] at (2,0) {$|\psi_0\rangle_f$};
    \node[below] at (3,0) {$I$};
    \node[below] at (4,0) {$r$};
    \node[below] at (6,0) {$R$};
    \node[below] at (5,-0.5) {$|\text{MAX}\rangle$};
    \end{tikzpicture}=
 |I|^2
    \begin{tikzpicture}[scale=0.8, baseline={([yshift=-.5ex]current bounding box.center)}]
    \draw[] (0,0) -- (0,1);
    \draw[] (2,0) -- (2,1);
    \draw[] (0, 1.5) -- (0, 2.75);
    \draw[] (2, 1.5) -- (2, 2.75);
    \draw[] (3,0) -- (3,1);
    \node[above] () at (0, 2.75) {$B$};
    \node[above] () at (2, 2.75) {$R$};
    
    \draw[fill=red!10!white] (-0.25, 0.75) -- (3.25, 0.75) -- (3.25, 1.5) -- (-0.25, 1.5) -- cycle;

    \node[] at (1.5,1.125) {$U'$};
    \node[below] at (0,0) {$|\psi\rangle_{\ell}$};
    \node[below] at (2,0) {$| \psi_0\rangle_f$};
    \node[below] at (9,0) {$I$};
    
    \draw[] (3, 0) -- ++ (0.75, -0.5) -- ++ (0.75, 0.5);
    \draw[] (6, 0) -- ++ (0, 2.5);
    \draw[] (7.5, 0) -- ++ (0, 2.5);
    \draw[] (9, 0) -- ++ (0, 2.5);
    \draw[] (4.5, 0) -- ++ (0, 2.5); 
    \draw[] (4.5, 2.5) -- ++ (0.75, 0.5) -- ++ (0.75, -0.5);
    \draw[] (7.5, 2.5) -- ++ (0.75, 0.5) -- ++ (0.75, -0.5);
    \draw[] (6, 0) -- ++ (0.75, -0.5) -- ++ (0.75, 0.5);

    \node[below] () at (4.125,-0.5) {$|\text{MAX}\rangle_{I_1 I_2}$};
    \node[below] () at (7.125,-0.5) {$|\text{MAX}\rangle_{I_3 I_4}$};

    \node[above] () at (5.3, 3) {$\langle \text{MAX}|_{I_2 I_3}$};
    \node[above] () at (8.3, 3) {$\langle \text{MAX}|_{I_4 I}$};
    
    \node[fill=blue, regular polygon, regular polygon sides=3, minimum size=0.35cm, inner sep=0pt] (u1) at (6, 2.25) {};
    \node[fill=blue, regular polygon, regular polygon sides=3, minimum size=0.35cm, inner sep=0pt] (u2) at (2, 2.25) {};
    \node[fill=red, regular polygon, regular polygon sides=3, minimum size=0.35cm, inner sep=0pt, rotate=180] (v1) at (7.5, 1.75) {};
    \node[fill=red, regular polygon, regular polygon sides=3, minimum size=0.35cm, inner sep=0pt, rotate=180] (v2) at (2, 1.75) {};
    
    \node[left] at (u2) {$u\phantom{l}$};
    \node[left] at (v2) {$v\phantom{l}$};
    
    \draw[blue, thick] (u1) -- (u2);
    \draw[red, thick] (v1) --  (v2);
    \end{tikzpicture}.
    \label{eq:circ_temp2}
\end{equation}
Here the dimensions of $I_1, I_2, I_3$, and $I_4$ are all chosen to be equal to that of $I$. Thus, we see that the postselection success probability is $|I|^{-2}$. Thus, using the QSVT approach in Section~\ref{sec:complexity}, we an bound the overall complexity of $V(U')$ as $O(|I|^2 \mathcal{C})$, where $\mathcal{C}$ is the complexity of $v, u, U'$, and the state preparation for $|\psi\rangle_{\ell}, |\psi_0\rangle_f, |\text{MAX}\rangle$ combined.

\section{Pseudorandom unitary: Candidate construction}
\label{appendix:pru_candidate}

In this Appendix, we review a candidate construction for pseudorandom unitary discussed in Ref.~\cite{Ji2018}. The basic idea is to apply a ``random phase'' over two complementary basis sets. Consider the unitary transformation
\begin{equation}
    U_{\text{PR}}(|x\rangle |y\rangle) = \omega_N^{F_x(y)}|x\rangle|y\rangle, \label{eq:pru_construction2}
\end{equation}
where $\omega_N= e^{2\pi i/N}$ and each $F_x(y): X\times Y \to Y$ is an efficiently computable function, and suppose that if $x$ is chosen uniformly over $X$, the resulting ensemble of functions cannot be distinguished from a uniformly random function by any adversary who is limited to computations with complexity polynomial in $\log |X|$. It was shown in Ref.~\cite{Ji2018} that Eq.~\eqref{eq:pru_construction2} can be used to generate an ensemble of states which is \emph{computationally indistinguishable} from a Haar-random ensemble. This ensemble of states can be realized by preparing a uniform superposition state in the $Y$ register, applying $U_{\text{PR}}$, and then tracing out the $Y$ register.

However, the corresponding ensemble of unitary transformations $\{ U_{\text{PR},x}: x\in X\}$, where $U_{\text{PR},x}|y\rangle = \omega_N^{F_x(y)}|y\rangle$, does not provide a pseudorandom unitary as prescribed in Definition \ref{definition:pru}, because its action on some states is easy to distinguish from the action of a Haar-random ensemble. In particular, for each $x$, the unitary $U_{\text{PR},x}$ maps computational basis states 
in the $Y$ register to 
computational basis states, while in contrast a typical unitary transformation sampled from the Haar-random ensemble maps such states to highly entangled states. 

Ref.~\cite{Ji2018} suggested a way to construct psuedorandom unitary transformations using $\{U_{\text{PR},x}\}$ combined with other quantum gates. The idea is to apply $\{ U_{\text{PR},x}H^{\otimes n}: x\in X\}$ several times in succession, where $H^{\otimes n}$ denotes a tensor product of Hadamard gates acting on all qubits, and $x$ is randomly sampled each time $U_{\text{PR},x}$ is applied.
Appending the extra Hadamard gates has the effect of rotating the basis set in which the pseudorandom phases are applied.
However, it has not been proven that the pseudorandomness of this modified ensemble follows from widely accepted computational assumptions.

\section{Black-hole-assisted quantum computation}
\label{appendix:bh_assisted_qc}

In this Appendix, we investigate the possibility of using a black hole to speed up a quantum computation. 
An observer, who stays outside the black hole, sends a robot into the black hole; this robot applies the unitaries $v$ and $u$. From the perspective of the observer, the resulting physical process is shown here:
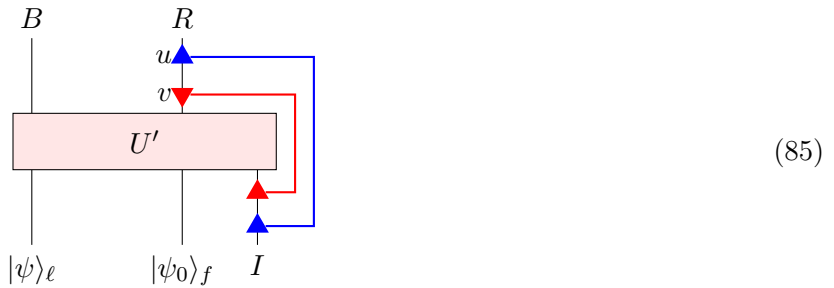
\begin{equation}
\begin{aligned}
    \begin{tikzpicture}
    \draw[] (0,0) -- (0,1);
    \draw[] (2,0) -- (2,1);
    \draw[] (0, 1.5) -- (0, 2.75);
    \draw[] (2, 1.5) -- (2, 2.75);
    \draw[] (3,0) -- (3,1);
    
    \draw[fill=red!10!white] (-0.25, 1) -- (3.25, 1) -- (3.25, 1.75) -- (-0.25, 1.75) -- cycle;
    
     \node[fill=blue, regular polygon, regular polygon sides=3, minimum size=0.35cm, inner sep=0pt] (u1) at (3, 0.25) {};
    \node[fill=blue, regular polygon, regular polygon sides=3, minimum size=0.35cm, inner sep=0pt] (u2) at (2, 2.5) {};
    \node[fill=red, regular polygon, regular polygon sides=3, minimum size=0.35cm, inner sep=0pt] (v1) at (3, 0.7) {};
    \node[fill=red, regular polygon, regular polygon sides=3, minimum size=0.35cm, inner sep=0pt, rotate=180] (v2) at (2, 2) {};
    
    \node[left] at (u2) {$u$};
    \node[left] at (v2) {$v$};
    
    \draw[blue, thick] (u1) -- ++ (0.75, 0) -- ++ (0,2.25) -- (u2);
    \draw[red, thick] (v1) -- ++ (0.5, 0) -- ++ (0, 1.3) --  (v2);
    
    \node[above] at (2, 2.75) {$R$};
    \node[above] at (0, 2.75) {$B$};
    \node[] at (1.5,1.375) {$U'$};
    \node[below] at (0,0) {$|\psi\rangle_{\ell}$};
    \node[below] at (2,0) {$|\psi_0\rangle_f$};
    \node[below] at (3,0) {$I$};
    \end{tikzpicture}
    \end{aligned}
\end{equation}
The observer can set the initial state of the robot ($I$) and later measure the radiation $R$ in a simple basis. The key question is: what computational problems can be solved efficiently by (i) initializing the state of $I$ and (ii) later measuring $R$? The dynamics of $U'$ is assumed to be pseudorandom, as in the main text.

Let $|\varphi\rangle_I$ be the initial state of $I$ and $|\Psi\rangle_R$ be the outcome when $R$ is measured, both assumed to have complexity at most polynomial in the black hole entropy. We also assume that the unitaries $u$ and $v$ --- the unitaries that describe the interaction between the robot and the radiation --- have polynomial complexity. Then the probability of obtaining this measurement outcome can be represented by the following diagram:
\begin{equation}
\begin{aligned}
    \begin{tikzpicture}[scale=0.8, baseline={([yshift=-.5ex]current bounding box.center)}]
    \draw[] (0,0) -- (0,1);
    \draw[] (2,0) -- (2,1);
    \draw[] (0, 1.5) -- (0, 3.75);
    \draw[] (2, 1.5) -- (2, 3);
    \draw[] (3,0) -- (3,1);
    
    \draw[fill=red!10!white] (-0.25, 1) -- (3.25, 1) -- (3.25, 1.75) -- (-0.25, 1.75) -- cycle;
    
     \node[fill=blue, regular polygon, regular polygon sides=3, minimum size=0.35cm, inner sep=0pt] (u1) at (3, 0.25) {};
    \node[fill=blue, regular polygon, regular polygon sides=3, minimum size=0.35cm, inner sep=0pt] (u2) at (2, 2.5) {};
    \node[fill=red, regular polygon, regular polygon sides=3, minimum size=0.35cm, inner sep=0pt] (v1) at (3, 0.7) {};
    \node[fill=red, regular polygon, regular polygon sides=3, minimum size=0.35cm, inner sep=0pt, rotate=180] (v2) at (2, 2) {};
    
    \node[left] at (u2) {$u\phantom{l}$};
    \node[left] at (v2) {$v\phantom{l}$};
    
    \node[above] () at (2, 3) {$\langle \Psi|_R$};
    
    \draw[blue, thick] (u1) -- ++ (0.75, 0) -- ++ (0,2.25) -- (u2);
    \draw[red, thick] (v1) -- ++ (0.5, 0) -- ++ (0, 1.3) --  (v2);
    
    \node[] at (1.5,1.375) {$U'$};
    \node[below] at (0,0) {$|\psi\rangle_{\ell}$};
    \node[below] at (2,0) {$|\psi_0\rangle_f$};
    \node[below] at (3,0) {$|\varphi\rangle_I$};
    \begin{scope}[yscale=-1, yshift=-7.5cm]
    
    \draw[] (0,0) -- (0,1);
    \draw[] (2,0) -- (2,1);
    \draw[] (0, 1.5) -- (0, 3.75);
    \draw[] (2, 1.5) -- (2, 3);
    \draw[] (3,0) -- (3,1);
    
    \node[below] () at (2, 3) {$| \Psi\rangle_R$};
    
    \draw[fill=red!10!white] (-0.25, 1) -- (3.25, 1) -- (3.25, 1.75) -- (-0.25, 1.75) -- cycle;
    
    \node[fill=blue, regular polygon, regular polygon sides=3, minimum size=0.35cm, inner sep=0pt, rotate=180] (u1) at (3, 0.25) {};
    \node[fill=blue, regular polygon, regular polygon sides=3, minimum size=0.35cm, inner sep=0pt, rotate=180] (u2) at (2, 2.5) {};
    \node[fill=red, regular polygon, regular polygon sides=3, minimum size=0.35cm, inner sep=0pt, rotate=180] (v1) at (3, 0.7) {};
    \node[fill=red, regular polygon, regular polygon sides=3, minimum size=0.35cm, inner sep=0pt] (v2) at (2, 2) {};
    
    \node[left] at (u2) {$u^{\dagger}$};
    \node[left] at (v2) {$v^{\dagger}$};
    
    \draw[blue, thick] (u1) -- ++ (0.75, 0) -- ++ (0,2.25) -- (u2);
    \draw[red, thick] (v1) -- ++ (0.5, 0) -- ++ (0, 1.3) --  (v2);
    
    \node[] at (1.5,1.375) {${U'}^{\dagger}$};
    \node[above] at (0,0) {$\langle\psi|_{\ell}$};
    \node[above] at (2,0) {$\langle\psi_0|_f$};
    \node[above] at (3,0) {$\langle\varphi|_I$};
    \end{scope}
    \end{tikzpicture} &= |I|^2
    \left(
    \begin{tikzpicture}[scale=0.8, baseline={([yshift=-.5ex]current bounding box.center)}]
    \draw[] (0,0) -- (0,1);
    \draw[] (2,0) -- (2,1);
    \draw[] (0, 1.5) -- (0, 4.25);
    \draw[] (2, 1.5) -- (2, 3.25);
    \draw[] (3,0) -- (3,1);
    \node[left] () at (0, 4.25) {$B$};
    \node[above] () at (2, 3.25) {$\langle \Psi|_R$};
    
    \draw[fill=red!10!white] (-0.25, 1) -- (3.25, 1) -- (3.25, 1.75) -- (-0.25, 1.75) -- cycle;

    \node[] at (1.5,1.375) {$U'$};
    \node[below] at (0,0) {$|\psi\rangle_{\ell}$};
    \node[below] at (2,0) {$| \psi_0\rangle_f$};
    
    \draw[] (3, 0) -- ++ (0.75, -0.5) -- ++ (0.75, 0.5);
    \draw[] (6, -0.5) -- ++ (0, 3.25);
    \draw[] (4.5, 0) -- ++ (0, 2.75); 
    \draw[] (4.5, 2.75) -- ++ (0.75, 0.5) -- ++ (0.75, -0.5);

    \node[below] () at (4.125,-0.5) {$|\text{MAX}\rangle_{I_1 I_2}$};
    \node[below] () at (6.125, -0.5) {$|\varphi\rangle_{I}$};
    
    \node[above] () at (5.3, 3.25) {$\langle \text{MAX}|_{I_2 I}$};
    
    \node[fill=blue, regular polygon, regular polygon sides=3, minimum size=0.35cm, inner sep=0pt] (u1) at (6, 2.375) {};
    \node[fill=blue, regular polygon, regular polygon sides=3, minimum size=0.35cm, inner sep=0pt] (u2) at (2, 2.375) {};
    \node[fill=red, regular polygon, regular polygon sides=3, minimum size=0.35cm, inner sep=0pt, rotate=180] (v1) at (4.5, 2) {};
    \node[fill=red, regular polygon, regular polygon sides=3, minimum size=0.35cm, inner sep=0pt, rotate=180] (v2) at (2, 2) {};
    
    \node[left] at (u2) {$u\phantom{l}$};
    \node[left] at (v2) {$v\phantom{l}$};
    
    \draw[blue, thick] (u1) -- (u2);
    \draw[red, thick] (v1) --  (v2);
    \begin{scope}[yshift = 8.5cm, yscale=-1]
    
    \draw[] (0,0) -- (0,1);
    \draw[] (2,0) -- (2,1);
    \draw[] (0, 1.5) -- (0, 4.25);
    \draw[] (2, 1.5) -- (2, 3.25);
    \draw[] (3,0) -- (3,1);
    \node[below] () at (2,3.25) {$|\Psi\rangle_R$};
    
    \draw[fill=red!10!white] (-0.25, 1) -- (3.25, 1) -- (3.25, 1.75) -- (-0.25, 1.75) -- cycle;

    \node[] at (1.5,1.375) {${U'}^{\dagger}$};
    \node[above] at (0,0) {$\langle\psi|_{\ell}$};
    \node[above] at (2,0) {$\langle \psi_0|_f$};
    
    \draw[] (3, 0) -- ++ (0.75, -0.5) -- ++ (0.75, 0.5);
    \draw[] (6, -0.5) -- ++ (0, 3.25);
    \draw[] (4.5, 0) -- ++ (0, 2.75); 
    \draw[] (4.5, 2.75) -- ++ (0.75, 0.5) -- ++ (0.75, -0.5);
    
 \node[above] () at (4.125,-0.5) {$\langle\text{MAX}|_{I_1 I_2}$};
    \node[above] () at (6.125, -0.5) {$\langle \varphi|_{I}$};
    
    \node[below] () at (5.3, 3.25) {$| \text{MAX}\rangle_{I_2 I}$};
    \node[fill=blue, regular polygon, regular polygon sides=3, minimum size=0.35cm, inner sep=0pt, rotate=180] (u1) at (6, 2.375) {};
    \node[fill=blue, regular polygon, regular polygon sides=3, minimum size=0.35cm, inner sep=0pt, rotate=180] (u2) at (2, 2.375) {};
    \node[fill=red, regular polygon, regular polygon sides=3, minimum size=0.35cm, inner sep=0pt] (v1) at (4.5, 2) {};
    \node[fill=red, regular polygon, regular polygon sides=3, minimum size=0.35cm, inner sep=0pt] (v2) at (2, 2) {};
    
    \node[left] at (u2) {$u^{\dagger}$};
    \node[left] at (v2) {$v^{\dagger}$};
    
    \draw[blue, thick] (u1) -- (u2);
    \draw[red, thick] (v1) --  (v2);
    
    \end{scope}
    \end{tikzpicture}\right),
    \end{aligned}
\end{equation}
where the right-hand side of the equation is obtained by straightening out the blue and red legs. One can view the right-hand side as a process in which one begins with a simple state $(|\psi\rangle_{\ell} |\psi_0\rangle_f |\textrm{MAX}\rangle_{I_1I_2} |\varphi\rangle_I)$, applies a pseudorandom unitary $U'$, then applies a polynomially complex unitary ($u$ and $v$) followed by a simple measurement. From the definition of the pseudorandom unitary, we can conclude the following:
\begin{equation}
    \begin{tikzpicture}[scale=0.8, baseline={([yshift=-.5ex]current bounding box.center)}]
    \draw[] (0,0) -- (0,1);
    \draw[] (2,0) -- (2,1);
    \draw[] (0, 1.5) -- (0, 4.25);
    \draw[] (2, 1.5) -- (2, 3.25);
    \draw[] (3,0) -- (3,1);
    \node[left] () at (0, 4.25) {$B$};
    \node[above] () at (2, 3.25) {$\langle \Psi|_R$};
    
    \draw[fill=red!10!white] (-0.25, 1) -- (3.25, 1) -- (3.25, 1.75) -- (-0.25, 1.75) -- cycle;

    \node[] at (1.5,1.375) {$U'$};
    \node[below] at (0,0) {$|\psi\rangle_{\ell}$};
    \node[below] at (2,0) {$| \psi_0\rangle_f$};
    
    \draw[] (3, 0) -- ++ (0.75, -0.5) -- ++ (0.75, 0.5);
    \draw[] (6, -0.5) -- ++ (0, 3.25);
    \draw[] (4.5, 0) -- ++ (0, 2.75); 
    \draw[] (4.5, 2.75) -- ++ (0.75, 0.5) -- ++ (0.75, -0.5);

    \node[below] () at (4.125,-0.5) {$|\text{MAX}\rangle_{I_1 I_2}$};
    \node[below] () at (6.125, -0.5) {$|\varphi\rangle_{I}$};
    
    \node[above] () at (5.3, 3.25) {$\langle \text{MAX}|_{I_2 I}$};
    
    \node[fill=blue, regular polygon, regular polygon sides=3, minimum size=0.35cm, inner sep=0pt] (u1) at (6, 2.375) {};
    \node[fill=blue, regular polygon, regular polygon sides=3, minimum size=0.35cm, inner sep=0pt] (u2) at (2, 2.375) {};
    \node[fill=red, regular polygon, regular polygon sides=3, minimum size=0.35cm, inner sep=0pt, rotate=180] (v1) at (4.5, 2) {};
    \node[fill=red, regular polygon, regular polygon sides=3, minimum size=0.35cm, inner sep=0pt, rotate=180] (v2) at (2, 2) {};
    
    \node[left] at (u2) {$u\phantom{l}$};
    \node[left] at (v2) {$v\phantom{l}$};
    
    \draw[blue, thick] (u1) -- (u2);
    \draw[red, thick] (v1) --  (v2);
    \begin{scope}[yshift = 8.5cm, yscale=-1]
    
    \draw[] (0,0) -- (0,1);
    \draw[] (2,0) -- (2,1);
    \draw[] (0, 1.5) -- (0, 4.25);
    \draw[] (2, 1.5) -- (2, 3.25);
    \draw[] (3,0) -- (3,1);
    \node[below] () at (2,3.25) {$|\Psi\rangle_R$};
    
    \draw[fill=red!10!white] (-0.25, 1) -- (3.25, 1) -- (3.25, 1.75) -- (-0.25, 1.75) -- cycle;

    \node[] at (1.5,1.375) {${U'}^{\dagger}$};
    \node[above] at (0,0) {$\langle\psi|_{\ell}$};
    \node[above] at (2,0) {$\langle \psi_0|_f$};
    
    \draw[] (3, 0) -- ++ (0.75, -0.5) -- ++ (0.75, 0.5);
    \draw[] (6, -0.5) -- ++ (0, 3.25);
    \draw[] (4.5, 0) -- ++ (0, 2.75); 
    \draw[] (4.5, 2.75) -- ++ (0.75, 0.5) -- ++ (0.75, -0.5);
    
 \node[above] () at (4.125,-0.5) {$\langle\text{MAX}|_{I_1 I_2}$};
    \node[above] () at (6.125, -0.5) {$\langle \varphi|_{I}$};
    
    \node[below] () at (5.3, 3.25) {$| \text{MAX}\rangle_{I_2 I}$};
    \node[fill=blue, regular polygon, regular polygon sides=3, minimum size=0.35cm, inner sep=0pt, rotate=180] (u1) at (6, 2.375) {};
    \node[fill=blue, regular polygon, regular polygon sides=3, minimum size=0.35cm, inner sep=0pt, rotate=180] (u2) at (2, 2.375) {};
    \node[fill=red, regular polygon, regular polygon sides=3, minimum size=0.35cm, inner sep=0pt] (v1) at (4.5, 2) {};
    \node[fill=red, regular polygon, regular polygon sides=3, minimum size=0.35cm, inner sep=0pt] (v2) at (2, 2) {};
    
    \node[left] at (u2) {$u^{\dagger}$};
    \node[left] at (v2) {$v^{\dagger}$};
    
    \draw[blue, thick] (u1) -- (u2);
    \draw[red, thick] (v1) --  (v2);
    
    \end{scope}
    \end{tikzpicture}
    \approx
    \int d \widetilde{V}
    \left(    
    \begin{tikzpicture}[scale=0.8, baseline={([yshift=-.5ex]current bounding box.center)}]
    \draw[] (0,0) -- (0,1);
    \draw[] (2,0) -- (2,1);
    \draw[] (0, 1.5) -- (0, 4.25);
    \draw[] (2, 1.5) -- (2, 3.25);
    \draw[] (3,0) -- (3,1);
    \node[left] () at (0, 4.25) {$B$};
    \node[above] () at (2, 3.25) {$\langle \Psi|_R$};
    
    \draw[fill=red!10!white] (-0.25, 1) -- (3.25, 1) -- (3.25, 1.75) -- (-0.25, 1.75) -- cycle;

    \node[] at (1.5,1.375) {$\widetilde{V}$};
    \node[below] at (0,0) {$|\psi\rangle_{\ell}$};
    \node[below] at (2,0) {$| \psi_0\rangle_f$};
    
    \draw[] (3, 0) -- ++ (0.75, -0.5) -- ++ (0.75, 0.5);
    \draw[] (6, -0.5) -- ++ (0, 3.25);
    \draw[] (4.5, 0) -- ++ (0, 2.75); 
    \draw[] (4.5, 2.75) -- ++ (0.75, 0.5) -- ++ (0.75, -0.5);

    \node[below] () at (4.125,-0.5) {$|\text{MAX}\rangle_{I_1 I_2}$};
    \node[below] () at (6.125, -0.5) {$|\varphi\rangle_{I}$};
    
    \node[above] () at (5.3, 3.25) {$\langle \text{MAX}|_{I_2 I}$};
    
    \node[fill=blue, regular polygon, regular polygon sides=3, minimum size=0.35cm, inner sep=0pt] (u1) at (6, 2.375) {};
    \node[fill=blue, regular polygon, regular polygon sides=3, minimum size=0.35cm, inner sep=0pt] (u2) at (2, 2.375) {};
    \node[fill=red, regular polygon, regular polygon sides=3, minimum size=0.35cm, inner sep=0pt, rotate=180] (v1) at (4.5, 2) {};
    \node[fill=red, regular polygon, regular polygon sides=3, minimum size=0.35cm, inner sep=0pt, rotate=180] (v2) at (2, 2) {};
    
    \node[left] at (u2) {$u\phantom{l}$};
    \node[left] at (v2) {$v\phantom{l}$};
    
    \draw[blue, thick] (u1) -- (u2);
    \draw[red, thick] (v1) --  (v2);
    \begin{scope}[yshift = 8.5cm, yscale=-1]
    
    \draw[] (0,0) -- (0,1);
    \draw[] (2,0) -- (2,1);
    \draw[] (0, 1.5) -- (0, 4.25);
    \draw[] (2, 1.5) -- (2, 3.25);
    \draw[] (3,0) -- (3,1);
    \node[below] () at (2,3.25) {$|\Psi\rangle_R$};
    
    \draw[fill=red!10!white] (-0.25, 1) -- (3.25, 1) -- (3.25, 1.75) -- (-0.25, 1.75) -- cycle;

    \node[] at (1.5,1.375) {$\widetilde{V}^{\dagger}$};
    \node[above] at (0,0) {$\langle\psi|_{\ell}$};
    \node[above] at (2,0) {$\langle \psi_0|_f$};
    
    \draw[] (3, 0) -- ++ (0.75, -0.5) -- ++ (0.75, 0.5);
    \draw[] (6, -0.5) -- ++ (0, 3.25);
    \draw[] (4.5, 0) -- ++ (0, 2.75); 
    \draw[] (4.5, 2.75) -- ++ (0.75, 0.5) -- ++ (0.75, -0.5);
    
 \node[above] () at (4.125,-0.5) {$\langle\text{MAX}|_{I_1 I_2}$};
    \node[above] () at (6.125, -0.5) {$\langle \varphi|_{I}$};
    
    \node[below] () at (5.3, 3.25) {$| \text{MAX}\rangle_{I_2 I}$};
    \node[fill=blue, regular polygon, regular polygon sides=3, minimum size=0.35cm, inner sep=0pt, rotate=180] (u1) at (6, 2.375) {};
    \node[fill=blue, regular polygon, regular polygon sides=3, minimum size=0.35cm, inner sep=0pt, rotate=180] (u2) at (2, 2.375) {};
    \node[fill=red, regular polygon, regular polygon sides=3, minimum size=0.35cm, inner sep=0pt] (v1) at (4.5, 2) {};
    \node[fill=red, regular polygon, regular polygon sides=3, minimum size=0.35cm, inner sep=0pt] (v2) at (2, 2) {};
    
    \node[left] at (u2) {$u^{\dagger}$};
    \node[left] at (v2) {$v^{\dagger}$};
    
    \draw[blue, thick] (u1) -- (u2);
    \draw[red, thick] (v1) --  (v2);
    
    \end{scope}
    \end{tikzpicture}
    \right), \label{eq:average_bh_assisted}
\end{equation}
with an approximation error superpolynomially small in $\log |B|$, which we denote as $\epsilon (|B|)$. The right-hand side of Eq.~\eqref{eq:average_bh_assisted} can be calculated using Eq.~\eqref{eq:haar_average}:
\begin{equation}
    \begin{aligned}
    \int d \widetilde{V}
    \left(    
    \begin{tikzpicture}[scale=0.8, baseline={([yshift=-.5ex]current bounding box.center)}]
    \draw[] (0,0) -- (0,1);
    \draw[] (2,0) -- (2,1);
    \draw[] (0, 1.5) -- (0, 4.25);
    \draw[] (2, 1.5) -- (2, 3.25);
    \draw[] (3,0) -- (3,1);
    \node[left] () at (0, 4.25) {$B$};
    \node[above] () at (2, 3.25) {$\langle \Psi|_R$};
    
    \draw[fill=red!10!white] (-0.25, 1) -- (3.25, 1) -- (3.25, 1.75) -- (-0.25, 1.75) -- cycle;

    \node[] at (1.5,1.375) {$\widetilde{V}$};
    \node[below] at (0,0) {$|\psi\rangle_{\ell}$};
    \node[below] at (2,0) {$| \psi_0\rangle_f$};
    
    \draw[] (3, 0) -- ++ (0.75, -0.5) -- ++ (0.75, 0.5);
    \draw[] (6, -0.5) -- ++ (0, 3.25);
    \draw[] (4.5, 0) -- ++ (0, 2.75); 
    \draw[] (4.5, 2.75) -- ++ (0.75, 0.5) -- ++ (0.75, -0.5);

    \node[below] () at (4.125,-0.5) {$|\text{MAX}\rangle_{I_1 I_2}$};
    \node[below] () at (6.125, -0.5) {$|\varphi\rangle_{I}$};
    
    \node[above] () at (5.3, 3.25) {$\langle \text{MAX}|_{I_2 I}$};
    
    \node[fill=blue, regular polygon, regular polygon sides=3, minimum size=0.35cm, inner sep=0pt] (u1) at (6, 2.375) {};
    \node[fill=blue, regular polygon, regular polygon sides=3, minimum size=0.35cm, inner sep=0pt] (u2) at (2, 2.375) {};
    \node[fill=red, regular polygon, regular polygon sides=3, minimum size=0.35cm, inner sep=0pt, rotate=180] (v1) at (4.5, 2) {};
    \node[fill=red, regular polygon, regular polygon sides=3, minimum size=0.35cm, inner sep=0pt, rotate=180] (v2) at (2, 2) {};
    
    \node[left] at (u2) {$u\phantom{l}$};
    \node[left] at (v2) {$v\phantom{l}$};
    
    \draw[blue, thick] (u1) -- (u2);
    \draw[red, thick] (v1) --  (v2);
    \begin{scope}[yshift = 8.5cm, yscale=-1]
    
    \draw[] (0,0) -- (0,1);
    \draw[] (2,0) -- (2,1);
    \draw[] (0, 1.5) -- (0, 4.25);
    \draw[] (2, 1.5) -- (2, 3.25);
    \draw[] (3,0) -- (3,1);
    \node[below] () at (2,3.25) {$|\Psi\rangle_R$};
    
    \draw[fill=red!10!white] (-0.25, 1) -- (3.25, 1) -- (3.25, 1.75) -- (-0.25, 1.75) -- cycle;

    \node[] at (1.5,1.375) {$\widetilde{V}^{\dagger}$};
    \node[above] at (0,0) {$\langle\psi|_{\ell}$};
    \node[above] at (2,0) {$\langle \psi_0|_f$};
    
    \draw[] (3, 0) -- ++ (0.75, -0.5) -- ++ (0.75, 0.5);
    \draw[] (6, -0.5) -- ++ (0, 3.25);
    \draw[] (4.5, 0) -- ++ (0, 2.75); 
    \draw[] (4.5, 2.75) -- ++ (0.75, 0.5) -- ++ (0.75, -0.5);
    
 \node[above] () at (4.125,-0.5) {$\langle\text{MAX}|_{I_1 I_2}$};
    \node[above] () at (6.125, -0.5) {$\langle \varphi|_{I}$};
    
    \node[below] () at (5.3, 3.25) {$| \text{MAX}\rangle_{I_2 I}$};
    \node[fill=blue, regular polygon, regular polygon sides=3, minimum size=0.35cm, inner sep=0pt, rotate=180] (u1) at (6, 2.375) {};
    \node[fill=blue, regular polygon, regular polygon sides=3, minimum size=0.35cm, inner sep=0pt, rotate=180] (u2) at (2, 2.375) {};
    \node[fill=red, regular polygon, regular polygon sides=3, minimum size=0.35cm, inner sep=0pt] (v1) at (4.5, 2) {};
    \node[fill=red, regular polygon, regular polygon sides=3, minimum size=0.35cm, inner sep=0pt] (v2) at (2, 2) {};
    
    \node[left] at (u2) {$u^{\dagger}$};
    \node[left] at (v2) {$v^{\dagger}$};
    
    \draw[blue, thick] (u1) -- (u2);
    \draw[red, thick] (v1) --  (v2);
    
    \end{scope}
    \end{tikzpicture}
    \right)
    &= \frac{1}{|I|^2}\int d\widetilde{V}    \left( \begin{tikzpicture}[scale=0.8, baseline={([yshift=-.5ex]current bounding box.center)}]
    \draw[] (0,0) -- (0,1);
    \draw[] (2,0) -- (2,1);
    \draw[] (0, 1.5) -- (0, 3.75);
    \draw[] (2, 1.5) -- (2, 3);
    \draw[] (3,0) -- (3,1);
    
    \draw[fill=red!10!white] (-0.25, 1) -- (3.25, 1) -- (3.25, 1.75) -- (-0.25, 1.75) -- cycle;
    
     \node[fill=blue, regular polygon, regular polygon sides=3, minimum size=0.35cm, inner sep=0pt] (u1) at (3, 0.25) {};
    \node[fill=blue, regular polygon, regular polygon sides=3, minimum size=0.35cm, inner sep=0pt] (u2) at (2, 2.5) {};
    \node[fill=red, regular polygon, regular polygon sides=3, minimum size=0.35cm, inner sep=0pt] (v1) at (3, 0.7) {};
    \node[fill=red, regular polygon, regular polygon sides=3, minimum size=0.35cm, inner sep=0pt, rotate=180] (v2) at (2, 2) {};
    
    \node[left] at (u2) {$u\phantom{l}$};
    \node[left] at (v2) {$v\phantom{l}$};
    
    \node[above] () at (2, 3) {$\langle \Psi|_R$};
    
    \draw[blue, thick] (u1) -- ++ (0.75, 0) -- ++ (0,2.25) -- (u2);
    \draw[red, thick] (v1) -- ++ (0.5, 0) -- ++ (0, 1.3) --  (v2);
    
    \node[] at (1.5,1.375) {$\widetilde{V}$};
    \node[below] at (0,0) {$|\psi\rangle_{\ell}$};
    \node[below] at (2,0) {$|\psi_0\rangle_f$};
    \node[below] at (3,0) {$|\varphi\rangle_I$};
    \begin{scope}[yscale=-1, yshift=-7.5cm]
    
    \draw[] (0,0) -- (0,1);
    \draw[] (2,0) -- (2,1);
    \draw[] (0, 1.5) -- (0, 3.75);
    \draw[] (2, 1.5) -- (2, 3);
    \draw[] (3,0) -- (3,1);
    
    \node[below] () at (2, 3) {$| \Psi\rangle_R$};
    
    \draw[fill=red!10!white] (-0.25, 1) -- (3.25, 1) -- (3.25, 1.75) -- (-0.25, 1.75) -- cycle;
    
    \node[fill=blue, regular polygon, regular polygon sides=3, minimum size=0.35cm, inner sep=0pt, rotate=180] (u1) at (3, 0.25) {};
    \node[fill=blue, regular polygon, regular polygon sides=3, minimum size=0.35cm, inner sep=0pt, rotate=180] (u2) at (2, 2.5) {};
    \node[fill=red, regular polygon, regular polygon sides=3, minimum size=0.35cm, inner sep=0pt, rotate=180] (v1) at (3, 0.7) {};
    \node[fill=red, regular polygon, regular polygon sides=3, minimum size=0.35cm, inner sep=0pt] (v2) at (2, 2) {};
    
    \node[left] at (u2) {$u^{\dagger}$};
    \node[left] at (v2) {$v^{\dagger}$};
    
    \draw[blue, thick] (u1) -- ++ (0.75, 0) -- ++ (0,2.25) -- (u2);
    \draw[red, thick] (v1) -- ++ (0.5, 0) -- ++ (0, 1.3) --  (v2);
    
    \node[] at (1.5,1.375) {$\widetilde{V}^{\dagger}$};
    \node[above] at (0,0) {$\langle\psi|_{\ell}$};
    \node[above] at (2,0) {$\langle\psi_0|_f$};
    \node[above] at (3,0) {$\langle\varphi|_I$};
    \end{scope}
    \end{tikzpicture} \right) \\
    &= 
    \frac{|B|}{|\ell| |I|^3 |f|}
        \begin{tikzpicture}[scale=0.8, baseline={([yshift=-.5ex]current bounding box.center)}]
    \draw[] (2, 1.5) --++ (-0.75, 0) --++ (0, 4.5) --++ (0.75, 0);
    \draw[] (2, 1.5) -- ++ (0, 1.5);
    \draw[] (2, 6) -- ++ (0, -1.5);
    
    \node[above] () at (2,3) {$\langle \Psi|_R$};
    \node[below] () at (2, 4.5) {$|\Psi\rangle_R$};
    
    \draw[] (3,0) -- (3,4);

     \node[fill=blue, regular polygon, regular polygon sides=3, minimum size=0.35cm, inner sep=0pt] (u1) at (3, 0.25) {};
    \node[fill=blue, regular polygon, regular polygon sides=3, minimum size=0.35cm, inner sep=0pt] (u2) at (2, 2.5) {};
    \node[fill=red, regular polygon, regular polygon sides=3, minimum size=0.35cm, inner sep=0pt] (v1) at (3, 0.7) {};
    \node[fill=red, regular polygon, regular polygon sides=3, minimum size=0.35cm, inner sep=0pt, rotate=180] (v2) at (2, 2) {};
    
    \node[left] at (u2) {$u\phantom{l}$};
    \node[left] at (v2) {$v\phantom{l}$};
    
    \draw[blue, thick] (u1) -- ++ (0.75, 0) -- ++ (0,2.25) -- (u2);
    \draw[red, thick] (v1) -- ++ (0.5, 0) -- ++ (0, 1.3) --  (v2);
    
    \node[below] at (3,0) {$|\varphi\rangle_I$};
    \begin{scope}[yscale=-1, yshift=-7.5cm]
    
    \draw[] (3,0) -- (3,4);
    
    \node[fill=blue, regular polygon, regular polygon sides=3, minimum size=0.35cm, inner sep=0pt, rotate=180] (u1) at (3, 0.25) {};
    \node[fill=blue, regular polygon, regular polygon sides=3, minimum size=0.35cm, inner sep=0pt, rotate=180] (u2) at (2, 2.5) {};
    \node[fill=red, regular polygon, regular polygon sides=3, minimum size=0.35cm, inner sep=0pt, rotate=180] (v1) at (3, 0.7) {};
    \node[fill=red, regular polygon, regular polygon sides=3, minimum size=0.35cm, inner sep=0pt] (v2) at (2, 2) {};
    
    \node[left] at (u2) {$u^{\dagger}$};
    \node[left] at (v2) {$v^{\dagger}$};
    
    \draw[blue, thick] (u1) -- ++ (0.75, 0) -- ++ (0,2.25) -- (u2);
    \draw[red, thick] (v1) -- ++ (0.5, 0) -- ++ (0, 1.3) --  (v2);
    
    \node[above] at (3,0) {$\langle \varphi|_I$};
    \end{scope}
    \end{tikzpicture}.
    \end{aligned}
    \end{equation}
Straightening the blue and red legs, we get 
    \begin{equation}
    \begin{aligned}    \int d \widetilde{V}
    \left(    
    \begin{tikzpicture}[scale=0.8, baseline={([yshift=-.5ex]current bounding box.center)}]
    \draw[] (0,0) -- (0,1);
    \draw[] (2,0) -- (2,1);
    \draw[] (0, 1.5) -- (0, 4.25);
    \draw[] (2, 1.5) -- (2, 3.25);
    \draw[] (3,0) -- (3,1);
    \node[left] () at (0, 4.25) {$B$};
    \node[above] () at (2, 3.25) {$\langle \Psi|_R$};
    
    \draw[fill=red!10!white] (-0.25, 1) -- (3.25, 1) -- (3.25, 1.75) -- (-0.25, 1.75) -- cycle;

    \node[] at (1.5,1.375) {$\widetilde{V}$};
    \node[below] at (0,0) {$|\psi\rangle_{\ell}$};
    \node[below] at (2,0) {$| \psi_0\rangle_f$};
    
    \draw[] (3, 0) -- ++ (0.75, -0.5) -- ++ (0.75, 0.5);
    \draw[] (6, -0.5) -- ++ (0, 3.25);
    \draw[] (4.5, 0) -- ++ (0, 2.75); 
    \draw[] (4.5, 2.75) -- ++ (0.75, 0.5) -- ++ (0.75, -0.5);

    \node[below] () at (4.125,-0.5) {$|\text{MAX}\rangle_{I_1 I_2}$};
    \node[below] () at (6.125, -0.5) {$|\varphi\rangle_{I}$};
    
    \node[above] () at (5.3, 3.25) {$\langle \text{MAX}|_{I_2 I}$};
    
    \node[fill=blue, regular polygon, regular polygon sides=3, minimum size=0.35cm, inner sep=0pt] (u1) at (6, 2.375) {};
    \node[fill=blue, regular polygon, regular polygon sides=3, minimum size=0.35cm, inner sep=0pt] (u2) at (2, 2.375) {};
    \node[fill=red, regular polygon, regular polygon sides=3, minimum size=0.35cm, inner sep=0pt, rotate=180] (v1) at (4.5, 2) {};
    \node[fill=red, regular polygon, regular polygon sides=3, minimum size=0.35cm, inner sep=0pt, rotate=180] (v2) at (2, 2) {};
    
    \node[left] at (u2) {$u\phantom{l}$};
    \node[left] at (v2) {$v\phantom{l}$};
    
    \draw[blue, thick] (u1) -- (u2);
    \draw[red, thick] (v1) --  (v2);
    \begin{scope}[yshift = 8.5cm, yscale=-1]
    
    \draw[] (0,0) -- (0,1);
    \draw[] (2,0) -- (2,1);
    \draw[] (0, 1.5) -- (0, 4.25);
    \draw[] (2, 1.5) -- (2, 3.25);
    \draw[] (3,0) -- (3,1);
    \node[below] () at (2,3.25) {$|\Psi\rangle_R$};
    
    \draw[fill=red!10!white] (-0.25, 1) -- (3.25, 1) -- (3.25, 1.75) -- (-0.25, 1.75) -- cycle;

    \node[] at (1.5,1.375) {$\widetilde{V}^{\dagger}$};
    \node[above] at (0,0) {$\langle\psi|_{\ell}$};
    \node[above] at (2,0) {$\langle \psi_0|_f$};
    
    \draw[] (3, 0) -- ++ (0.75, -0.5) -- ++ (0.75, 0.5);
    \draw[] (6, -0.5) -- ++ (0, 3.25);
    \draw[] (4.5, 0) -- ++ (0, 2.75); 
    \draw[] (4.5, 2.75) -- ++ (0.75, 0.5) -- ++ (0.75, -0.5);
    
 \node[above] () at (4.125,-0.5) {$\langle\text{MAX}|_{I_1 I_2}$};
    \node[above] () at (6.125, -0.5) {$\langle \varphi|_{I}$};
    
    \node[below] () at (5.3, 3.25) {$| \text{MAX}\rangle_{I_2 I}$};
    \node[fill=blue, regular polygon, regular polygon sides=3, minimum size=0.35cm, inner sep=0pt, rotate=180] (u1) at (6, 2.375) {};
    \node[fill=blue, regular polygon, regular polygon sides=3, minimum size=0.35cm, inner sep=0pt, rotate=180] (u2) at (2, 2.375) {};
    \node[fill=red, regular polygon, regular polygon sides=3, minimum size=0.35cm, inner sep=0pt] (v1) at (4.5, 2) {};
    \node[fill=red, regular polygon, regular polygon sides=3, minimum size=0.35cm, inner sep=0pt] (v2) at (2, 2) {};
    
    \node[left] at (u2) {$u^{\dagger}$};
    \node[left] at (v2) {$v^{\dagger}$};
    
    \draw[blue, thick] (u1) -- (u2);
    \draw[red, thick] (v1) --  (v2);
    
    \end{scope}
    \end{tikzpicture}
    \right)
    = \frac{1}{|R| |I|^2}
        \begin{tikzpicture}[scale=0.8, baseline={([yshift=-.5ex]current bounding box.center)}]
    \draw[] (2, 1.5) --++ (-0.75, 0) --++ (0, 4.5) --++ (0.75, 0);
    \draw[] (2, 1.5) -- ++ (0, 1.5);
    \draw[] (2, 6) -- ++ (0, -1.5);
    
    \node[above] () at (2,3) {$\langle \Psi|_R$};
    \node[below] () at (2, 4.5) {$|\Psi\rangle_R$};
    
    \draw[] (3,1) -- (3,4);

     \node[fill=blue, regular polygon, regular polygon sides=3, minimum size=0.35cm, inner sep=0pt] (u1) at (3, 2) {};
    \node[fill=blue, regular polygon, regular polygon sides=3, minimum size=0.35cm, inner sep=0pt] (u2) at (2, 2) {};
    \node[fill=red, regular polygon, regular polygon sides=3, minimum size=0.35cm, inner sep=0pt] (v1) at (3, 2.5) {};
    \node[fill=red, regular polygon, regular polygon sides=3, minimum size=0.35cm, inner sep=0pt] (v2) at (1.25, 2.5) {};
    
    \node[left] at (u2) {$u\phantom{l}$};
    \node[left] at (v2) {$v\phantom{l}$};
    
    \draw[blue, thick] (u1) -- (u2);
    \draw[red, thick] (v1) --(v2);
    
    \node[below] at (3,1) {$|\varphi\rangle_I$};
    \begin{scope}[yscale=-1, yshift=-7.5cm]
    
    \draw[] (3,1) -- (3,4);
    
   \node[fill=blue, regular polygon, regular polygon sides=3, minimum size=0.35cm, inner sep=0pt, rotate=180] (u1) at (3, 2) {};
    \node[fill=blue, regular polygon, regular polygon sides=3, minimum size=0.35cm, inner sep=0pt, rotate=180] (u2) at (2, 2) {};
    \node[fill=red, regular polygon, regular polygon sides=3, minimum size=0.35cm, inner sep=0pt, rotate=180] (v1) at (3, 2.5) {};
    \node[fill=red, regular polygon, regular polygon sides=3, minimum size=0.35cm, inner sep=0pt, rotate=180] (v2) at (1.25, 2.5) {};
    
    \node[left] at (u2) {$u^{\dagger}$};
    \node[left] at (v2) {$v^{\dagger}$};
    
    \draw[blue, thick] (u1) -- (u2);
    \draw[red, thick] (v1) --  (v2);
    
    \node[above] at (3,1) {$\langle \varphi|_I$};
    \end{scope}
    \end{tikzpicture}
    = \frac{1}{|I|^2}
        \begin{tikzpicture}[scale=0.8, baseline={([yshift=-.5ex]current bounding box.center)}]
    \draw[] (2, 1.5) --++ (-0.75, -0.25) -- ++ (-0.75, 0.25) --++ (0, 4.5) --++ (0.75, 0.25) -- ++ (0.75, -0.25);
    \draw[] (2, 1.5) -- ++ (0, 1.5);
    \draw[] (2, 6) -- ++ (0, -1.5);
    
    \node[above] () at (2,3) {$\langle \Psi|_R$};
    \node[below] () at (2, 4.5) {$|\Psi\rangle_R$};
    \node[below] () at (1.25, 1.25) {$|\textrm{MAX}\rangle_{R'R}$};
    \node[above] () at (1.25, 6.25) {$\langle \textrm{MAX}|_{R'R}$};
    
    \draw[] (3,1) -- (3,4);

     \node[fill=blue, regular polygon, regular polygon sides=3, minimum size=0.35cm, inner sep=0pt] (u1) at (3, 2) {};
    \node[fill=blue, regular polygon, regular polygon sides=3, minimum size=0.35cm, inner sep=0pt] (u2) at (2, 2) {};
    \node[fill=red, regular polygon, regular polygon sides=3, minimum size=0.35cm, inner sep=0pt] (v1) at (3, 2.5) {};
    \node[fill=red, regular polygon, regular polygon sides=3, minimum size=0.35cm, inner sep=0pt] (v2) at (0.5, 2.5) {};
    
    \node[left] at (u2) {$u\phantom{l}$};
    \node[left] at (v2) {$v\phantom{l}$};
    
    \draw[blue, thick] (u1) -- (u2);
    \draw[red, thick] (v1) --(v2);
    
    \node[below] at (3,1) {$|\varphi\rangle_I$};
    \begin{scope}[yscale=-1, yshift=-7.5cm]
    
    \draw[] (3,1) -- (3,4);
    
   \node[fill=blue, regular polygon, regular polygon sides=3, minimum size=0.35cm, inner sep=0pt, rotate=180] (u1) at (3, 2) {};
    \node[fill=blue, regular polygon, regular polygon sides=3, minimum size=0.35cm, inner sep=0pt, rotate=180] (u2) at (2, 2) {};
    \node[fill=red, regular polygon, regular polygon sides=3, minimum size=0.35cm, inner sep=0pt, rotate=180] (v1) at (3, 2.5) {};
    \node[fill=red, regular polygon, regular polygon sides=3, minimum size=0.35cm, inner sep=0pt, rotate=180] (v2) at (0.5, 2.5) {};
    
    \node[left] at (u2) {$u^{\dagger}$};
    \node[left] at (v2) {$v^{\dagger}$};
    
    \draw[blue, thick] (u1) -- (u2);
    \draw[red, thick] (v1) --  (v2);
    
    \node[above] at (3,1) {$\langle \varphi|_I$};
    \end{scope}
    \end{tikzpicture},
    \end{aligned}
\end{equation}
where $R'$ is an auxiliary subsystem whose dimension is equal to the dimension of $R$. Therefore, we conclude 
\begin{equation}
        \begin{tikzpicture}[scale=0.8, baseline={([yshift=-.5ex]current bounding box.center)}]
    \draw[] (0,0) -- (0,1);
    \draw[] (2,0) -- (2,1);
    \draw[] (0, 1.5) -- (0, 3.75);
    \draw[] (2, 1.5) -- (2, 3);
    \draw[] (3,0) -- (3,1);
    
    \draw[fill=red!10!white] (-0.25, 1) -- (3.25, 1) -- (3.25, 1.75) -- (-0.25, 1.75) -- cycle;
    
     \node[fill=blue, regular polygon, regular polygon sides=3, minimum size=0.35cm, inner sep=0pt] (u1) at (3, 0.25) {};
    \node[fill=blue, regular polygon, regular polygon sides=3, minimum size=0.35cm, inner sep=0pt] (u2) at (2, 2.5) {};
    \node[fill=red, regular polygon, regular polygon sides=3, minimum size=0.35cm, inner sep=0pt] (v1) at (3, 0.7) {};
    \node[fill=red, regular polygon, regular polygon sides=3, minimum size=0.35cm, inner sep=0pt, rotate=180] (v2) at (2, 2) {};
    
    \node[left] at (u2) {$u\phantom{l}$};
    \node[left] at (v2) {$v\phantom{l}$};
    
    \node[above] () at (2, 3) {$\langle \Psi|_R$};
    
    \draw[blue, thick] (u1) -- ++ (0.75, 0) -- ++ (0,2.25) -- (u2);
    \draw[red, thick] (v1) -- ++ (0.5, 0) -- ++ (0, 1.3) --  (v2);
    
    \node[] at (1.5,1.375) {$U'$};
    \node[below] at (0,0) {$|\psi\rangle_{\ell}$};
    \node[below] at (2,0) {$|\psi_0\rangle_f$};
    \node[below] at (3,0) {$|\varphi\rangle_I$};
    \begin{scope}[yscale=-1, yshift=-7.5cm]
    
    \draw[] (0,0) -- (0,1);
    \draw[] (2,0) -- (2,1);
    \draw[] (0, 1.5) -- (0, 3.75);
    \draw[] (2, 1.5) -- (2, 3);
    \draw[] (3,0) -- (3,1);
    
    \node[below] () at (2, 3) {$| \Psi\rangle_R$};
    
    \draw[fill=red!10!white] (-0.25, 1) -- (3.25, 1) -- (3.25, 1.75) -- (-0.25, 1.75) -- cycle;
    
    \node[fill=blue, regular polygon, regular polygon sides=3, minimum size=0.35cm, inner sep=0pt, rotate=180] (u1) at (3, 0.25) {};
    \node[fill=blue, regular polygon, regular polygon sides=3, minimum size=0.35cm, inner sep=0pt, rotate=180] (u2) at (2, 2.5) {};
    \node[fill=red, regular polygon, regular polygon sides=3, minimum size=0.35cm, inner sep=0pt, rotate=180] (v1) at (3, 0.7) {};
    \node[fill=red, regular polygon, regular polygon sides=3, minimum size=0.35cm, inner sep=0pt] (v2) at (2, 2) {};
    
    \node[left] at (u2) {$u^{\dagger}$};
    \node[left] at (v2) {$v^{\dagger}$};
    
    \draw[blue, thick] (u1) -- ++ (0.75, 0) -- ++ (0,2.25) -- (u2);
    \draw[red, thick] (v1) -- ++ (0.5, 0) -- ++ (0, 1.3) --  (v2);
    
    \node[] at (1.5,1.375) {${U'}^{\dagger}$};
    \node[above] at (0,0) {$\langle\psi|_{\ell}$};
    \node[above] at (2,0) {$\langle\psi_0|_f$};
    \node[above] at (3,0) {$\langle\varphi|_I$};
    \end{scope}
    \end{tikzpicture} \,\,\, \approx  \,\,\,
        \begin{tikzpicture}[scale=0.8, baseline={([yshift=-.5ex]current bounding box.center)}]
    \draw[] (2, 1.5) --++ (-0.75, -0.25) -- ++ (-0.75, 0.25) --++ (0, 4.5) --++ (0.75, 0.25) -- ++ (0.75, -0.25);
    \draw[] (2, 1.5) -- ++ (0, 1.5);
    \draw[] (2, 6) -- ++ (0, -1.5);
    
    \node[above] () at (2,3) {$\langle \Psi|_R$};
    \node[below] () at (2, 4.5) {$|\Psi\rangle_R$};
    \node[below] () at (1.25, 1.25) {$|\textrm{MAX}\rangle_{R'R}$};
    \node[above] () at (1.25, 6.25) {$\langle \textrm{MAX}|_{R'R}$};
    
    \draw[] (3,1) -- (3,4);

     \node[fill=blue, regular polygon, regular polygon sides=3, minimum size=0.35cm, inner sep=0pt] (u1) at (3, 2) {};
    \node[fill=blue, regular polygon, regular polygon sides=3, minimum size=0.35cm, inner sep=0pt] (u2) at (2, 2) {};
    \node[fill=red, regular polygon, regular polygon sides=3, minimum size=0.35cm, inner sep=0pt] (v1) at (3, 2.5) {};
    \node[fill=red, regular polygon, regular polygon sides=3, minimum size=0.35cm, inner sep=0pt] (v2) at (0.5, 2.5) {};
    
    \node[left] at (u2) {$u\phantom{l}$};
    \node[left] at (v2) {$v\phantom{l}$};
    
    \draw[blue, thick] (u1) -- (u2);
    \draw[red, thick] (v1) --(v2);
    
    \node[below] at (3,1) {$|\varphi\rangle_I$};
    \begin{scope}[yscale=-1, yshift=-7.5cm]
    
    \draw[] (3,1) -- (3,4);
    
   \node[fill=blue, regular polygon, regular polygon sides=3, minimum size=0.35cm, inner sep=0pt, rotate=180] (u1) at (3, 2) {};
    \node[fill=blue, regular polygon, regular polygon sides=3, minimum size=0.35cm, inner sep=0pt, rotate=180] (u2) at (2, 2) {};
    \node[fill=red, regular polygon, regular polygon sides=3, minimum size=0.35cm, inner sep=0pt, rotate=180] (v1) at (3, 2.5) {};
    \node[fill=red, regular polygon, regular polygon sides=3, minimum size=0.35cm, inner sep=0pt, rotate=180] (v2) at (0.5, 2.5) {};
    
    \node[left] at (u2) {$u^{\dagger}$};
    \node[left] at (v2) {$v^{\dagger}$};
    
    \draw[blue, thick] (u1) -- (u2);
    \draw[red, thick] (v1) --  (v2);
    
    \node[above] at (3,1) {$\langle \varphi|_I$};
    \end{scope}
    \end{tikzpicture}, \label{eq:main_result_black_hole_assisted}
\end{equation}
with an approximation error bounded by $|I|^2\epsilon(|B|)$. Eq.~\eqref{eq:main_result_black_hole_assisted} is the key result of this Appendix. If $U'$ is pseudorandom, then the expression on the left-hand side is equal to expression on the right-hand side, up to an error bounded above by $|I|^2 \epsilon(|B|)$. 

Recall now the interpretation of these two diagrams. In the case of the diagram on the left, we envision a black-hole-assisted quantum computation which is initialized by preparing the state $|\varphi\rangle_I$ of the infalling robot $I$. The robot interacts with radiation both outside and inside the black hole, where both interactions have a computational complexity which is polynomial in $\log |B|$, the size of the black hole. Then the radiation system $R$ is measured and the outcome $|\Psi\rangle_R$ is obtained; this measurement also has complexity polynomial in $\log |B|$. The diagram represents the probability for this particular outcome of the measurement of $R$. Because of the postselection inherent in the holographic map, the computational complexity of the overall process is not immediately apparent. 

Now consider the diagram on the right. Again a quantum computation is initialized by preparing $|\varphi\rangle_I$. In addition, the maximally entangled state $|\textrm{MAX}\rangle_{R'R}$ is prepared. Then $I$ interacts with $R'R$, where these interactions have a computational complexity polynomial in $\log |B|$. Finally, system $R'$ is discarded, and a low-complexity measurement is performed on $R$, yielding the outcome $|\Psi\rangle_R$. The diagram represents the probability for this particular measurement outcome. In this case, it is manifest that the entire process has computational complexity which is polynomial in $\log |B|$.

We conclude that if $U'$ is pseudorandom, and the error $|I|^2 \epsilon(|B|)$ is small, then the nonunitary postselected quantum computation on the left can be accurately simulated by the efficient unitary quantum computation on the right. The error is exponentially small in $\log |B|$ if $\epsilon(|B|)= e^{-\alpha\log|B|}$ and $|I|= e^{\beta\log |B|}$ where $\beta$ is strictly less than $\alpha/2$. Under these assumptions, the postselection in the holographic map does not result in superpolynomial computational power. 

\end{document}